\documentclass[12pt,a4paper]{article}

\usepackage{epsfig,graphicx,bbm}
\usepackage{amsmath,amsfonts,amssymb}
\usepackage{citesort}
 \usepackage[footnotesize]{caption}

\newcommand{\unit}{\leavevmode\hbox{\small1\kern-3.6pt\normalsize1}}

\begin{document}

\begin{flushright}
FTUAM-07/10\\
IFT-UAM/CSIC-07-28\\
LPT-ORSAY-07-51\\

\vspace*{3mm}
{\today}
\end{flushright}

\vspace*{5mm}
\begin{center}
{\Large \textbf{$\mu \,-\, e$ conversion in nuclei within the CMSSM seesaw:
    universality versus non-universality } }

\vspace{0.5cm} 
{\large
E.~Arganda\,$^{a}$, M.~J.~Herrero\,$^{a}$ and A.~M.~Teixeira\,$^{b}$}\\[0.4cm]

{$^{a}$\textit{Departamento de F\'{\i }sica Te\'{o}rica C-XI 
and Instituto de F\'{\i }sica Te\'{o}rica C-XVI, \\[0.1cm]
Universidad Aut\'{o}noma de Madrid,
Cantoblanco, E-28049 Madrid, Spain}}\\[0.3cm]

{$^{b}$\textit {Laboratoire de Physique Th\'eorique, UMR 8627}}\\[0.1cm]

{\textit{Universit\'e de
      Paris-Sud XI, B\^atiment 201, F-91405 Orsay Cedex, France}}\\[0pt] 
\vspace*{0.3cm} 

\begin{abstract}
In this paper we study $\mu-e$ conversion in nuclei within the context of 
the Constrained Minimal Supersymmetric Standard Model, enlarged by three right
handed neutrinos and their supersymmetric partners, and where the neutrino
masses are generated via a seesaw mechanism. Two different scenarios with either 
universal or non-universal 
soft supersymmetry breaking Higgs masses at the
gauge coupling unification scale are considered. In the first part we present a complete one-loop computation of the
conversion rate for this process that includes the photon-, $Z$-boson, and Higgs-boson
penguins, as well as box diagrams, and compare their size in
the two considered scenarios. Then, in these two scenarios we analyse the relevance of the
various parameters on the conversion rates, particularly emphasising
the role played by the  
heavy neutrino masses, 
$\tan \beta$, and especially $\theta_{13}$. In the case of hierachical heavy neutrinos, 
an extremely high
sensitivity of the rates to $\theta_{13}$ is indeed found.
The last part of this work is devoted to the study of the interesting loss 
of correlation between
the $\mu-e$ conversion and $\mu \to e \gamma$ rates that occurs in the
non-universal scenario. In the case of large $\tan \beta$ and light $H^0$ Higgs boson,
an enhanced ratio of the $\mu-e$ to $\mu \to e \gamma$ rates, with respect to the 
universal case is found, and this could be tested with the future experimental sensitivities.       
   
\end{abstract}

\end{center}

\newpage 
\section{Introduction}\label{intro}

Neutrino physics, in particular neutrino oscillations and 
the measured neutrino mass differences, strongly manifest that 
Nature does not conserve the lepton flavour quantum
number in the neutrino sector~\cite{neutrinodata,Yao:2006px}. 
However, it is not known yet if lepton
flavour violation (LFV) 
also occurs in the charged lepton sector. If such is the case, one
still has to address if LFV in the neutral and charged lepton sectors
arises from a common or different origin.
It is well known that if the Standard Model
of Particle Physics (SM) is minimally extended in order to accommodate 
the present data on neutrino masses and mixings,
the corresponding loop induced LFV in the charged lepton sector 
(exclusively induced from neutrino oscillations) 
is extremely tiny and hopeless to be experimentally observed.
Therefore, a potential future measurement of LFV
in the charged lepton sector will provide a unique insight into the 
nature of new physics beyond the SM (for a review, see~\cite{Kuno:1999jp}).

Among the various candidates for physics beyond the SM that produce 
potentially observable effects in LFV processes, 
one of the most appealing are Supersymmetric (SUSY) 
extensions of the SM, where a seesaw mechanism~\cite{seesaw:I,seesaw:II} is implemented to 
generate neutrino masses.
In these SUSY-seesaw models a new source of LFV appears in the
off-diagonal elements of the slepton mass matrices, which can be 
radiatively 
generated. The size of these elements is governed by the
strength of the neutrino Yukawa couplings and, in the case of Majorana
neutrinos, the latter
can be large, of the order of one.  
The LFV effects in the charged lepton processes are then induced by 
flavour violating slepton-lepton interactions, appearing in 
SUSY-loop diagrams mediated by sleptons~\cite{Borzumati:1986qx}.

Concerning the LFV processes, in our work we will focus on those
which involve flavour transitions between the first and second
generation of charged leptons. At present, the most relevant $\mu-e$
flavour violating processes are $\mu \to e
\gamma$, $\mu \to 3 e$ and $\mu-e$ conversion in nuclei. The current
experimental bounds on the muon decays are BR$(\mu \to e \gamma)<  1.2
\times 10^{-11}$~\cite{Brooks:1999pu} and BR$(\mu \to 3 e)<  1.0\times
10^{-12}$~\cite{Bellgardt:1987du}. Regarding $\mu-e$ conversion 
in heavy nuclei, the most stringent constraints arise for Titanium and
Gold, respectively with 
CR($\mu - e$, Ti)$ < 4.3  \times 10^{-12}$~\cite{Dohmen:1993mp} and  
CR($\mu - e$, Au)$ < 7 \times 10^{-13}$~\cite{Bertl:2001fu}.
In the future, one expects significant improvements in the sentitivies 
to these LFV rates. For instance, MEG aims at reaching a
sensitivity for $\mu \to e \gamma$ of $10^{-13}$~\cite{Ritt:2006cg} in
the very near future, which could further be improved to 
$10^{-14}$ in the next 4-5 years~\cite{Ritt:private}.
Although the situation for BR$(\mu \to 3 e)$ is
less certain, one does not expect the sensitivities to better 
$10^{-13}-10^{-14}$~\cite{Ritt:private}.
Undoubtedly, the most challenging prospects concern the experimental
sensitivities to $\mu-e$ conversion in Titanium nuclei. The 
dedicated J-PARC experiment
PRISM/PRIME has anounced a remarkable improvement, albeit in a  
farer future, of $10^{-18}$~\cite{PRIME}.  

In this paper we will focus on $\mu-e$ conversion in nuclei, working
in the context of the Minimal Supersymmetric Standard Model (MSSM)
enlarged by  three right handed neutrinos and their corresponding SUSY 
partners, where a type-I seesaw mechanism~\cite{seesaw:I} is implemented.
 To reduce the number of unknown parameters in the SUSY
sector, we choose to work in the so-called constrained MSSM
(CMSSM)(for a review see for instance~\cite{Kane:1993td}), assuming universality of the soft-SUSY breaking parameters at
the scale of gauge coupling unification, 
$M_X \sim 2 \times 10^{16}$ GeV. An interesting departure from the
CMSSM-seesaw can be obtained by relaxing the universality hypothesis 
for the soft SUSY breaking masses of the Higgs sector. This partially
constrained MSSM is commonly referred to as the Non Universal Higgs Mass
(NUHM) scenario~\cite{NUHMrefs}, and its enlarged version (including 
right handed neutrinos and sneutrinos) will be here designated  
NUHM-seesaw.

Within the context of the  CMSSM- and NUHM-seesaw, we conduct here a 
thourough
analysis of the predictions for the  $\mu-e$ conversion
rates in nuclei. The present computation is the first
to include
the {\it full} set of SUSY one-loop diagrams (photon, $Z$- and
Higgs-boson mediated, as well as box diagrams), and to be strictly
done in terms of physical eigenstates for the exchanged SUSY
particles.
For all scenarios here addressed, we obtain the low-energy parameters 
by numerically solving the full renormalisation 
group equations (RGEs), including the neutrino and sneutrino sectors.
The photon and $Z$-boson mediated penguins,
as well as the vector contributions from box diagrams were first computed 
in the CMSSM-seesaw
in~\cite{Hisano:1995cp}. Here we have confirmed their analytical results  
for the photon-mediated and box contributions, correcting the 
analytical expressions for the $Z$-boson mediated processes.
We further added the scalar contributions from box diagrams and the Higgs-mediated 
contributions, and improved 
the computation, by considering in the numerical analysis the
possibility of either degenerate or hierarchical heavy neutrino
spectrum, and by fitting the light neutrino parameters to the present
data. 

The effects of the Higgs-mediated contribution on $\mu-e$ conversion
rates were firstly investigated in~\cite{Kitano:2003wn}, in the
context of a SUSY-seesaw with degenerate heavy neutrinos, working in 
the effective Lagrangian approximation and in the $SU(2)_L \times
U(1)_Y$ limit. It was observed that the Higgs-mediated LFV diagrams 
could provide the dominant contribution for the large 
$\tan \beta$ regime and for small masses of the heavy Higgs
scalar, owing to a $\tan^6{\beta}$ enhancement and $(m_{H^0})^{-4}$ dependence of
the conversion rates. By comparing the latter with the corresponding
$\mu \to e \gamma$ rates for large universal SUSY-breaking mass values, 
$M_0,M_{1/2} \sim \mathcal{O}(1$ TeV), they further 
showed that the ratio of observables CR$(\mu-e,$ Al)/BR$(\mu \to e 
\gamma)$
could be enhanced from a value of $\mathcal{O}(\alpha)$ (within the usual dominant 
photon-mediated case) to $\mathcal{O}(10^{-1})$ for extreme values of 
$\tan \beta =60$, $M_R= 10^{14}$ GeV and $m_{H^0} \sim 100$ GeV.   

In the present work we will explore in full detail the various
contributions to the $\mu-e$ conversion rates and study the dependence
on all parameters entering in the considered MSSM-seesaw framework. 
In addition to the relevant role played by 
the mass of the right handed neutrinos, $m_{N_i}$,
the soft masses $M_0$, $M_{1/2}$ (and $M_{H_{1,2}}$ for the
NUHM-seesaw), and  $\tan \beta$, we will show that 
the light neutrino mixing angle $\theta_{13}$ has an important impact on
CR$(\mu-e, \text{Nucleus}$). The conversion rates 
turn out to be very sensitive to this angle, 
varying in many orders of magnitude
(up to five in the case of hierarchical heavy neutrinos) for
$\theta_{13}$ values within the present experimentally allowed region 
$0^\circ\leq \theta_{13} \leq 10^\circ$~\cite{neutrinodata_fits}. 
We will further verify that with the future sensitivity of JPARC 
($\mathcal{O}(10^{-18})$)~\cite{PRIME} most of the parameter space could be
covered. 

On the other hand, the comparison between the predictions obtained for the 
CMSSM-seesaw and the NUHM-seesaw cases will allow us to 
draw interesting conclusions about the departure from the strongly
correlated behaviour of CR$(\mu-e, \text{Nucleus})$ and BR$(\mu \to e \gamma)$, as
predicted in photon-dominated scenarios (as is the case of the 
CMSSM-seesaw).
In the latter scenario, the ratio of the two rates was found 
to be at most $1/160$ for $\tan \beta =50$~\cite{Yaguna:2005qn}.
In contrast, we will discuss here particular scenarios in the
NUHM-seesaw, where the ratio CR$(\mu-e$, Ti)/BR$(\mu \to e \gamma)$ is 
indeed enhanced with respect to the universal case, by as much as one
order of magnitude, in agreement with the approximate 
results of~\cite{Kitano:2003wn}. 

One of the most challenging tasks in this $\mu-e$ conversion process 
will be
to disantangle between the different scenarios for new physics if 
a measurement is finally obtained. Indeed, it has been already noticed in
early works~\cite{stringent_mue} that $\mu-e$ conversion could
constrain new physics more stringently than $\mu \to e \gamma$. We will see
here that this is the case in the NUHM scenario. Furthermore, we will also show that, with the expected sensitivities for Titanium of $\mathcal{O}(10^{-18}$), one could distinguish
CMSSM- from NUHM- seesaw scenarios by extracting the scalar contribution to the 
CR$(\mu-e$, Ti) rates.

The paper is organised as follows. In Section~\ref{susy:seesaw} we review the
most relevant features of the SUSY-seesaw scenario. The analytical results of
the $\mu - e$ conversion rates are presented in
Section~\ref{CR:prediction}. Section~\ref{results} is devoted to the numerical
results for both CMSSM-seesaw and NUHM-seesaw scenarios. An extensive
discussion about the sensitivities to the various parameters in
these two scenarios is also included. Finally, the conclussions are summarised
in Section~\ref{concs}.
     
\section{The SUSY-seesaw scenario}\label{susy:seesaw}

The leptonic superpotential containing the relevant terms to describe a 
type-I SUSY seesaw is given by
\begin{equation}\label{W:Hl:def}
W\,=\,\hat N^c\,Y_\nu\,\hat L \, \hat H_2 \,+\,
\hat E^c\,Y_l\,\hat L \, \hat H_1 \,+\,
\frac{1}{2}\,\hat N^c\,m_M\,\hat N^c\,,
\end{equation}
where $\hat N^c$ are the additional superfields that contain the three
right-handed  
neutrinos $\nu_{R_i}$ and their scalar partners $\tilde \nu_{R_i}$.
The lepton Yukawa couplings $Y_{l,\nu}$ and the
Majorana mass $m_M$ are $3\times 3$ matrices in lepton flavour
space. From now on, we will assume that we are in a basis where 
$Y_l$ and $m_M$ are diagonal.

After electroweak (EW) symmetry breaking, the charged lepton and 
Dirac neutrino mass matrices
can be written as
\begin{equation}
m_l\,=\,Y_l\,\,v_1\,, \quad \quad
m_D\,=\,Y_\nu\,v_2\,,
\end{equation}
where $v_i$ are the vacuum expectation values (VEVs) of the neutral Higgs
scalars, with $v_{1(2)}= \,v\,\cos (\sin) \beta$ and $v=174$
GeV.

The $6\times 6$ neutrino mass matrix is given by
\begin{equation}\label{seesaw:def}
M^\nu\,=\,\left(
\begin{array}{cc}
0 & m_D^T \\
m_D & m_M
\end{array} \right)\,. 
\end{equation}
The eigenvalues of $M^\nu$ are the masses of the six physical 
Majorana neutrinos. In the seesaw limit, the three right-handed 
masses are much heavier than the EW scale,  
$m_{M_i}\,\gg\,v$, and one obtains three light and three heavy
states, $\nu_i$ and $N_i$, respectively. 

Block-diagonalisation of the neutrino mass matrix of
Eq.~(\ref{seesaw:def}), leads (at lowest order in the $(m_D/m_M)^n$
expansion) to the standard seesaw equation 
for the light neutrino mass matrix,
\begin{equation}\label{seesaw:light}
m_\nu\,=\, - m_D^T m_M^{-1} m_D  \,. 
\end{equation}
Since we are working in a basis where $m_M$ is diagonal, the heavy
eigenstates are then given by 
\begin{equation}\label{def:Ndiag}
m_N^\text{diag}\,=\,m_M\,=\, \text{diag}\,(m_{N_1},m_{N_2},m_{N_3})\,.
\end{equation}
The matrix $m_\nu$ can be diagonalised by the
Maki-Nakagawa-Sakata unitary matrix
$U_{\text{MNS}}$~\cite{Maki:1962mu,Pontecorvo:1957cp}, leading
to the following masses for the light physical states
\begin{align}\label{physicalmasses}
m_{\nu}^\text{diag}&
\,=\,U_\text{MNS}^T \,m_{\nu}\, U_\text{MNS} 
\,=\, \text{diag}\,(m_{\nu_1},m_{\nu_2},m_{\nu_3})\,.
\end{align}
Here we use the standard parameterisation for $U_\mathrm{MNS}$ 
given by
\begin{equation}
U_\text{MNS}=
\left( 
\begin{array}{ccc} 
c_{12} \,c_{13} & s_{12} \,c_{13} & s_{13} \, e^{-i \delta} \\ 
-s_{12}\, c_{23}\,-\,c_{12}\,s_{23}\,s_{13}\,e^{i \delta} 
& c_{12} \,c_{23}\,-\,s_{12}\,s_{23}\,s_{13}\,e^{i \delta} 
& s_{23}\,c_{13} \\ 
s_{12}\, s_{23}\,-\,c_{12}\,c_{23}\,s_{13}\,e^{i \delta} 
& -c_{12}\, s_{23}\,-\,s_{12}\,c_{23}\,s_{13}\,e^{i \delta} 
& c_{23}\,c_{13}
\end{array} \right) \,.\, V\,,
\label{Umns}
\end{equation}
with
\begin{equation}
 V\,=\,\text{diag}\,(e^{-i\frac{\phi_1}{2}},e^{-i\frac{\phi_2}{2}},1)\,,
\end{equation}
where $c_{ij} \equiv \cos \theta_{ij}$, $s_{ij} \equiv \sin \theta_{ij}$.
$\theta_{ij}$ are the neutrino flavour mixing angles, $\delta$ is the Dirac
phase and $\phi_{1,2}$ are the Majorana phases. 

In view of the above, 
the seesaw equation (\ref{seesaw:light}) can be solved for $m_D$
as~\cite{Casas:2001sr} 
\begin{equation}\label{seesaw:casas}
m_D\,=\, i \sqrt{m^\text{diag}_N}\, R \,
\sqrt{m^\text{diag}_\nu}\,  U^\dagger_{\text{MNS}}\,,
\end{equation}
where $R$ is a generic complex orthogonal $3 \times 3$ matrix that
encodes the possible extra neutrino mixings (associated with the
right-handed sector) in addition to the ones in
$U_{\text{MNS}}$. $R$ can be parameterised 
in terms of three complex angles, $\theta_i$ $(i=1,2,3)$ as~\cite{Casas:2001sr}
\begin{equation}\label{Rcasas}
R\, =\, 
\left( 
\begin{array}{ccc} 
c_{2}\, c_{3} & -c_{1}\, s_{3}\,-\,s_1\, s_2\, c_3
& s_{1}\, s_3\,-\, c_1\, s_2\, c_3 \\ 
c_{2}\, s_{3} & c_{1}\, c_{3}\,-\,s_{1}\,s_{2}\,s_{3} 
& -s_{1}\,c_{3}\,-\,c_1\, s_2\, s_3 \\ 
s_{2}  & s_{1}\, c_{2} & c_{1}\,c_{2}
\end{array} 
\right)\,,
\end{equation}
with $c_i\equiv \cos \theta_i$, $s_i\equiv \sin\theta_i$. 
Eq.~(\ref{seesaw:casas}) is a convenient means of parameterising our ignorance 
of the full neutrino Yukawa couplings, while at the same time allowing
to accommodate the experimental data.
Notice that it is only valid at the 
right-handed neutrino scales $m_M$, so that the quantities appearing
in Eq.~(\ref{seesaw:casas}) are the renormalised ones, 
$m^\text{diag}_\nu\,(m_M)$ and $U_{\text{MNS}}\,(m_M)$.

We shall focus on the scenario where 
the light neutrinos are hierarchical, and we will 
assume a normal hierarchy,
\begin{align}
&m_{\nu_1}\, \ll\, m_{\nu_2}\, \ll\, m_{\nu_3}\,. 
\end{align}  
The masses $m_{\nu_{2,3}}$ 
can be written in terms of the lightest mass $m_{\nu_{1}}$, and of  
the solar and atmospheric mass-squared differences as
\begin{align}
& m_{\nu_2}^2\,=\, \Delta m_\text{sol}^2 \, + \, m_{\nu_1}^2\,, \nonumber\\  
& m_{\nu_3}^2\,=\, \Delta m_\text{atm}^2 \, + \, m_{\nu_1}^2\,.
\end{align}

Regarding the heavy neutrinos, we will consider the two following cases,  
\begin{align}
\text{Degenerate:}\  &m_{N_1}=m_{N_2}=m_{N_3} \equiv m_N \,,\nonumber \\
\text{Hierarchical:}\  &m_{N_1}\, \ll\, m_{N_2}\, \ll \,m_{N_3}\,. \nonumber 
\end{align}

Concerning the SUSY parameters, and since we are working within an
extended MSSM, with enlarged neutrino and sneutrino sectors,
there will be new soft SUSY breaking parameters associated to the
latter sectors.
Thus, in addition to the usual soft breaking parameters for the
gauginos ($M_{1,2,3}$), Higgs bosons ($M_{H_{1,2}}$), squarks ($m_{\tilde Q}$, 
$m_{\tilde U}$, $m_{\tilde D}$, $A_q$) 
and sleptons ($m_{\tilde L}$, $m_{\tilde E}$, $A_l$), there will also be
the sneutrino soft breaking masses $m_{\tilde M}$, the sneutrino trilinear
couplings $A_\nu$, and the new bilinear parameter $B_M$.
As already mentioned in the introduction, we will work in a constrained MSSM,
where the number of input parameters is reduced by assuming partial universality of the soft parameters at the gauge
coupling unification scale, $M_X= 2 \times 10^{16}$ GeV. Specifically, we 
will work in two scenarios, the CMSSM-seesaw with universal scalar masses, 
trilinear couplings and gaugino masses, and the NUHM-seesaw 
with non-universal soft masses for the Higgs bosons. 
Therefore, when specifying the parameters of these two
constrained MSSM scenarios we will fix, in addition to the seesaw
parameters, 
the following soft SUSY breaking parameters at the scale $M_X$:

\begin{align}
\text{CMSSM-seesaw:} \ & M_0\,, M_{1/2}\,,A_0\,, \tan \beta\,, 
\text{sign}(\mu)\,,
\nonumber \\
\text{NUHM-seesaw:}\ & M_0\,, M_{1/2}\,,A_0 \,, \tan \beta\, ,
\text{sign}(\mu)\,,M_{H_1}, M_{H_2}.
\end{align} 
The departure from universality in the NUHM-seesaw  
will be parameterised 
in terms of the non-vanishing parameters $\delta_1$ and $\delta_2$, 
\begin{align}
\text{Non-universality:} \ &
M^2_{H_1}\,=\,M^2_0\,(1+\delta_1)\,,\ \ M^2_{H_2}\,=\,M^2_0\,(1+\delta_2)\,.
\end{align}
For simplicity, and to further reduce the number of input parameters,
in this case we will also impose $M_0= M_{1/2} \equiv M_{\rm SUSY}$.

Once the above set of parameters is fixed at $M_X$,  
the predictions for the low-energy parameters are
obtained by solving the full one-loop RGEs, 
including the extended neutrino and sneutrino sectors. 
Due to the existence of intermediate scales $m_M$ introduced by the seesaw
mechanism, the running must be carried out in two steps.
The full set of equations is first run down from $M_X$ to
$m_M$. At the seesaw scales, 
the right-handed neutrinos as well as their SUSY partners
decouple, and the new RGEs (without the equations and terms for 
$\nu_R$ and $\tilde \nu_R$) are then run down from $m_M$ to the EW
scale, where the couplings and mass matrices are finally computed.

Working in constrained MSSM scenarios, all 
flavour mixing originates solely from the neutrino Yukawa couplings,
which induce flavour violation 
in the slepton sector by the RGE running from
$M_X$ down to the EW scale $m_Z$.
Flavour mixing is then manifest in
the values of the off-diagonal elements of the charged slepton squared
mass matrix.
The $LL$, $RR$, $LR$ and $RL$ elements of the latter
$M_{\tilde{l}}^2$ matrix can be summarised as follows:
\begin{align}
M_{LL}^{ij \, 2} & \,=  \,
m_{\tilde{L}, ij}^2 \, + \, v_1^2  \,\left( Y_l^{\dagger} \, Y_l 
\right)_{ij} \, + \, 
m_Z^2  \,\cos 2 \beta \, \left(-\frac{1}{2} \,+ \, \sin^2 \theta_{W}
\right)  \,  \delta_{ij} \,, \nonumber \\
M_{RR}^{ij \, 2} & \,=  \,
m_{\tilde{E}, ij}^2 \, + \, v_1^2 \, \left( Y_l^{\dagger} \, Y_l
\right)_{ij} \, -  \, 
m_Z^2 \, \cos 2 \beta  \,\sin^2 \theta_{W}  \,\delta_{ij} \,, \nonumber \\
M_{LR}^{ij \, 2} & \,=  \,
v_1  \,\left(A_l^{ij}\right)^{*}  \,- \,\mu \, Y_l^{ij} \, v_2 \,, \nonumber \\
M_{RL}^{ij \, 2} & \,=  \,\left(M_{LR}^{ji \, 2}\right)^{*} \, .
\end{align}
In the above, $m_Z$ denotes the $Z$-boson mass, 
$\theta_W$ the weak mixing angle, and
$i,j=1,2,3$ are flavour indices.
Given that below $m_M$ the right-handed sneutrinos decouple, 
the low-energy sneutrino mass eigenstates are dominated by the 
$\tilde \nu_L$ components. 
Thus, sneutrino flavour
mixing is confined to the left-handed sector, and described by the
following $3 \times 3 $ matrix:
\begin{equation}
M_{\tilde{\nu}}^{ij \, 2}
\,=\,
m_{\tilde{L}, ij}^2  + \frac{1}{2}\, m_Z^2 \,\cos 2 \beta \, 
\delta_{ij}\,.
\end{equation} 
The physical masses and states are obtained by diagonalising the
previous mass matrices, leading to
\begin{align}
{M_{\tilde l}^2}^\text{diag} & \,=\, 
R^{(l)}  \,M_{\tilde l}^2  \,R^{(l)\,\dagger} \, = \,
\text{diag} \,(m_{\tilde l_1}^2,..,m_{\tilde l_6}^2) \,,
\nonumber \\
{M_{\tilde \nu}^2}^\text{diag}  & \,=  \,
R^{(\nu)}  \,M_{\tilde \nu}^2  \,R^{(\nu)\,\dagger} \, = \,
\text{diag}\,(m_{\tilde \nu_1}^2, \,
m_{\tilde \nu_2}^2, \,m_{\tilde \nu_3}^2)\,,
\end{align}
where $R^{(l,\nu)}$ are unitary rotation matrices.

After having introduced our scenario, in the next section we will summarise
some of the more relevant details leading to the computation of $\mu-e$
conversion rates in nuclei.

\section{Analytical results of the $\mu-e$ conversion 
rates}\label{CR:prediction}
In this section we report the analytical results for the $\mu-e$ conversion
rates in terms of the parameters introduced in
Section 2. We emphasise that all the results are obtained in terms of physical mass eigenstates (with full
propagators) for all MSSM particles entering in the computation, namely,
charginos $\tilde{\chi}_A^- (A=1, 2)$, neutralinos $\tilde{\chi}_A^0 (A=1,...,
4)$, charged sleptons $\tilde{l}_X^- (X=1,...,6)$, sneutrinos $\tilde{\nu}_X^-
(X=1, 2, 3)$ and the neutral Higgs bosons, $h^0$ and $H^0$.

\begin{figure}[t]
  \begin{center} 
        \psfig{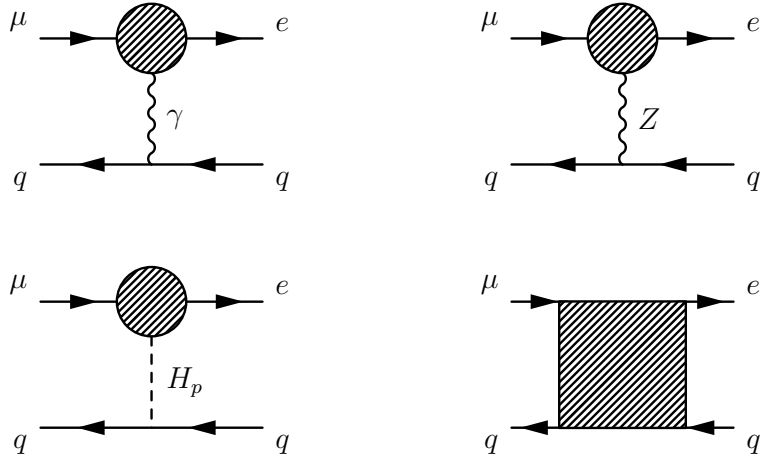}
    \caption{Photon-, $Z$-, $H$-penguin and box diagrams contributing to $\mu - e$ conversion in nuclei.
    }\label{CR_diagrams} 
  \end{center}
\end{figure}

For the presentation of the results we closely follow the general
parameterisation (and approximations) of~\cite{Kuno:1999jp}.
One starts with the most general effective
Lagrangian for four-fermion interactions which describes
coherent $\mu-e$ conversion. At the quark level, this is given by 
\begin{equation}\label{Leff}
{\cal L}_{\rm eff} = -\frac{G_F}{\sqrt{2}} \sum_q 
\left\{ \left[ g_{LS(q)} \bar{e}_L \mu_R + g_{RS(q)} \bar{e}_R \mu_L \right] 
\bar{q} q + \left[ g_{LV(q)} \bar{e}_L \gamma^{\mu} \mu_L + g_{RV(q)} 
\bar{e}_R 
\gamma^{\mu} \mu_R \right] \bar{q} \gamma_{\mu} q \right\}\,,
\end{equation}
where $G_F$ is the Fermi coupling. Notice that only scalar (S) and vector (V)
effective operators do contribute, with couplings given
by $g_{LS(q)}, g_{RS(q)}$ and  $g_{LV(q)}, g_{RV(q)}$ 
(respectively left and right, in both cases).
This effective Lagrangian at the quark
level is then converted into an effective Lagrangian at
the nucleon level, by means of the appropriate nucleon form 
factors~\cite{nucleon_level}. In the limit 
of negligible momentum dependence of the nucleon form factors,
(a reasonable approximation given the small momentum transfer in the
$\mu-e$ process), 
the quark matrix elements can be simply replaced by the nucleon 
matrix elements as follows:
\begin{align}
\langle p| \, \bar{q} \, \Gamma_K \, q \, 
|p \rangle &= G_K^{(q, p)} \, \bar{p} \,\Gamma_K \,p \,,\nonumber \\
\langle n| \, \bar{q} \, \Gamma_K \, q \, 
|n \rangle &= G_K^{(q, n)} \, \bar{n}\, \Gamma_K \,n\,,
\end{align}
where $\Gamma_K = (1, \gamma_\mu)$ respectively for $K = (S, V)$. The 
numerical values of the relevant $G_K$'s are~\cite{Kuno:1999jp,Kosmas:2001mv}:
\begin{align}
&
G_V^{(u, p)}\, =\, G_V^{(d, n)\,} =\, 2 \,;\, \ \ \ \ 
G_V^{(d, p)}\, =\, G_V^{(u, n)}\, = 1\,; \nonumber \\
&
G_S^{(u, p)}\, =\, G_S^{(d, n)}\, =\, 5.1\,;\, \ \ 
G_S^{(d, p)}\, =\, G_S^{(u, n)}\, = \,4.3 \,;\, \nonumber \\
&
G_S^{(s, p)}\,=\, G_S^{(s, n)}\, = \,2.5\,.
\end{align}   
The conversion rates are then predicted in terms of the relevant 
isoscalar, $g_{XK}^{(0)}$, and isovector couplings, $g_{XK}^{(1)}$ 
(with $X = L, R$ and $K = S, V$), which are given by:
\begin{align}
g_{XK}^{(0)} &= \frac{1}{2} \sum_{q = u,d,s} \left( g_{XK(q)} G_K^{(q,p)} +
g_{XK(q)} G_K^{(q,n)} \right)\,, \nonumber \\
g_{XK}^{(1)} &= \frac{1}{2} \sum_{q = u,d,s} \left( g_{XK(q)} G_K^{(q,p)} - 
g_{XK(q)} G_K^{(q,n)} \right)\,.
\end{align}

Further working under the approximation of equal proton and neutron
densities in the nucleus, and of a non-relativistic muon wave function for 
the $1\,s$ state, the final formula for the $\mu-e$ conversion rate,
relative to the the muon capture rate, can be finally written as   
\begin{align}
{\rm CR} (\mu- e, {\rm Nucleus}) &= 
\frac{p_e \, E_e \, m_\mu^3 \, G_F^2 \, \alpha^3 
\, Z_{\rm eff}^4 \, F_p^2}{8 \, \pi^2 \, Z}  \nonumber \\
&\times \left\{ \left| (Z + N) \left( g_{LV}^{(0)} + g_{LS}^{(0)} \right) + 
(Z - N) \left( g_{LV}^{(1)} + g_{LS}^{(1)} \right) \right|^2 + 
\right. \nonumber \\
& \ \ \ 
 \ \left. \,\, \left| (Z + N) \left( g_{RV}^{(0)} + g_{RS}^{(0)} \right) + 
(Z - N) \left( g_{RV}^{(1)} + g_{RS}^{(1)} \right) \right|^2 \right\} 
\frac{1}{\Gamma_{\rm capt}}\,,
\end{align}  
where $Z$ and $N$ are the number of protons and neutrons in the nucleus, 
while $Z_{\rm eff}$ is an effective atomic charge, obtained by 
averaging the muon wave function over the nuclear density~\cite{Chiang:1993xz}. 
$F_p$ is the nuclear matrix element and $\Gamma_{\rm capt}$ denotes 
the total muon capture rate.   
The other quantities in the above formula correspond to the muon
mass, $m_\mu$, the momentum and energy of the electron, $p_e$ and $E_e$ (which
are set to $m_\mu$ in the numerical evaluation), and the
electromagnetic coupling constant, $\alpha$.

We have computed the full set of one-loop diagrams leading into the 
quantity ${\rm CR} (\mu- e, {\rm Nucleus})$: 
$\gamma$-penguins, $Z$- and Higgs-boson
penguins and box diagrams. These are schematically drawn at the quark
level in Fig.~\ref{CR_diagrams}, and receive contributions from
several diagrams, mediated by SUSY particles, which are collected in 
Appendix~\ref{apendice1}.
The analytical results of the computation are summarised in terms of
the contributions of these diagrams to the vector and scalar
couplings,
\begin{align}
g_{LV(q)} &= g_{LV(q)}^{\gamma} + g_{LV(q)}^{Z} + 
g_{LV(q)}^{\rm B}\,, \nonumber \\
g_{LS(q)} &= g_{LS(q)}^{H} + g_{LV(q)}^{\rm B}\,. 
\end{align}
In the above, the photon couplings $g_{LX(q)}^{\gamma}$, the $Z$-boson
couplings $g_{LX(q)}^{Z}$, the $H$-boson couplings 
$g_{LS(q)}^{H}$, and the couplings arising from the boxes 
$g_{LX(q)}^{\rm B}$ (with $X=V,S$) are respectively given by
\begin{align}
g_{LV(q)}^{\gamma} &= \frac{\sqrt{2}}{G_F} e^2 Q 
\left(A_1^L - A_2^R \right)\,, \nonumber \\
g_{LV(q)}^{Z} &= -\frac{\sqrt{2}}{G_F} \, \frac{Z_L^q + Z_R^q}{2} \, 
\frac{F_L}{m_Z^2}\,, \nonumber \\
g_{LV(q)}^{\rm B} &= -\frac{\sqrt{2}}{G_F} \left( B_q^{(n)LV} + 
B_q^{(c)LV} \right)\,, \nonumber \\
g_{LS(q)}^{H} &= -\frac{\sqrt{2}}{G_F} \frac{1}{2} 
\sum_p \frac{1}{m_{H_p}^2} H_L^{(p)}  
\left( S_{L,q}^{(p)} + S_{R,q}^{(p)} \right) \,,\nonumber \\
g_{LS(q)}^{\rm B} &= -\frac{\sqrt{2}}{G_F} 
\left( B_q^{(n)LS} + B_q^{(c)LS} \right)\,. 
\end{align}
Likewise, for the right-handed couplings we find
\begin{align}
g_{RV(q)}&=\left. g_{LV(q)} \right|_{L \leftrightarrow R}\,,
\nonumber \\		
g_{RS(q)} &= \left. g_{LS(q)} \right|_{L \leftrightarrow R}\,.
\end{align}
The explicit formulae for the form factors of the photon 
($A_{(1,2)}^{(L,R)}$), of the $Z$-boson ($F_{(L,R)}$), of the Higgs-boson
($H_{(L,R)}^{(p)}$, where $p = 1,2,3$ corresponds to 
$H_p = h^0, H^0, A^0$),
and of the box diagrams ($B_q^{(n,c)(L,R)(V,S)}$) are listed in 
Appendix~\ref{apendice1}. In each case, the relevant couplings 
$Z_{(L,R)}^q$,  $S_{(L,R)q}^{(p)}$ etc., can be found 
in Appendix~\ref{apendice2}.  

It is important to stress that 
$S_{L,q}^{(3)} + S_{R,q}^{(3)}$ vanishes and therefore there are no
contributions from the CP-odd Higgs boson $A^0$. 
This is a consequence of working in the
approximation of coherent $\mu-e$ conversion, in which case 
the initial and final nucleus state is the same, thus leading to vanishing
matrix elements for pseudoscalar currents like 
$\langle{\rm Nucleus}\,|\,{\bar q}\,\gamma_5 \,q\,|\,{\rm Nucleus}\rangle$.
Also notice that from the values of the $S_{(L,R)q}^{(p)}$ Higgs 
couplings, one can anticipate that in the large $\tan \beta$ and small
Higgs mass regime, the dominant Higgs contribution
will be that of $H^0$.  

When compared to the results obtained in~\cite{Hisano:1995cp},
our expressions coincide in the formulae for the photon-penguins. 
Up to a global sign, the  vector contributions from boxes also agree.
Divergences occur regarding the $Z$-penguins, and the differences can
be read by comparing our expressions in 
Eqs.~(44-48) 
of Appendix~\ref{apendice1}, with those of 
Eqs.(22-29) in~\cite{Hisano:1995cp}.
As previously mentioned, we have included in addition scalar 
contributions from boxes and Higgs-mediated diagrams not considered in~\cite{Hisano:1995cp}.

A connection between our results for the Higgs contributions and 
those reported in~\cite{Kitano:2003wn} can be established in 
the large $\tan\beta$ limit, writing the output in the 
mass-insertion approximation format. 
Under these conditions, and considering the limit of a common mass
for all SUSY particles involved, which is much larger than the SM
particle masses,  $M_{\rm soft} \sim \mu \sim M_{\rm SUSY}>>m_W$, 
one arrives at the following simple expression for the dominant $H^0$
form factor~\cite{Arganda:2004bz} 
\begin{equation}\label{HL2}
H_L^{(2)}\,=\,-\frac{1}{(4\pi)^2}\,
\frac{m_\mu}{12\,m_W}\,\delta_{21}^{l}\,\tan^2 \beta\, 
\left[1+\frac{1}{2}(1-3\tan^2\theta_W)\right]\,,
\end{equation}
where the first term arises
from chargino mediated loops, while the second stems from neutralino
mediated contributions.
In the mass insertion approximation, the dominant slepton mixing
effects are associated with $\delta^l_{21}$, which can be written as:
\begin{equation}
\delta^l_{21}\,=\,\frac{(\Delta m^2_{\tilde L})_{21}}{M_{\rm SUSY}^2}\,.
\end{equation} 
From the above, one can finally obtain a 
simple expression for the $H^0$ contribution to 
the conversion rate, which is clearly dominated by the strange quark coupling,
due to the enhancement in the coupling by $m_s$. 
This arises via
$g^{(0)}_{LS} \simeq g_{LS(s)}^{H^0}G_S^{(s,p)}$ with
\begin{equation}
g_{LS(s)}^{H^0} \,=\, \frac{\sqrt{2}}{G_F}\, 
\frac{1}{2} \,\frac{1}{m_{H^0}^2}\, H_L^{(2)}\,
\frac{g m_s}{m_W}\,\tan \beta\,.
\end{equation}
Plugging this simplified result for the $g_{LS}^{(0)}$ coupling into the
approximate conversion rate for the Higgs-dominated case,
\begin{align}
{\rm CR} (\mu- e, {\rm Nucleus}) &\simeq\, 
\frac{p_e \,E_e\, m_\mu^3\, G_F^2\, \alpha^3 \,Z_{\rm eff}^4\, 
F_p^2}{8 \pi^2 \,Z}\, 
\left\{ \left| (Z + N)\,  g_{LS}^{(0)} \right|^2 \right\}\, 
\frac{1}{\Gamma_{\rm capt}}\,
\end{align}  
we obtain the expected  
$\tan^6 \beta$ enhancement of the $H^0$ contribution. Moreover, the
dependence on the Higgs mass ($\frac{1}{m_{H^0}^4}$), as
well as the typical prefactor $|\delta_{21}^l|^2$ accounting for the
lepton flavour changing effect are equally recovered. 
Within this approximation, and taking a specific value of
$\delta_{21}^l= 10^{-3}$, allows to obtain an order-of-magnitude
estimate for the conversion rate in the case of Titanium nuclei, 
\begin{equation}
 {\rm CR} (\mu- e, {\rm Ti})\, \simeq \,
\mathcal{O}(10^{-12})\, \left( \frac {115 \,\text{GeV}}{m_{H^0}}\right)^4\,
 \left(\frac {\tan \beta}{50}\right)^6 \,,
\end{equation}
in agreement with the estimate of~\cite{Kitano:2003wn}.

Finally, it is worth mentioning that the heavy SUSY particles do not 
decouple in the Higgs
contributions to the $\mu-e$ conversion rates.
This can be understood from the previous result of $H_L^{(2)}$ in 
Eq.~(\ref{HL2}), 
which is constant in the large $M_{\rm SUSY}$ limit. 
This SUSY non-decoupling effect has also been noticed in association
to other Higgs-mediated LFV 
processes~\cite{Arganda:2004bz,Brignole_Rossi,Paradisi:2005tk,Babu:2002et}.

\section{Numerical results and discussion}\label{results}
In this section we present the numerical results of the $\mu-e$ conversion
rates in nuclei within the SUSY-seesaw context described in 
Section~\ref{susy:seesaw}.
We begin by addressing the CMSSM-seesaw, and then proceed to the
NUHM-seesaw. In both scenarios, we consider the dependence of the
theoretical predictions for the conversion rates on the most relevant
SUSY-seesaw parameters. In our discussion, we will give a particular
emphasis to the most significant differences between the CMSSM- and 
NUHM-seesaw scenarios.

The numerical results presented in this section are mainly devoted to 
the particular case of Titanium nuclei, given that one expects a notable improvement of  future experimental
sensitivities in that case~\cite{PRIME}. 
However, some additional predictions for other nuclei are also included here,
for comparison. The case of Gold nuclei is of
particular interest, since at present the most stringent bound is that
of CR($\mu - e$, Au)~\cite{Bertl:2001fu}.

For the purpose of numerical evaluation, we begin by defining the
input parameters at the gauge coupling unification scale, $M_X$. 
In the case of a CMSSM-seesaw scenario, 
and instead of scanning over the full ($M_{1/2},\,M_0,\,A_0,\,\tan
\beta,\,\text{sign} \mu$) parameter space, we study specific
points, each exhibiting distinct characteristics from
the low-energy phenomenology point of view. We specify these
parameters by means of the ``Snowmass Points and Slopes''
(SPS)~\cite{Allanach:2002nj} cases defined in Table~\ref{SPS:def:15}.
\begin{center}
\begin{table}[h]\hspace*{25mm}
\begin{tabular}{|c|c|c|c|c|c|}
\hline
SPS & $M_{1/2}$ (GeV) & $M_0$ (GeV) & $A_0$ (GeV) & $\tan \beta$ & 
 $\mu$ \\\hline
 1\,a & 250 & 100 & -100 & 10 &  $>\,0 $ \\
 1\,b & 400 & 200 & 0 & 30 &   $>\,0 $ \\
 2 &  300 & 1450 & 0 & 10 &  $>\,0 $ \\
 3 &  400 & 90 & 0 & 10 &    $>\,0 $\\
 4 &  300 & 400 & 0 & 50 &   $>\,0 $ \\
 5 &  300 & 150 & -1000 & 5 &   $>\,0 $\\\hline
\end{tabular} 
\caption{Values of $M_{1/2}$, $M_0$, $A_0$, $\tan \beta$, 
and sign $\mu$ for the SPS points considered in the analysis.}
\label{SPS:def:15}
\end{table}
\end{center} 

In the case of the NUHM-seesaw scenario, and in order to reduce the number of
input parameters, we set $M_0=M_{1/2}\equiv M_{\rm SUSY}$, and explore the 
($M_{\rm SUSY}\,,A_0,\,\tan \beta,\,\text{sign}\,\mu, \delta_1,\delta_2$)
parameter space considering the following intervals: 
\begin{align}
250 \,{\rm GeV} \, < \, & M_{\rm SUSY} \, < \, 
1000 \,{\rm GeV} \,, \nonumber \\ 
-500 \,{\rm GeV} \, < \, & A_0 \, < \, 500 \,{\rm GeV} \,, \nonumber \\ 
5  \,< &  \,\tan \beta  \,< \,50\,, \nonumber \\ 
-2 \, < \,& \delta_1\,,\,\delta_2 \, < \,2\,. 
\end{align}
In addition, we also consider the two possibilities, 
$\text{sign} (\mu) =\pm 1$.

To obtain the low-energy parameters of the model (and thus compute the
relevant physical masses and couplings), 
the full one-loop RGEs (including  the 
neutrino and sneutrino sectors) are firstly run down from $M_X$ to the right
handed neutrino scale $m_{M}$. 
At this scale we 
impose the boundary condition of
Eq.~(\ref{seesaw:casas}). After the decoupling of the heavy neutrinos
and sneutrinos, the new RGEs 
are then run down from $m_{M}$ to 
the EW scale, at which the conversion rates are computed. Notice that, 
in the case of hierarchical heavy neutrinos, 
the sequential running is done from $M_X$
down to $m_{N_3}$ and from $m_{N_1}$ down to the EW scale. 
This implies that we do not take into account the running effects from  
the intermediate right handed neutrino scales, i.e. from $m_{N_3}$ to  
$m_{N_2}$ and from  $m_{N_2}$ to  $m_{N_1}$. 
We have estimated these threshold effects by means of 
the leading logarithmic (LLog) approximation, verifying that they are
indeed negligible for the numerical values chosen 
in the present work.

The numerical implementation of the above procedure is achieved by
means of the public Fortran code {\tt SPheno2.2.2}~\cite{Porod:2003um}. The
value of $M_X$ is derived from the unification condition of the
$SU(2)$ and $U(1)$ gauge couplings (systematically leading 
to a value of $M_X$ very close to $2 \times 10^{16}$ GeV
throughout the numerical analysis), while $|\mu|$ is derived from  
the requirement of obtaining the correct radiative EW symmetry
breaking. The code {\tt SPheno2.2.2} has been
adapted~\cite{Arganda:2005ji} in order to fully incorporate the right-handed
neutrino (and sneutrino) sectors, as well as the full lepton flavour 
structure. The computation of the $\mu-e$ conversion 
rates in nuclei, as well as of other LFV observables, has
been implemented into the code by means of additional subroutines. 
 
Regarding the light neutrino masses
and the $U_\text{MNS}$ matrix elements, we take the following input values:
\begin{align}
& 
\Delta\, m^2_\text{sol} \,=\,8\,\times 10^{-5}\,\,\text{eV}^2\,,
\quad 
\Delta \, m^2_\text{atm} \,=\,2.5\,\times 10^{-3}\,\,\text{eV}^2\,,
\quad 
m_{{\nu}_1}\,=\,10^{-3}\,\,\text{eV}\,,
\nonumber \\
& 
\theta_{12}\,=\,30^\circ\,, 
\quad 
\theta_{23}\,=\,45^\circ\,,
\quad 
\theta_{13}\,\lesssim\,10^\circ\,,
\quad \quad 
\delta\,=\,\phi_1\,=\,\phi_2\,=\,0\,,
\end{align}
compatible with present experimental data (see, for
instance, the analyses 
of~\cite{neutrinodata_fits}).
We do not address the impact of
non-vanishing $U_\text{MNS}$ phases (Dirac or Majorana) in the $\mu-e$
conversion rates.

Finally, although not used in this work, it is clarifying to recall that a
simplified estimation of the generated flavour mixing in the slepton sector can be obtained by
means of the LLog approximation. Using the latter, the relevant
off-diagonal slepton mass matrix element for the processes involving 
lepton flavour violation in the $\mu-e$ sector (as is the case of
$\mu-e$ conversion in nuclei) can be given as
\begin{align}\label{misalignment_sleptons}
(\Delta m_{\tilde{L}}^2)_{21}&\,=\,
-\frac{1}{8\, \pi^2}\, (3\, M_0^2+ A_0^2)\, (Y_{\nu}^\dagger\, 
L\, Y_{\nu})_{21}\,\,;\, 
L_{kl}\, \equiv \,\log \left( \frac{M_X}{m_{N_k}}\right) \delta_{kl}\,. 
\end{align}
Writing $( Y_\nu^\dagger L Y_\nu )_{21}$ using the parameterisation of 
Eqs.~(\ref{def:Ndiag}-\ref{Rcasas}), and considering the limit of 
$m_{\nu_1}=0$, $\phi_{1,2}=\delta=0$ (which is appropriate for the
subsequent discussion), one obtains the following expression: 
\begin{align}\label{Y21:LLog}
\lefteqn{v_2^2 \,( Y_\nu^\dagger \,L\, Y_\nu )_{21}\, = \,
} \nonumber \\
& \,\,\,\,\,\,\, 
{L_{33}}\,{m_{N_3}}\,\left[ {c_{13}}\,
      {\sqrt{{m_{\nu_2}}}}\,{c_2^*}\,
         {s_1^*}\,{s_{12}}   + 
     {\sqrt{{m_{\nu_3}}}}\,{c_1^*}\,{c_2^*}\,{s_{13}}
     \right] \nonumber \\ 
&\,\,\,\,\,\,\,\,\,\,\,\,\,\,\,\,\,\,\,\,
\left[ {\sqrt{{m_{\nu_3}}}}\,{c_1}\,{c_2}\,{c_{13}}\,
      {s_{23}}  + \right.  
\left. 
     {{\sqrt{{m_{\nu_2}}}}\,{c_2}\,{s_1}\,
        \left( {c_{12}}\,{c_{23}} - 
          \,{s_{12}}\,{s_{13}}\,{s_{23}}
          \right) } \right] \nonumber \\
& 
+ {L_{22}}\,m_{N_2}\,\left[ 
       {\sqrt{{m_{\nu_2}}}}\,{c_{13}}\,
      \left( {c_1^*}\,{c_3^*} - {s_1^*}\,{s_2^*}\,{s_3^*} \right) \,
{s_{12}}    
- \,{\sqrt{{m_{\nu_3}}}}\,
      \left( {c_3^*}\,{s_1^*} + {c_1^*}\,{s_2^*}\,{s_3^*} \right) \,{s_{13}}
     \right] \, \nonumber \\
&\,\,\,\,\,\,\,\,\,\,\,\,\,\,\,\,\,\,\,\,    
\left[ - {\sqrt{{m_{\nu_3}}}}\,{c_{13}}\,
        \left( {c_3}\,{s_1} + {c_1}\,{s_2}\,{s_3} \right) \,
        {s_{23}}  +
     {{\sqrt{{m_{\nu_2}}}}\,
        \left( {c_1}\,{c_3} - {s_1}\,{s_2}\,{s_3} \right) \,
        \left( {c_{12}}\,{c_{23}} - 
          \,{s_{12}}\,{s_{13}}\,{s_{23}}
          \right) } \right]\nonumber \\
& + {L_{11}}\,m_{N_1}\,\left[  - 
     \,{\sqrt{{m_{\nu_2}}}}\,{c_{13}}\,
      \left( {c_3^*}\,{s_1^*}\,{s_2^*} + {c_1^*}\,{s_3^*} \right) \,{s_{12}}   
+ \,{\sqrt{{m_{\nu_3}}}}\,
      \left( - {c_1^*}\,{c_3^*}\,{s_2^*}   + 
        {s_1^*}\,{s_3^*} \right) \,{s_{13}} \right] \,\nonumber \\
&\,\,\,\,\,\,\,\,\,\,\,\,\,\,\,\,\,\,\,\, 
 \left[ {\sqrt{{m_{\nu_3}}}}\,{c_{13}}\,
      \left( - {c_1}\,{c_3}\,{s_2}   + 
        {s_1}\,{s_3} \right) \,{s_{23}}
    - {{\sqrt{{m_{\nu_2}}}}\,
        \left( {c_3}\,{s_1}\,{s_2} + {c_1}\,{s_3} \right) \,
        \left( {c_{12}}\,{c_{23}} - 
          \,{s_{12}}\,{s_{13}}\,{s_{23}}
          \right) } \right]\,.  	   
\end{align}  

In what follows, we begin by investigating the theoretical predictions
for the $\mu-e$ conversion rates in Titanium nuclei within the CMSSM-seesaw.

\subsection{Universality: CMSSM-seesaw}\label{CMSSM:seesaw}
\begin{figure}[t!]
  \begin{center} 
    \begin{tabular}{cc} \hspace*{-12mm}
        \psfig{file=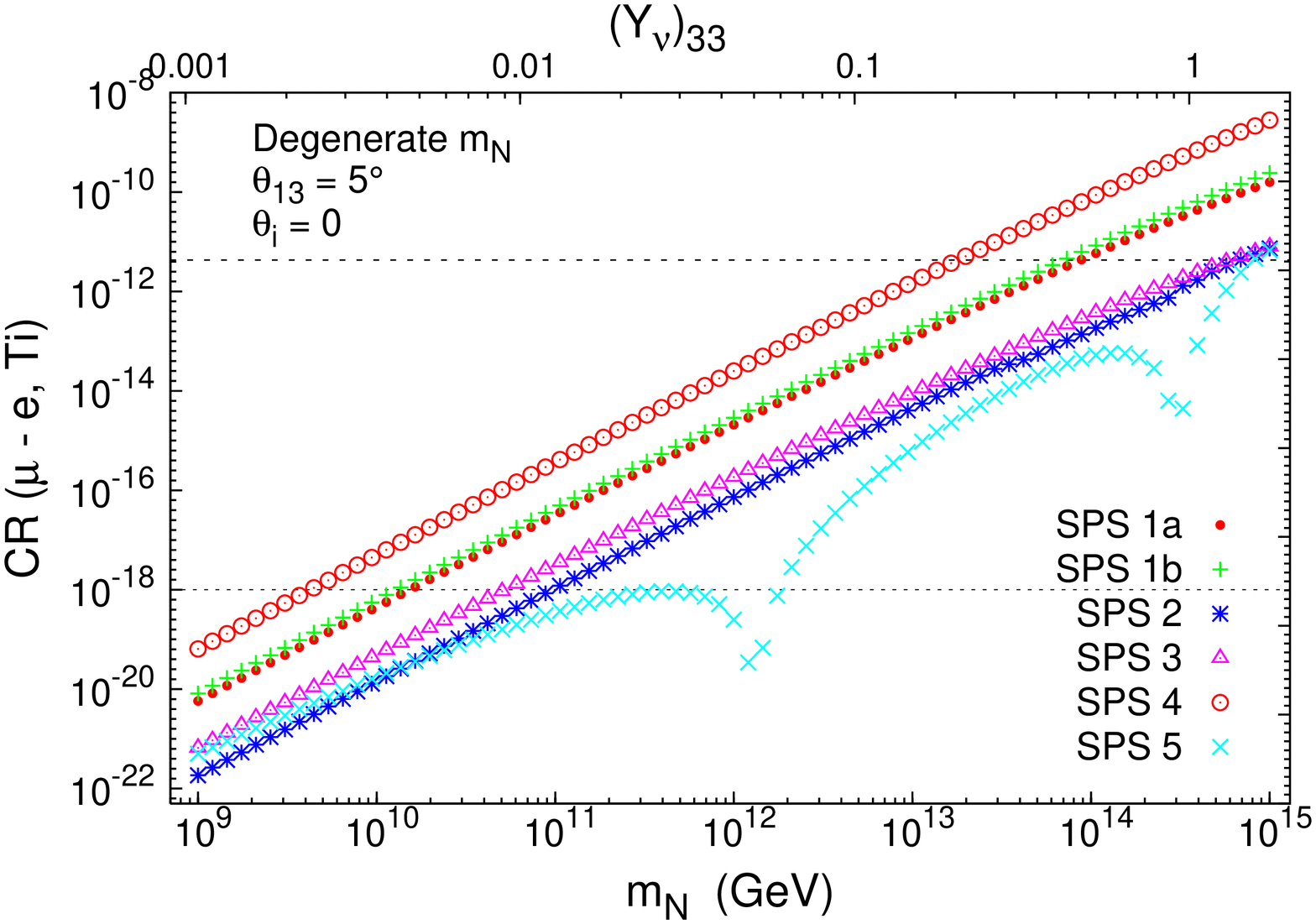,width=60mm,angle=270,clip=} 
&
 	\psfig{file=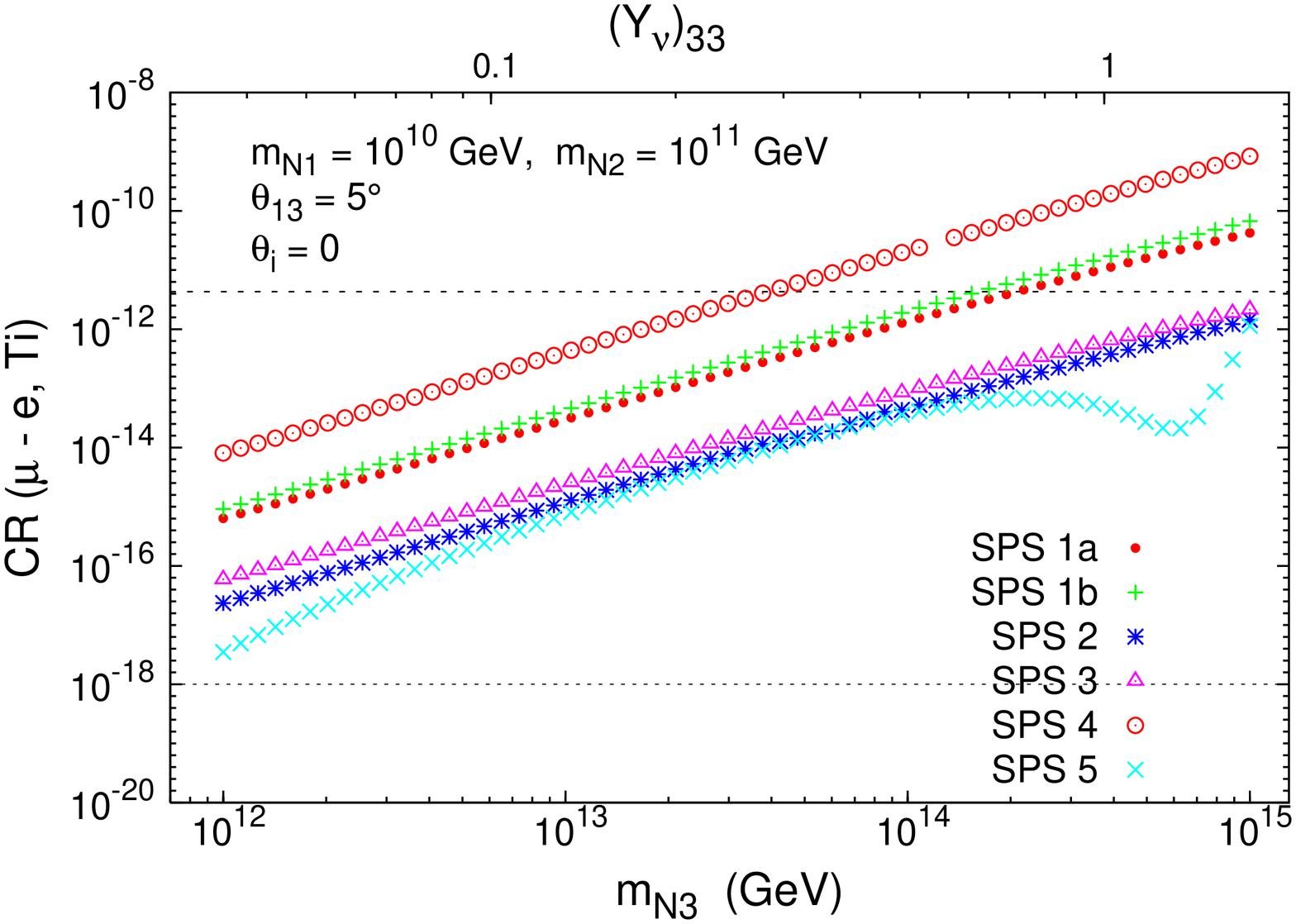,width=60mm,angle=270,clip=}   
    \end{tabular}
    \caption{CR($\mu -e$, Ti) as a function of the relevant heavy
      neutrino mass: $m_N$ (on the left) and $m_{N_3}$ (on the right),
      respectively associated with the degenerate and hierarchical
      cases. The predictions for SPS 1a
      (dots), 1b (crosses), 2 (asterisks), 3 (triangles), 4 (circles) and 5
      (times) are included. On the upper horizontal axis we display the associated 
      value of $(Y_\nu)_{33}$.
      In each case, we set $\theta_{13}=5^\circ$, and
      consider the limit where $R=1$ ($\theta_i=0$). A dashed (dotted)
      horizontal line denotes the present experimental bound (future
      sensitivity).
    }\label{fig:CR:MRMR3:SPSX} 
  \end{center}
\end{figure}

\begin{figure}[t!]
\vspace{-3cm}
  \begin{center} 
        \begin{tabular}{cc}\hspace*{-10mm}
        \psfig{file=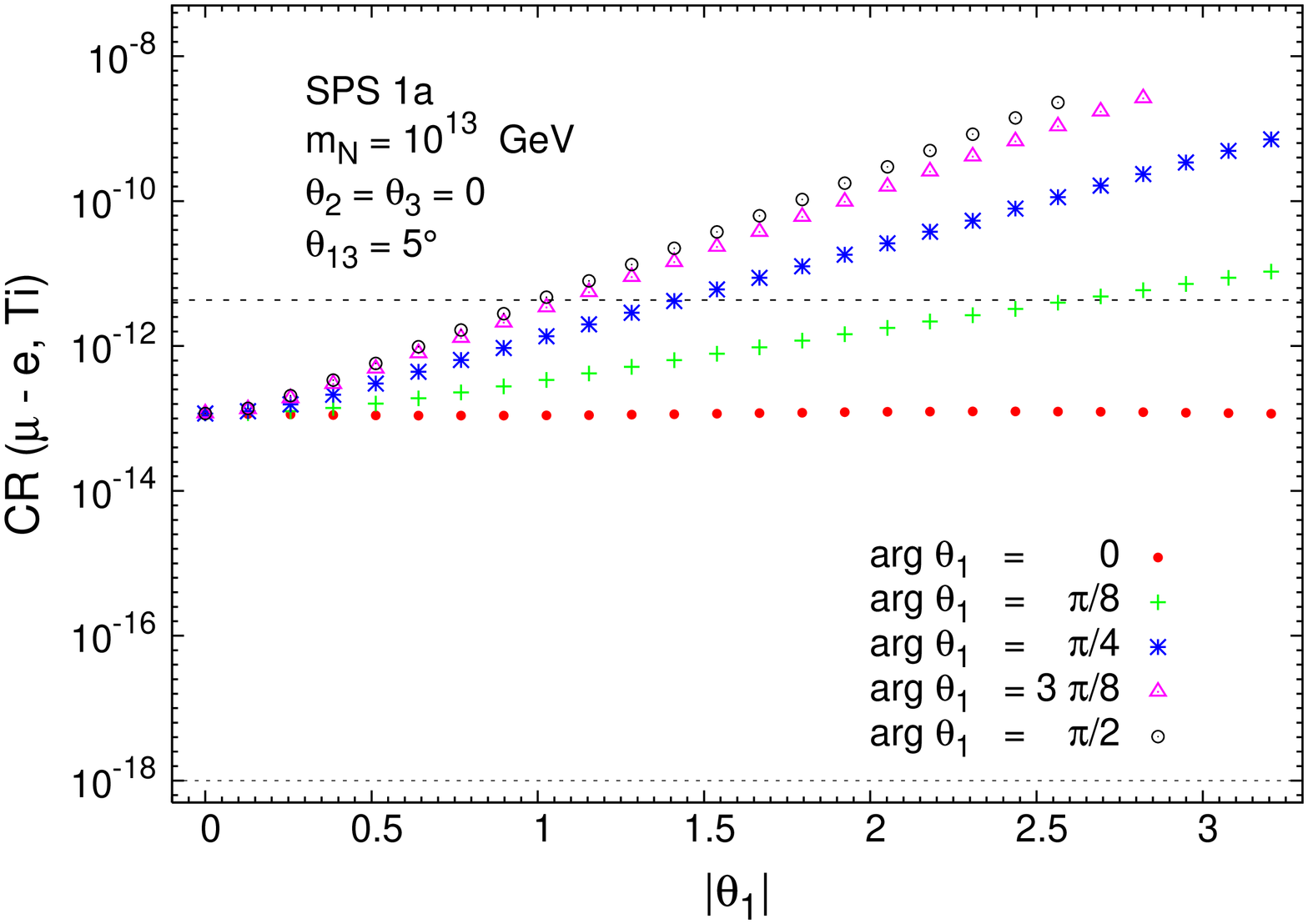,width=60mm,angle=270,clip=} &
	\psfig{file=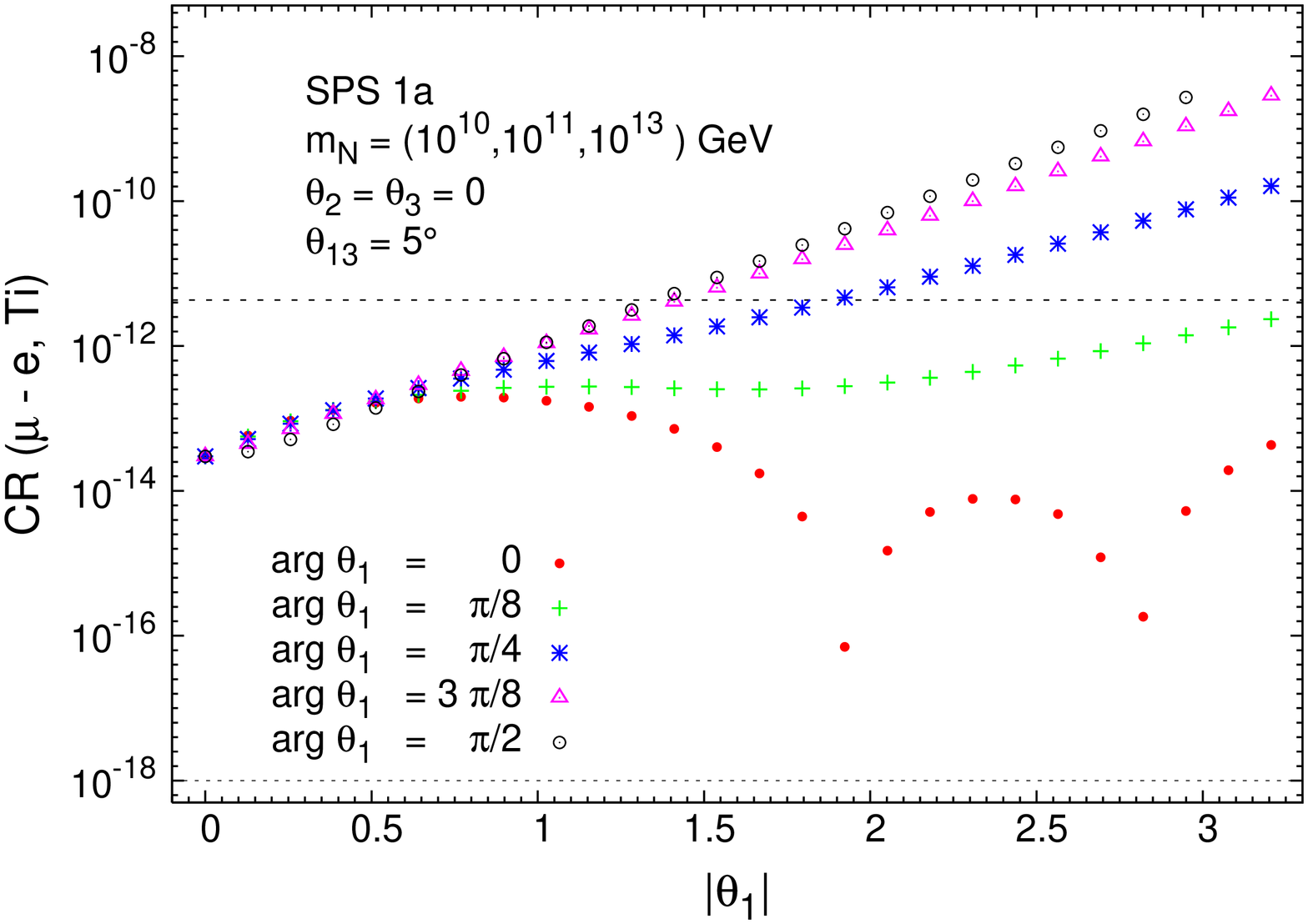,width=60mm,angle=270,clip=} \\  \hspace*{-10mm}
	\psfig{file=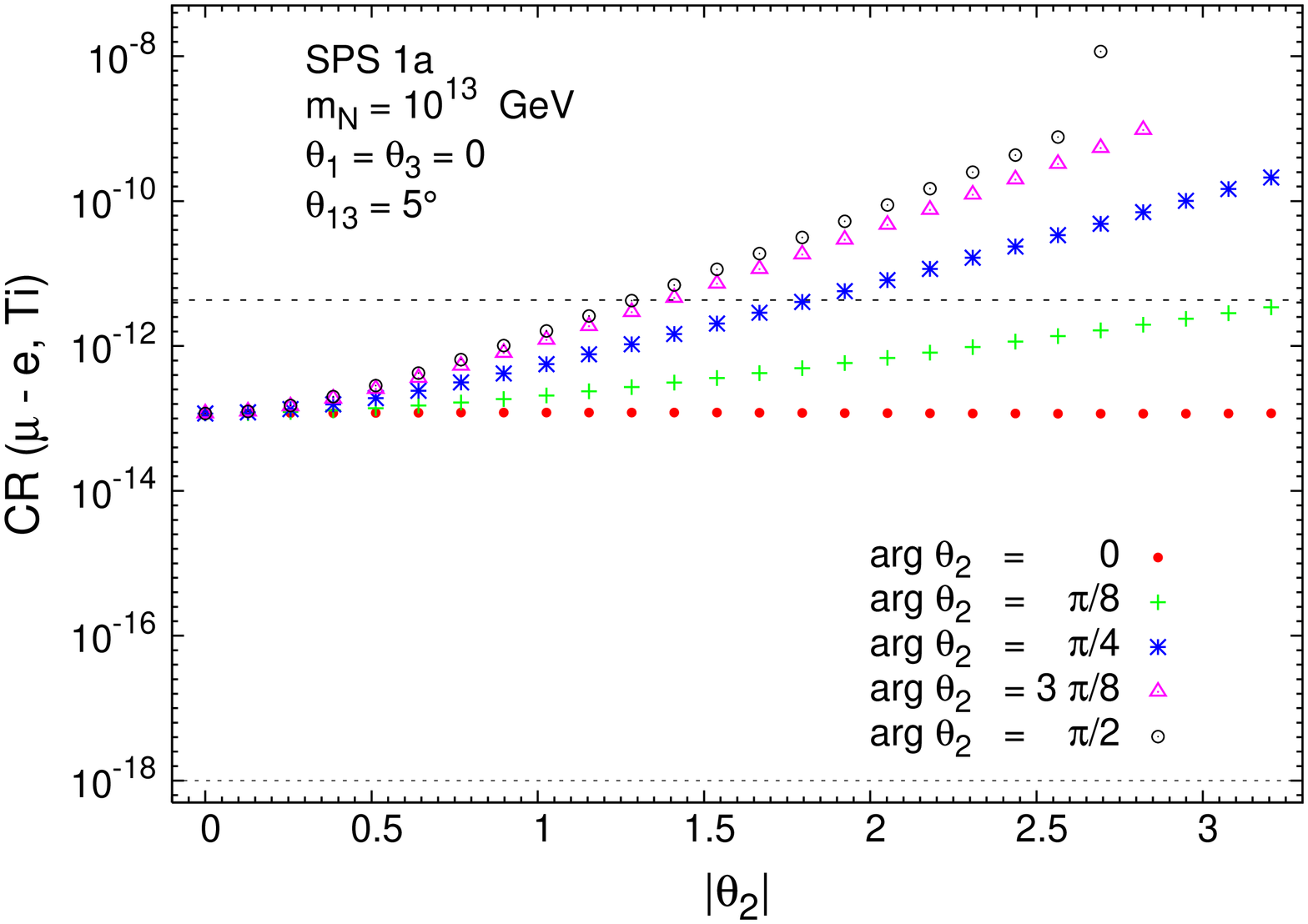,width=60mm,angle=270,clip=} &
	\psfig{file=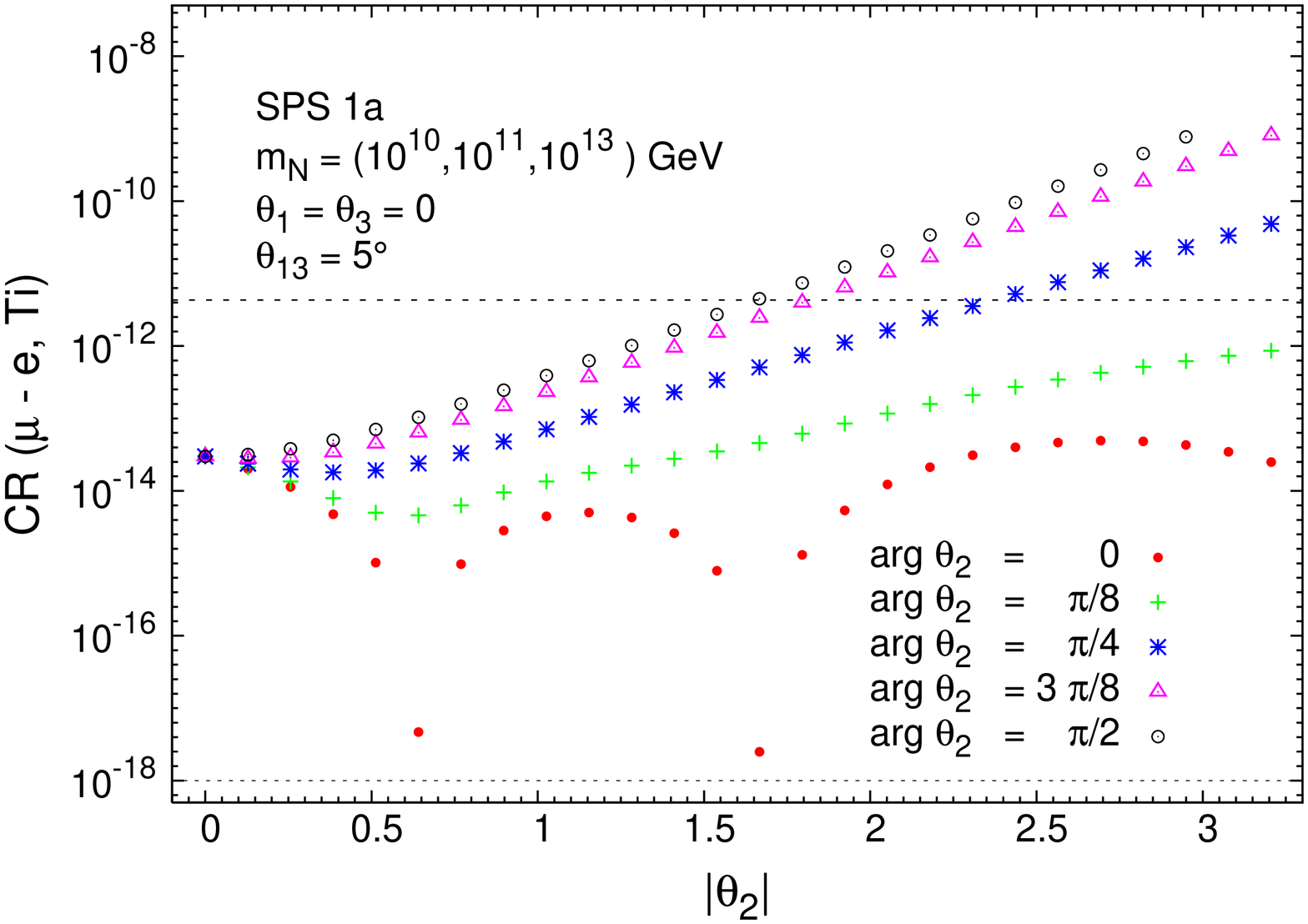,width=60mm,angle=270,clip=} \\ \hspace*{-10mm}
	\psfig{file=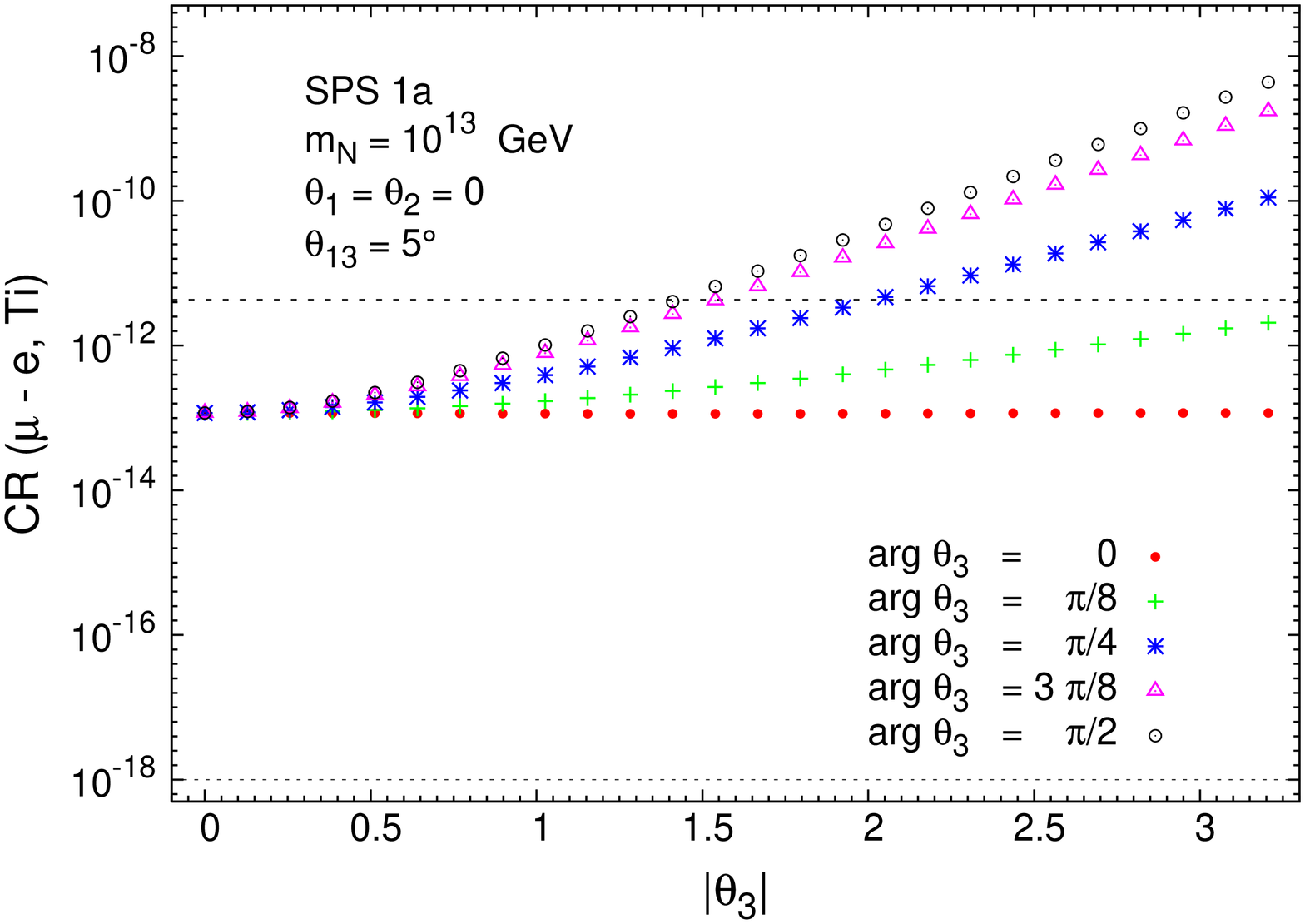,width=60mm,angle=270,clip=} &
	\psfig{file=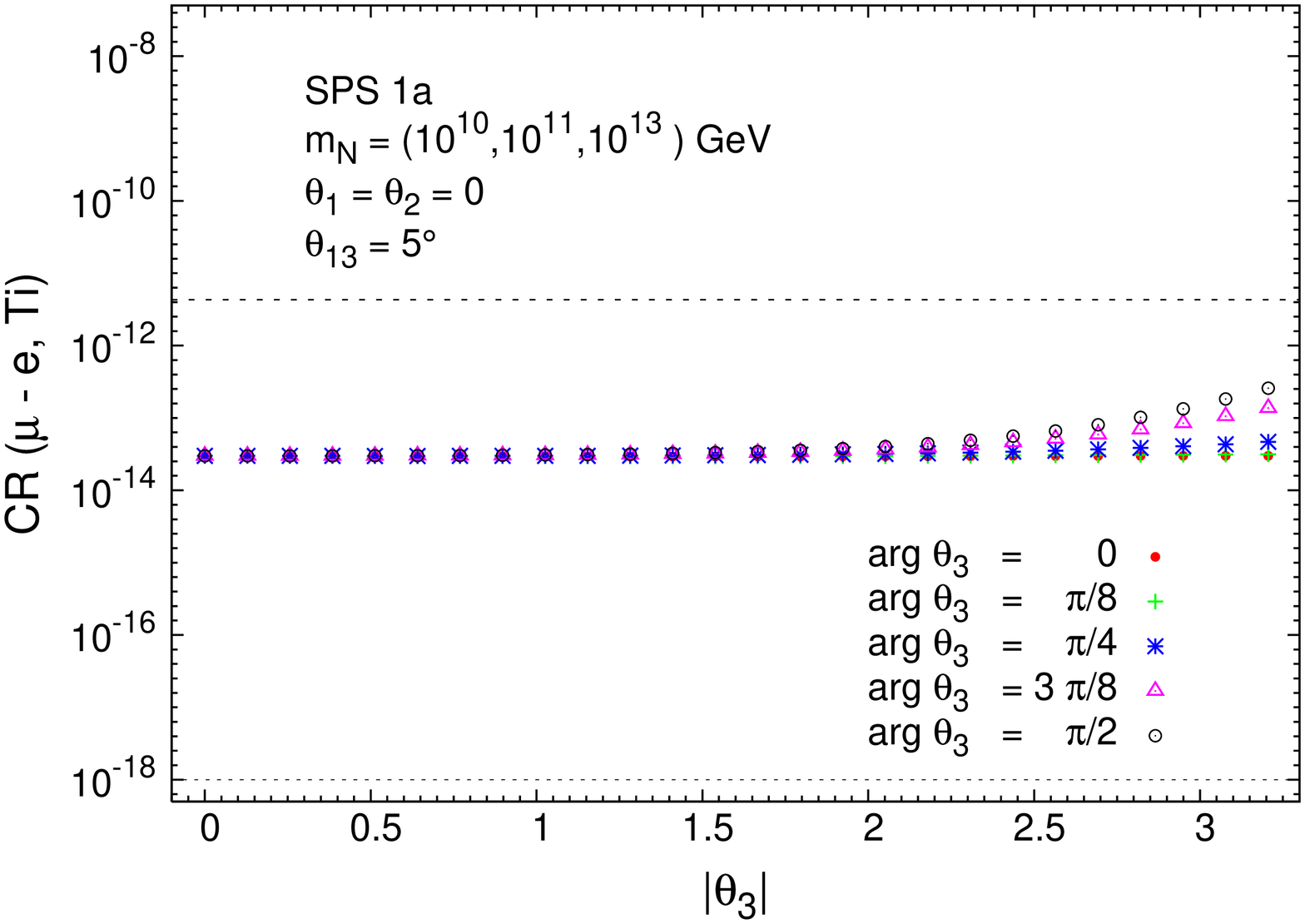,width=60mm,angle=270,clip=} 
    \end{tabular}
    \caption{From top to bottom, CR($\mu -e$, Ti) as a function of 
      $|\theta_i|$ ($i=1,2,3$), for 
      $\arg \theta_i\,=\,\{0,\, \pi/8\,,\,\pi/4\,,\,3\pi/8, \,\pi/2
      \}$ (dots, crosses, asterisks, triangles and circles,
      respectively). Both $|\theta_i|$ and $\arg{\theta_i}$ are given in
      radians. On the left we consider degenerate 
      heavy neutrinos (with $m_N = 10^{13}$ GeV), while on the right the
      hierarchical case is displayed (with $m_{N_i} =
      (10^{10},10^{11},10^{13})$ GeV). 
      In all cases we take $\theta_{13}=5^\circ$, 
      and set the CMSSM parameters to the SPS 1a case.
      A dashed (dotted) horizontal line denotes the present
      experimental bound (future sensitivity).
    }\label{fig:CR:theta123:SPS1a:MRMR3} 
  \end{center}
\end{figure}

\begin{figure}[h]
  \begin{center} 
    \begin{tabular}{cc}\hspace*{-10mm}
      \psfig{file=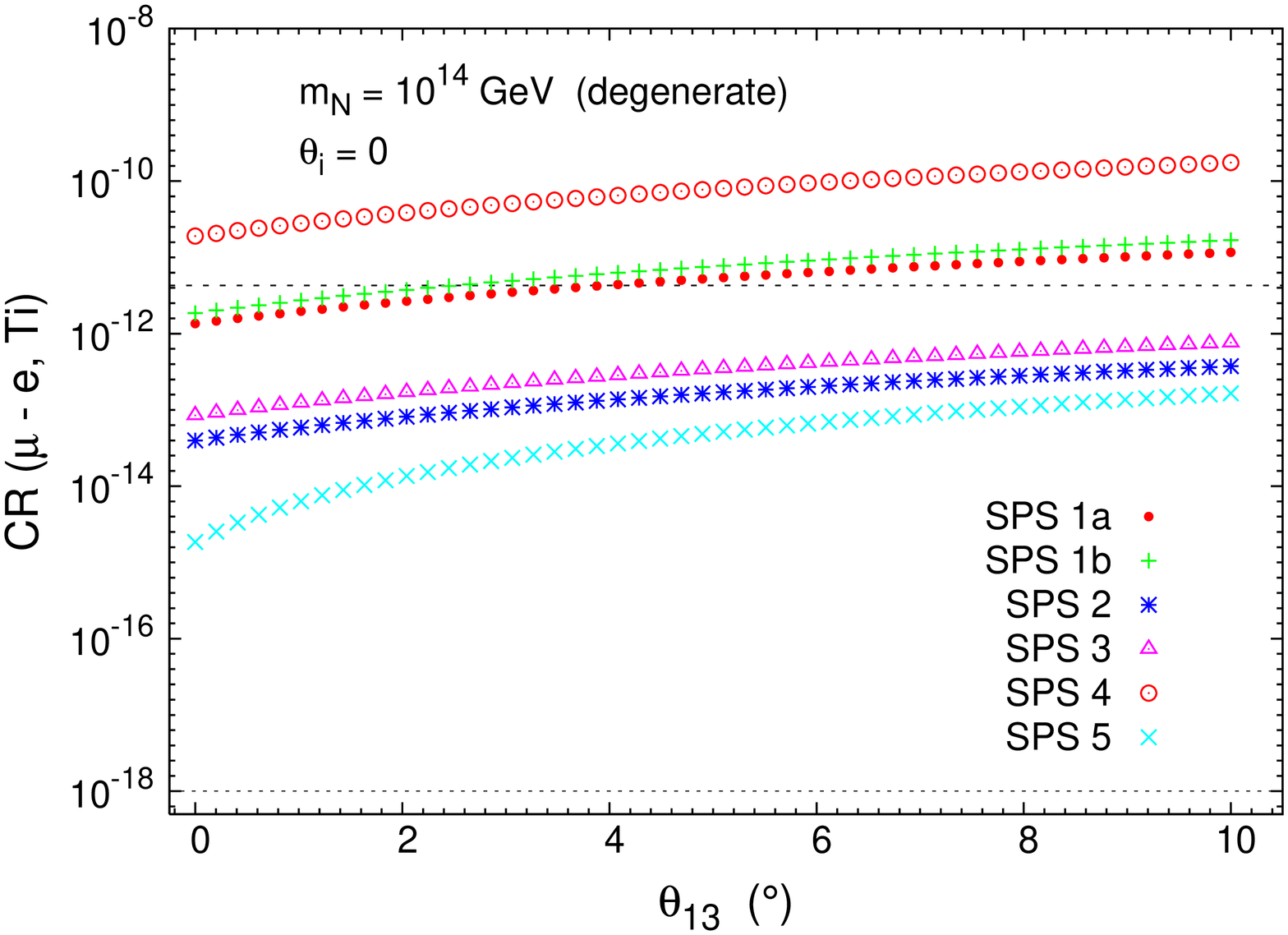,width=60mm,angle=270,clip=} &
      \psfig{file=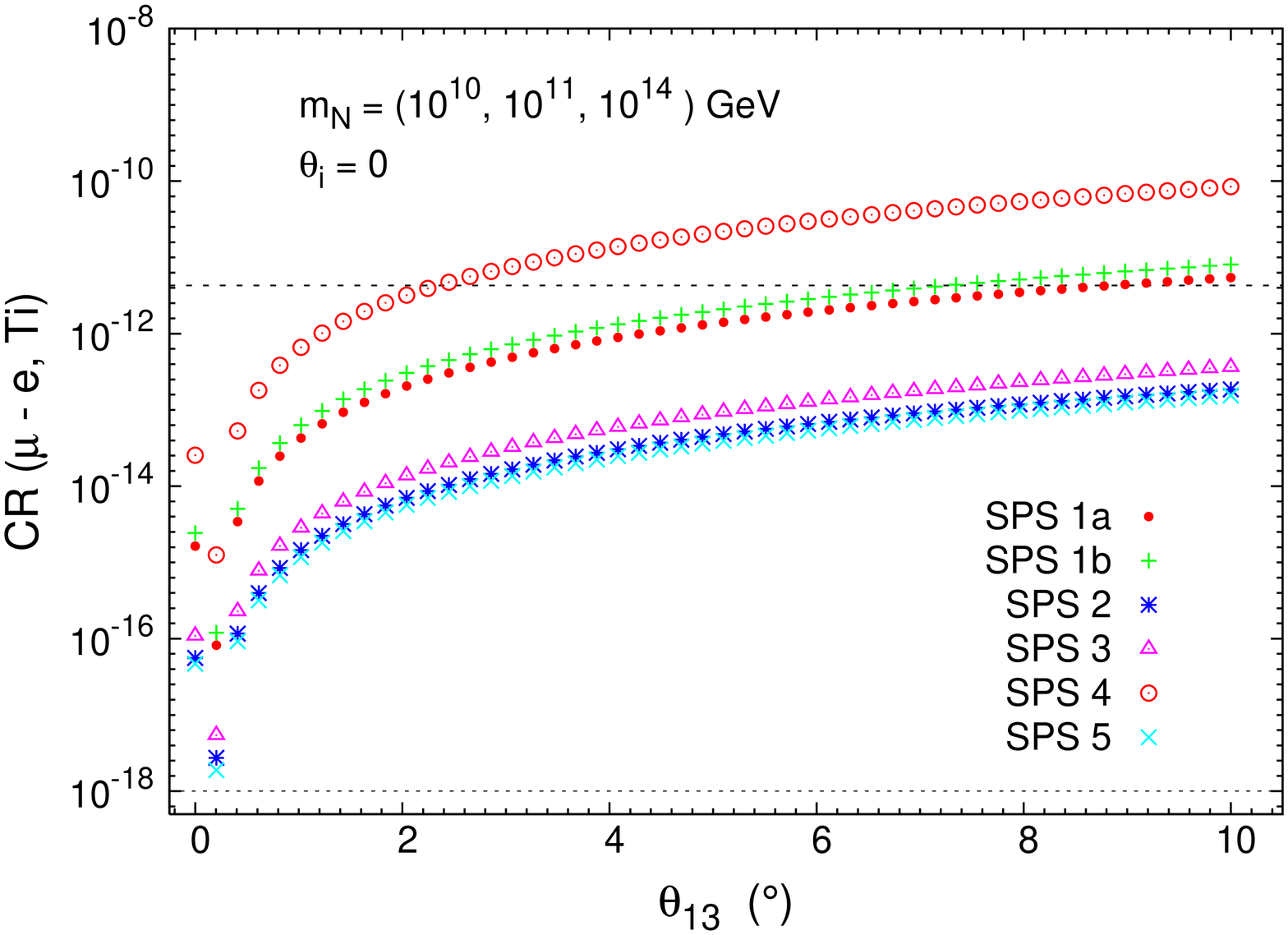,width=60mm,angle=270,clip=}   
    \end{tabular}
    \caption{CR($\mu -e$, Ti) as a function of $\theta_{13}$ (in
      degrees), for SPS 1a
      (dots), 1b (crosses), 2 (asterisks), 3 (triangles), 4 (circles) and 5
      (times). On the left we consider degenerate
      heavy neutrinos (with $m_N = 10^{14}$ GeV), while on the right the
      hierarchical case is displayed (with $m_{N_i} =
      (10^{10},10^{11},10^{14})$ GeV). In both cases we choose
      $R=1$ ($\theta_i=0$). A dashed (dotted)
      horizontal line denotes the present experimental bound (future
      sensitivity).
    }\label{fig:CR:theta13:SPSX:MRMR3} 
  \end{center}
\end{figure}

\begin{figure}[h]
  \begin{center} 
    \begin{tabular}{cc}\hspace*{-10mm}
      \psfig{file=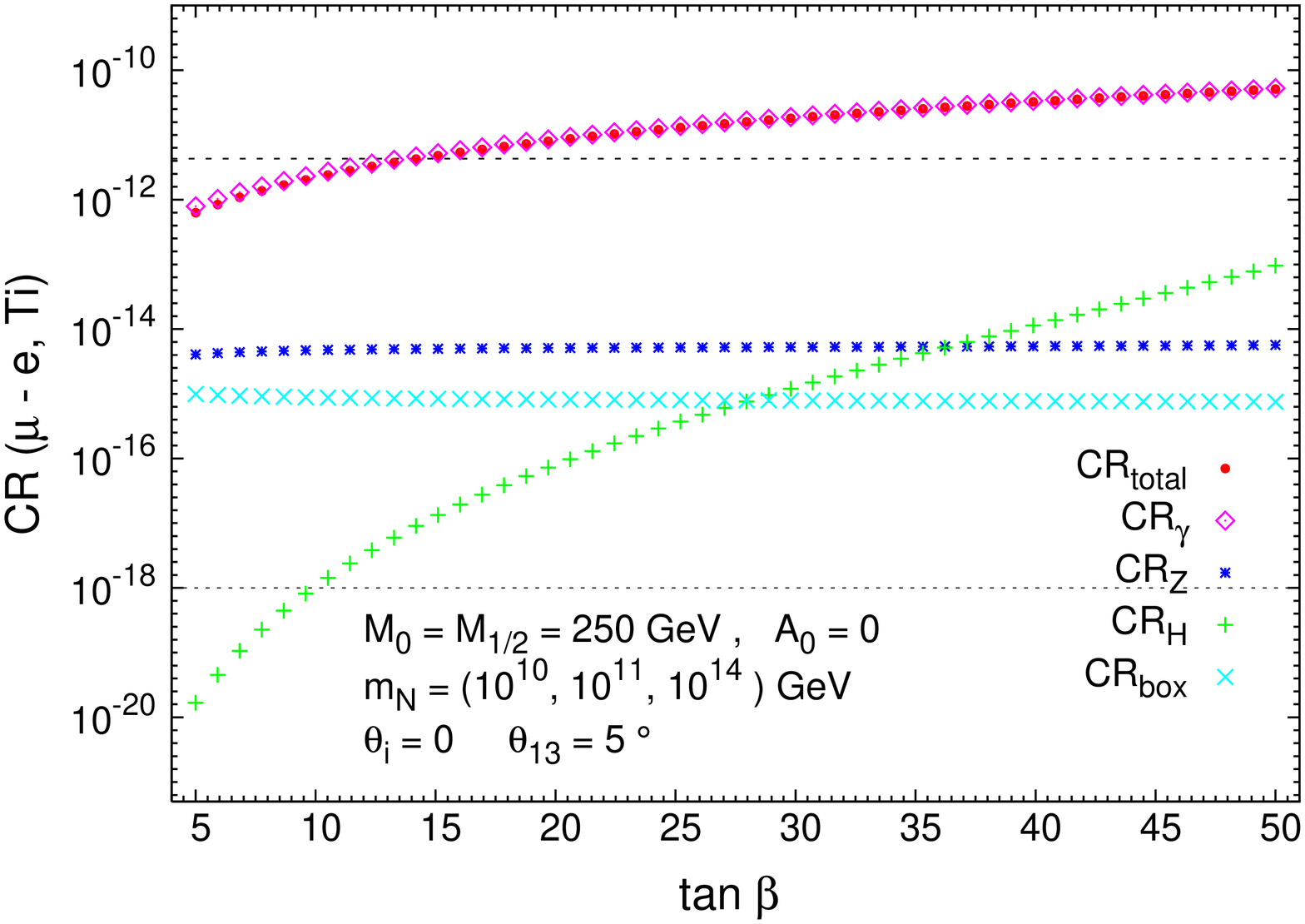,width=60mm,angle=270,clip=} &
      \psfig{file=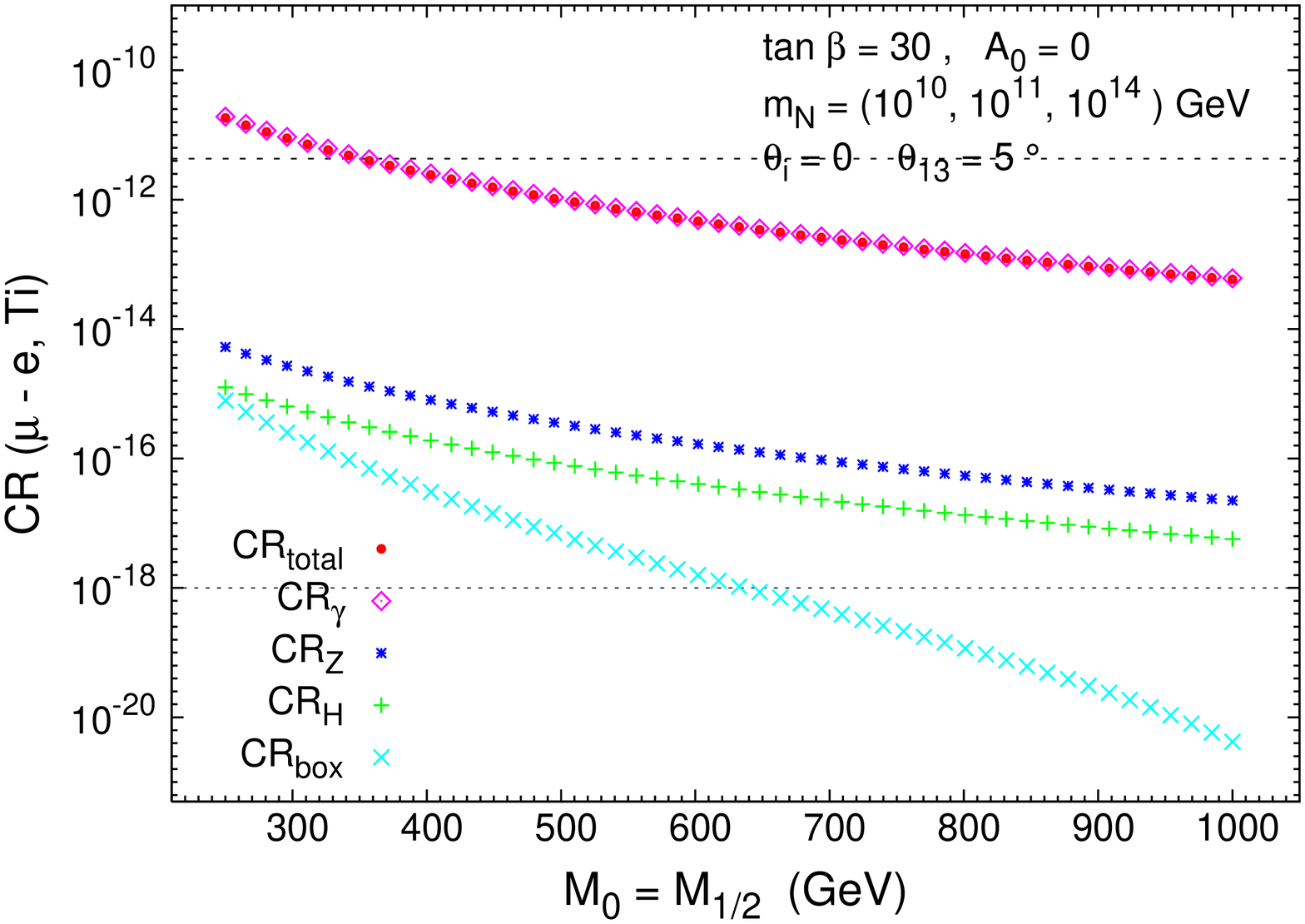,width=60mm,angle=270,clip=}   
    \end{tabular}
    \caption{Contributions to CR($\mu -e$, Ti): 
    total (dots), $\gamma$-penguins (diamonds),
    $Z$-penguins (asterisks), $H$-penguins (crosses) and
    box diagrams (times). On the left we present the dependence on
    $\tan \beta$, for $M_0=M_{1/2}=250$ GeV and $A_0=0$. On the right,
    we exhibit the evolution as a function of $M_0(=M_{1/2})$, for
    $\tan \beta=30$ and $A_0=0$. In either case, we consider
    hierarchical heavy neutrinos with $m_{N_i} = (10^{10},10^{11},10^{14})$
    GeV, and set $\theta_{13}=5^\circ$, and 
    $R=1$ ($\theta_i=0$). 
    A dashed (dotted) horizontal line denotes the present
    experimental bound (future sensitivity).
    }\label{fig:CR:several:tanb:M0} 
  \end{center}
\end{figure}

The numerical results of the CR($\mu-e$, Ti) 
within the CMSSM-seesaw scenario are displayed in
figures~\ref{fig:CR:MRMR3:SPSX} through~\ref{fig:CR:several:tanb:M0}.
The following discussion is focused on the most relevant parameters, namely
$m_{N_i}$, $\theta_{1,2,3}$, $\theta_{13}$, $\tan \beta$,
$M_0$ and $M_{1/2}$. 
 
In Fig.~\ref{fig:CR:MRMR3:SPSX}, we display 
the prediction of CR($\mu -e$, Ti) as a function of the heavy neutrino 
masses for the various SPS points,
and for the particular choice $\theta_i=0$ ($i=1,2,3$) 
and $\theta_{13}=5^\circ$. We also consider the case of degenerate and
hierarchical heavy neutrino spectra (respectively left and right panels).
In both scenarios for degenerate and hierarchical heavy neutrinos, we find 
a strong dependence on the the heavy neutrino masses. 
We also see that the rates for 
the various SPS points exhibit the following hierarchy,
BR$_{4}$~$>$~BR$_\text{1b}$~$\gtrsim$~BR$_\text{1a}$~$>
$~BR$_{3}$~$\gtrsim$~BR$_{2}$~$>$~BR$_{5}$. 
This behaviour can be understood in terms of the growth of the CRs with $\tan
\beta$, and from the different mass spectra associated with each point.  

In the case of degenerate heavy neutrinos, 
we find the expected fast growing behaviour of CR($\mu -e$, Ti) as a
function of the common neutrino mass $m_N$. For the values of $m_N$
within the studied interval 
$\left[10^9 \, \text{GeV},10^{15} \, \text{GeV}\right]$, the predictions for the CR($\mu -e$,
Ti) range over ten orders of magnitude. 
We also see that, for the chosen input parameter values, the predicted rates 
cross the experimental bound for 
the large $m_N$ region. In the latter,  
the Yukawa couplings can be large (for instance, $Y_{33}^\nu$
and $Y_{32}^\nu$ can be $\mathcal{O}(1)$, 
while $Y_{22}^\nu$ and $Y_{21}^\nu$ can be of $\mathcal{O}(10^{-3})$), 
leading to excessively large rates, so that these large 
$m_N$ values are disfavoured by data. 
The experimental bound is saturated for  
$m_N$ values 
ranging from $2 \times 10^{13}$ GeV for SPS 4 
up to about $10^{15}$ GeV for SPS 5. 
In the case of hierarchical heavy neutrinos a similar
behaviour of the predicted rates is found with respect to the heaviest
neutrino mass, 
$m_{N_3}$. 
We have also checked that the conversion 
rates do not significantly depend on 
$m_{N_1}$ and $m_{N_2}$, provided that their values are kept well 
below $m_{N_3}$.
With the planned future sensitivity of $10^{-18}$ it will be possible
to reach into wider regions of the heavy neutrino spectrum. 
Heavy neutrino masses above $10^{12}$ GeV can be probed for the
several considered scenarios.  

For most of the studied points, 
the previously illustrated dependence of the rates 
on the heavy neutrino masses is in agreement
with the expected behaviour 
$~|m_N \log m_N|^2$ obtained in the LLog approximation (as derived 
from Eq.~(\ref{Y21:LLog})). However, a clear departure from this
approximation is found for some points, the most remarkable being 
the case of SPS 5. 
This failure of the LLog approximation has been known to happen in some
scenarios, for instance those with either large
$A_0$, or low $M_0$ and large $M_{1/2}$~\cite{Antusch:2006vw}.   
         
The predictions for CR($\mu -e$, Ti) as a function of the 
$R$-matrix angles, $\theta_{1,2,3}$,
are displayed in Fig.~\ref{fig:CR:theta123:SPS1a:MRMR3}. 
In this case we have fixed the
other relevant parameters as $\theta_{13}=5^\circ$, $m_N=10^{13}$ GeV and
$m_{N_{i}}=(10^{10},10^{11},10^{13})$ GeV (degenerate and hierarchical
heavy neutrinos, respectively) and chosen SPS 1a. To fully 
explore the variation of the rates with the complex angles\footnote{
Complex $\theta_i$ may imply the presence of CP violation in the neutrino
Yukawa couplings. In addition to affecting the LFV rates, these phases 
will induce contributions to flavour-conserving CP violating observables, 
as is the case of charged lepton electric dipole moments (EDMs). 
Throughout the present study we have verified that the associated
predictions for the charged lepton EDMs are in agreement with current
experimental bounds~\cite{Yao:2006px}.} $\theta_i$, 
we have scanned the intervals $0<|\theta_i|<\pi $ rad 
and $0\leq \arg \theta_i \leq \frac{\pi}{2}$ rad. 
From this figure we see that the dependence on the three 
$\theta_i$ is very similar in the degenerate case, 
whereas the same does not occur for hierarchical heavy neutrinos. 
In the former, the rates smoothly grow with both modulus and
argument, and are independent of $\theta_i$ in the real case. 
In the latter, the rates are almost independent of $\theta_3$,
and present a different minima pattern regarding $\theta_1$ and $\theta_2$.
The deep minima occuring in the real case are a consequence
of the corresponding minima appearing 
in the relevant elements of the
Yukawa couplings (as given by Eq.~(\ref{seesaw:casas})).
Notice that the observed behaviour of CR($\mu -e$, Ti) as a function of 
$\theta_i$ can be indeed easily understood from the simple
analytical expression obtained in the LLog approximation (cf. 
Eq.~(\ref{Y21:LLog})).

The most important outcome from Fig.~\ref{fig:CR:theta123:SPS1a:MRMR3}
is that for both cases of degenerate and hierarchical heavy neutrinos,
complex values of $\theta_i$ can increase the $\mu-e$ conversion rates 
by almost five orders of magnitude with respect to the $\theta_i=0$ case.
Only for a few specific choices of $\theta_i$ (for instance real $\theta_1$
or $\theta_2$, in the hierarchical case) can we observe a strong decrease
with respect to the $\theta_i=0$ case, but clearly this is not a
generic situation. 

In the following, and in order to simplify the analysis with respect to
the other parameters, we will set $\theta_i=0$, and assume that the 
corresponding predictions for the CR($\mu -e$, Ti) 
will constitute a 
representative case for the lowest conversion rates.  

In Fig.~\ref{fig:CR:theta13:SPSX:MRMR3} we show
the dependence of the $\mu-e$
conversion rates on the light 
neutrino mixing angle $\theta_{13}$. 
The other parameters are set to 
$m_N=10^{14}$ GeV, 
$m_{N_{i}} =(10^{10},10^{11},10^{14})$ GeV, 
( respectively for degenerate and hierarchical
heavy neutrinos) and $\theta_i=0$.
All the SPS points in Table~\ref{SPS:def:15} have been considered. 
For degenerate heavy neutrinos, the
dependence on $\theta_{13}$ is softer than what is observed for the 
hierarchical case, leading to a
variation in the rates of at most one order of magnitude 
in the studied range of 
$0^\circ \leq \theta_{13} \leq 10^\circ$ (the only exception being SPS 5,
where the  variation can reach up to two orders of magnitude). 
In contrast, this figure clearly manifests the very strong sensitivity
of the CR($\mu -e$, Ti) to the $\theta_{13}$ mixing angle for
hierarchical heavy neutrinos. 
In the hierarchical case, a variation of $\theta_{13}$ in the studied interval 
leads to an increase in the conversion rates by as much as five orders
of magnitude.
This huge variation is due to the strong decrease of this observable for very
small $\theta_{13}$ angles, as can be easily understood from the dependence on
this angle of the dominant $(L_{33} m_{N_3} \sqrt{m_{\nu_3}} c_1^* c_2^*
s_{13})$ term in Eq.~(\ref{Y21:LLog}). Furthermore, the minimum of CR$(\mu - e,
\text{Ti})$ is expected to occur at a vanishing mixing angle, but this being the
value at the seesaw scale, i.e., $\theta_{13}(m_M) = 0$. The deep minima in
Fig.~\ref{fig:CR:theta13:SPSX:MRMR3} are at $\theta_{13}(m_Z) \simeq 0.2^{\circ}$,
which is precisely the RGE shifted value at $m_Z$ from $\theta_{13}(m_M) =
0$. As $\theta_{13}$ grows, the predictions for 
SPS 4, SPS 1a and SPS 1b cross the present experimental bound. 
In particular, notice that for SPS 4,
and for the present choice of input parameters, $\theta_{13}$ values
larger than $2^\circ$ would be excluded by present data. 

An equally remarkable sensitivity to $\theta_{13}$ has been found in
other $\mu-e$ violating processes, 
like $\mu \to e \gamma$ and $\mu \to 3 e$, and also in
tau decays as is the case of 
$\tau \to e \gamma$ and $\tau \to 3e$~\cite{Antusch:2006vw}. 
This interesting behaviour with $\theta_{13}$ has been 
proposed in~\cite{Antusch:2006vw} as a powerful 
tool to test the seesaw-I hypothesis for neutrino mass generation and, 
in case of a 
measurement of these branching ratios, as a unique
way to derive some hints on the seesaw parameters, 
especially on the value of $m_{N_3}$.
The $\mu-e$ conversion rates here presented will certainly add new 
interesting information on this type of analysis. 
Fig.~\ref{fig:CR:theta13:SPSX:MRMR3} also
shows that with the expected future sensitivity of $10^{-18}$, the full  
$0^\circ \leq \theta_{13}\leq 10^\circ$ interval can be thoroughly covered.

In the following study we will restrict ourselves to the 
hierarchical case where we have found this strong sensitivity to
$\theta_{13}$. For
definiteness, we will also 
fix the heavy neutrino masses and $\theta_{13}$ to ``reference'' values of  
$m_{N_i} =(10^{10},10^{11},10^{14})$ GeV and $\theta_{13}=5^\circ$. 

Fig.~\ref{fig:CR:several:tanb:M0} illustrates the predictions for
the  CR($\mu -e$, Ti) as a function of $\tan \beta$, $M_0$ and
$M_{1/2}$. Here we have separately displayed
the various contributions to the $\mu-e$ conversion rates in order to
conclude about their relative importance in this CMSSM-seesaw
scenario. We set the values of the remaining CMSSM
parameters to $M_0=M_{1/2}=250$ GeV in the study with $\tan \beta$
(left panel) and to $\tan \beta=30$ in the study with 
$M_{\rm SUSY} \equiv M_0=M_{1/2}$ (right panel), taking 
$A_0=0$ in both cases.
We choose our ``reference'' values of 
$m_{N_{1,2,3}} =(10^{10},10^{11},10^{14})$ GeV, $\theta_{13}=5^\circ$, and
$\theta_i=0$. 

In both panels of Fig.~\ref{fig:CR:several:tanb:M0} 
we clearly observe the dominance of the photon-mediated
contributions, which are in fact indistinguishable from the total CR, 
for all the explored  
parameter ranges. The dependence of the various
contributions on $\tan \beta$ illustrates the expected fast growing
behaviour with
$\tan^6 \beta$ of the Higgs-mediated contributions, and the milder  
$\tan^2 \beta$ dependence of the photon-mediated ones. In addition, we see 
that the $Z$ boson-mediated and the box diagram contributions
are almost independent of $\tan \beta$. 
Although not displayed in this plot, we have also verified that the 
Higgs-mediated contribution is largely dominated by the exchange of
$H^0$,  which is indeed
the Higgs boson with enhanced couplings to charged leptons 
in the large $\tan \beta$ regime.  
  
The decoupling behaviour for large $M_{\rm SUSY}$ 
of each of these contributions (${\rm CR}_\gamma$, ${\rm CR}_Z$,  
${\rm CR}_H$ and ${\rm CR}_{\rm box}$) is clearly  
manifested in the right panel of Fig.~\ref{fig:CR:several:tanb:M0}.
The most important conclusion from this figure is that, within
a CMSSM-seesaw scenario, the $\gamma$-penguin diagrams completely dominate
the conversion rates, even for the largest $\tan \beta$ considered 
($\tan \beta=$50). Therefore, the total CR($\mu -e$, Ti) does not
manifest the Higgs contributions, so that in this universal scenario 
there is no chance for the $\mu-e$ conversion process to provide any 
information on the Higgs sector. 
We will see next that the situation is remarkably different in the 
non-universal case, where the Higgs contributions turn
out to be much larger than in the universal case. 

\subsection{Non-universality: NUHM-seesaw}
\begin{figure}[h]
  \begin{center}  
    \begin{tabular}{cc}\hspace*{-10mm}
      \psfig{file=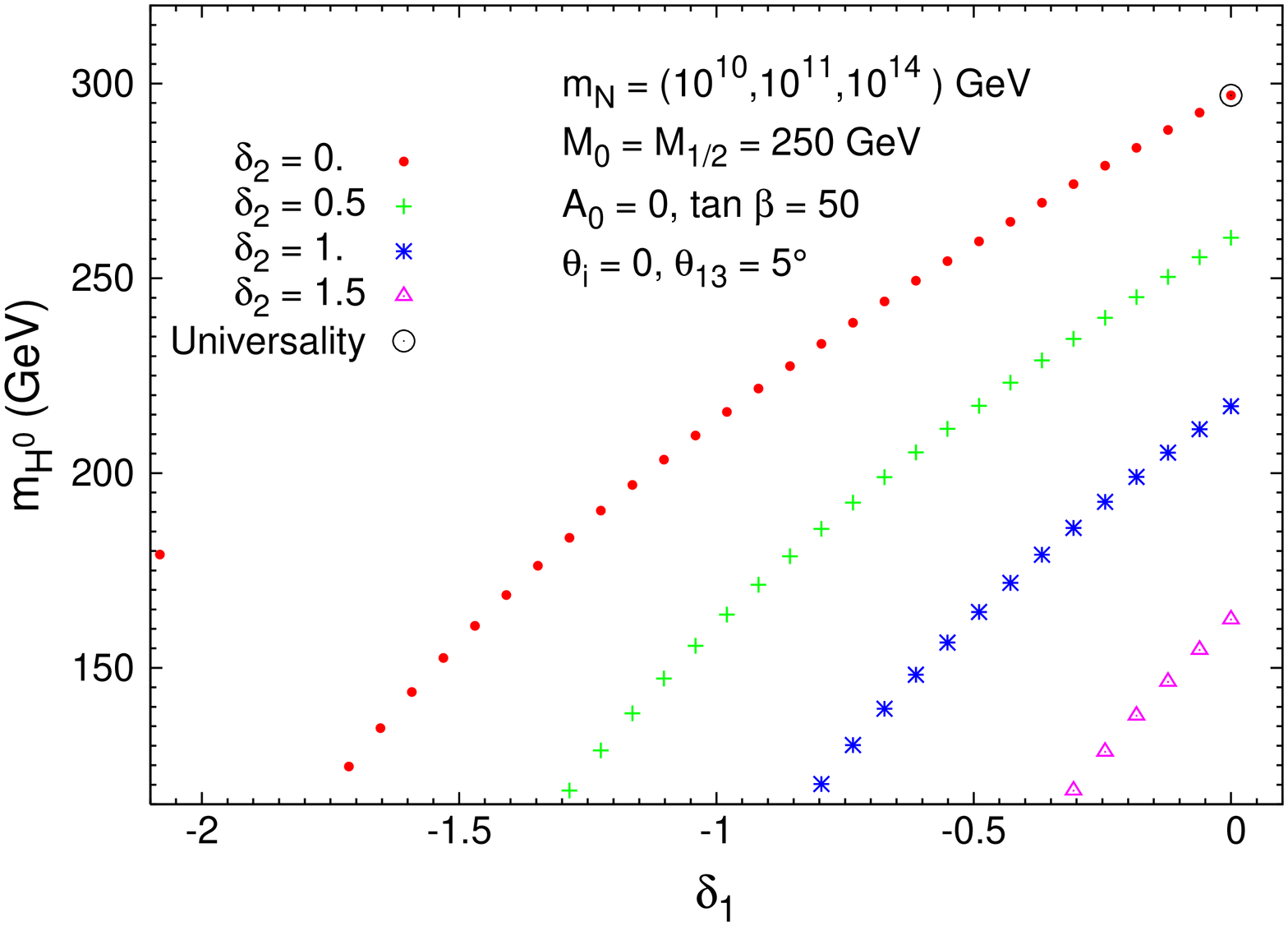,width=60mm,angle=270,clip=} &
      \psfig{file=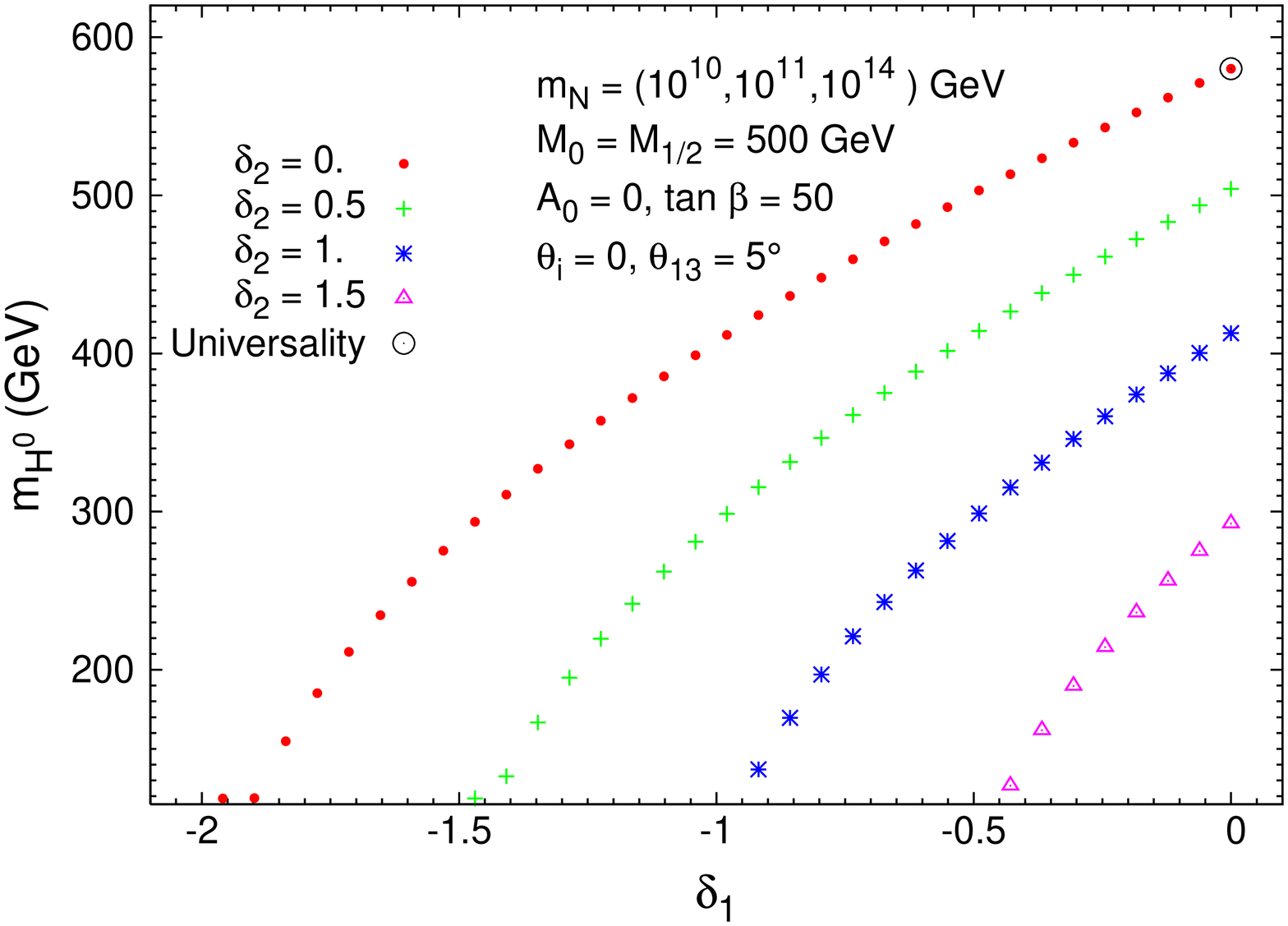,width=60mm,angle=270,clip=} \\\hspace*{-10mm}
      \psfig{file=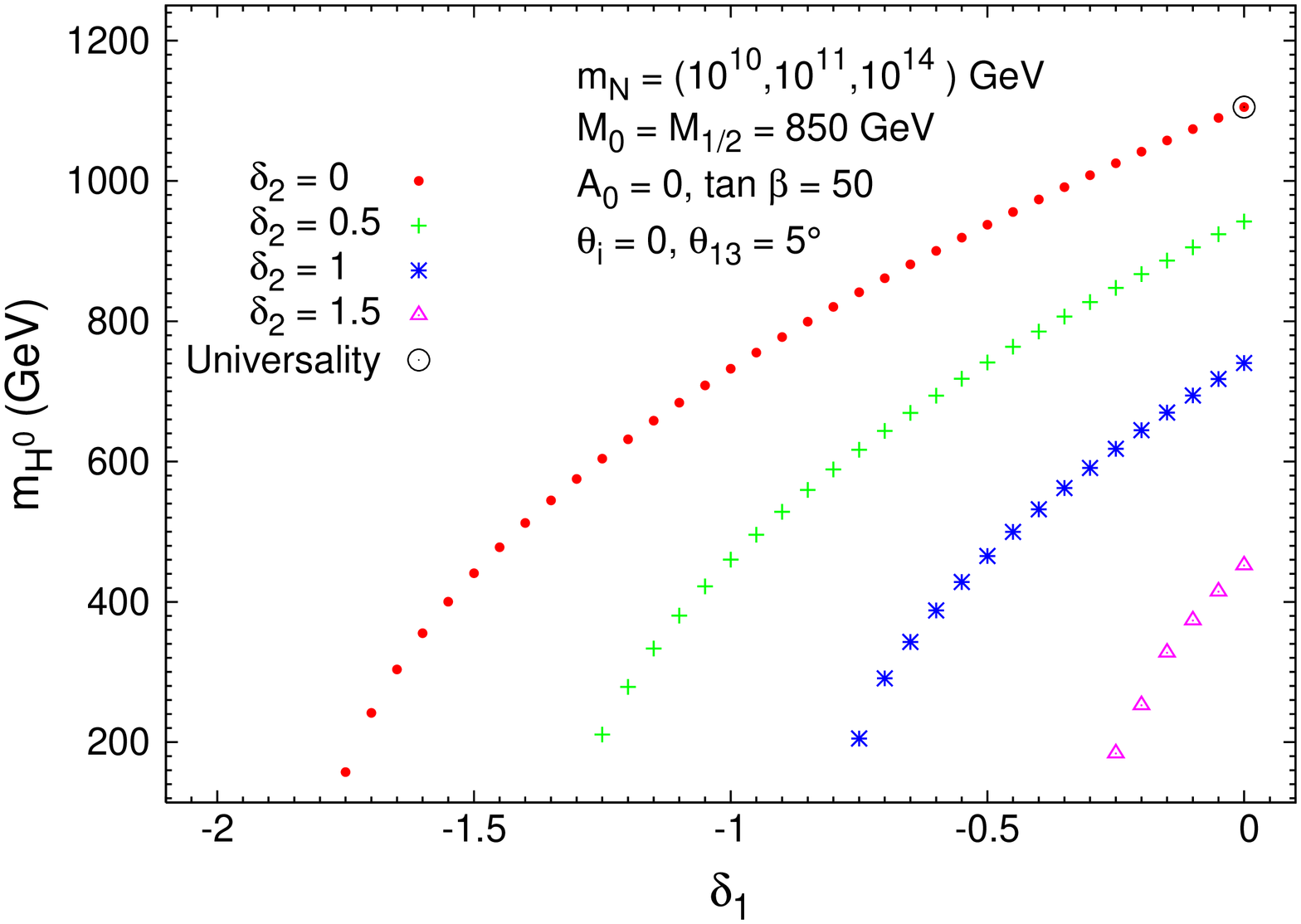,width=60mm,angle=270,clip=} &
      \psfig{file=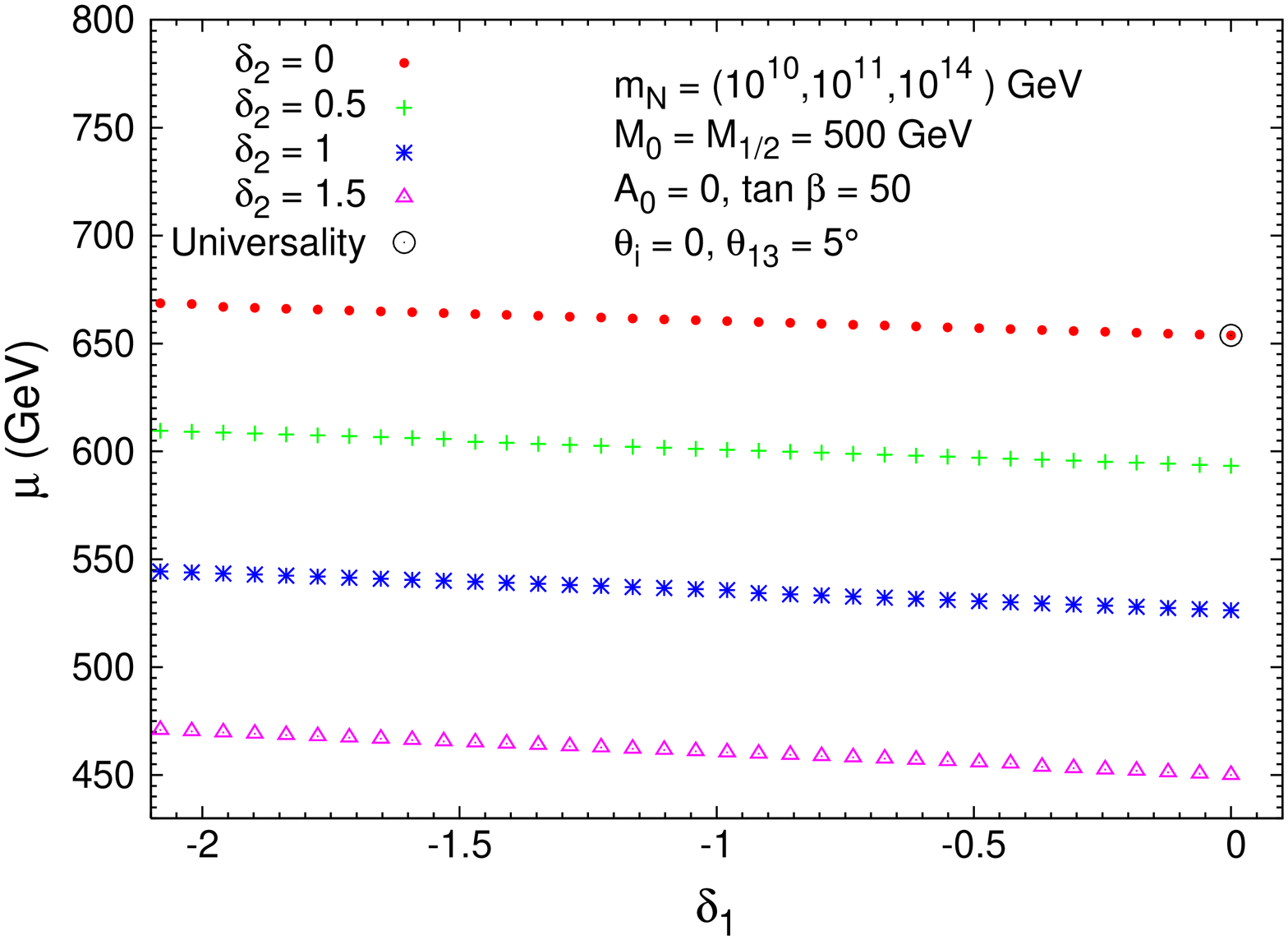,width=60mm,angle=270,clip=}
    \end{tabular}
    \caption{Mass of the heaviest Higgs scalar ($m_{H^0}$) as a function of 
      the non-universality parameter $\delta_1$, for fixed values of
      $\delta_2= \{0,\,0.5,\,1,\,1.5\}$ (respectively 
      dots, crosses, asterisks, triangles). The universality case
      $\delta_{1,2} = 0$ is
      represented by a large circle. We also take
      $m_{N_i} = (10^{10},10^{11},10^{14})$
      GeV, set $\theta_i=0$, $A_0=0$, $\tan \beta=50$ and
      impose the relation $M_0=M_{1/2}$. The first three plots
      correspond to $M_0=250,\,500$ and 850 GeV, respectively.
      On the fourth plot, we display the $\mu$ parameter as a function of 
      the non-universality parameter $\delta_1$, for fixed values of
      $\delta_2$, and for $M_0=M_{1/2}=500$ GeV.
        }\label{fig:MH0andmu:delta1delta2} 
  \end{center}
\end{figure}

\begin{figure}[t]
  \begin{center} 
    \begin{tabular}{cc}\hspace*{-10mm}
      \psfig{file=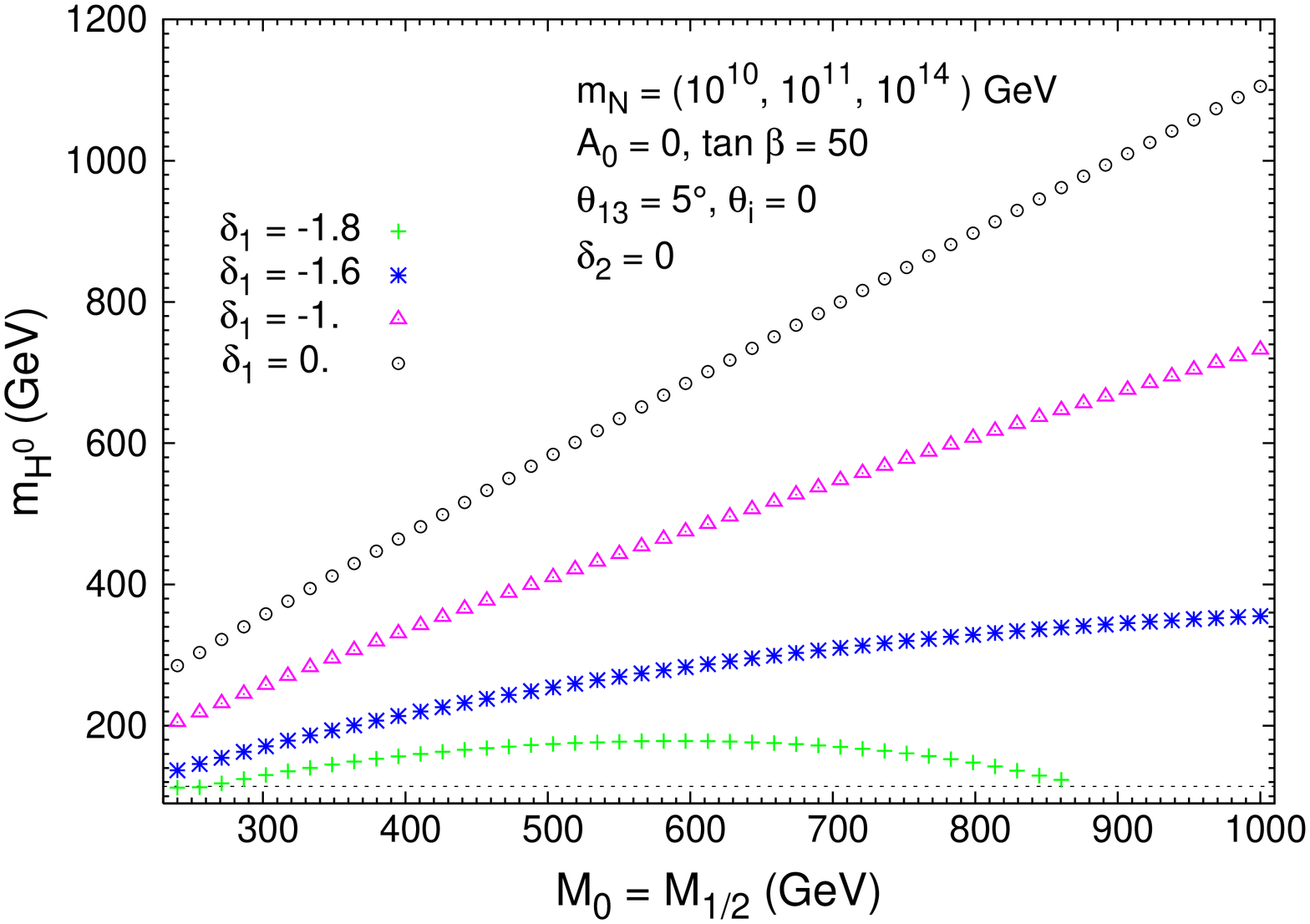,width=60mm,angle=270,clip=} &
      \psfig{file=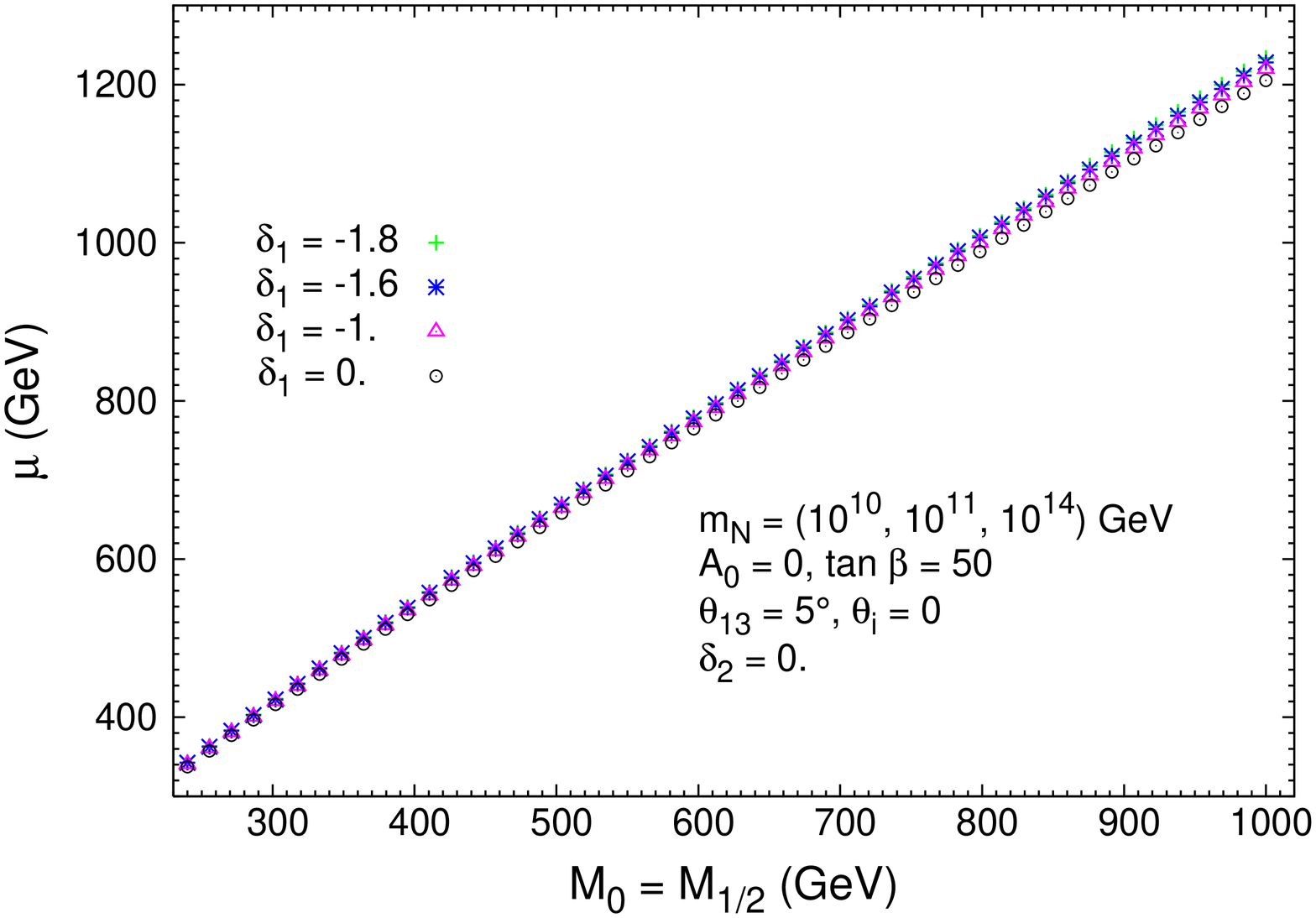,width=60mm,angle=270,clip=}
    \end{tabular}
    \caption{On the left, mass of the heaviest Higgs scalar ($m_{H^0}$) 
      as a function of the SUSY scale ($M_\text{SUSY}=M_0=M_{1/2}$), 
      for fixed values of $\delta_1=\{-1.8,\,-1.6,\,-1,\,0\}$ 
      (respectively crosses, asterisks, triangles and circles). 
      We take $m_{N_i} = (10^{10},10^{11},10^{14})$
      GeV, and set $\theta_i=0$, $A_0=0$, $\tan \beta=50$ with
      $\theta_{13}=5^\circ$. On the right, the $\mu$-parameter is displayed as a
      function of $M_\text{SUSY}$, for the same choices of SUSY-seesaw
      parameters. 
        }\label{fig:MH0andmu:M0M12} 
  \end{center}
\end{figure}

\begin{figure}[h]
  \begin{center}
    \begin{tabular}{cc}\hspace*{-10mm}
    \psfig{file=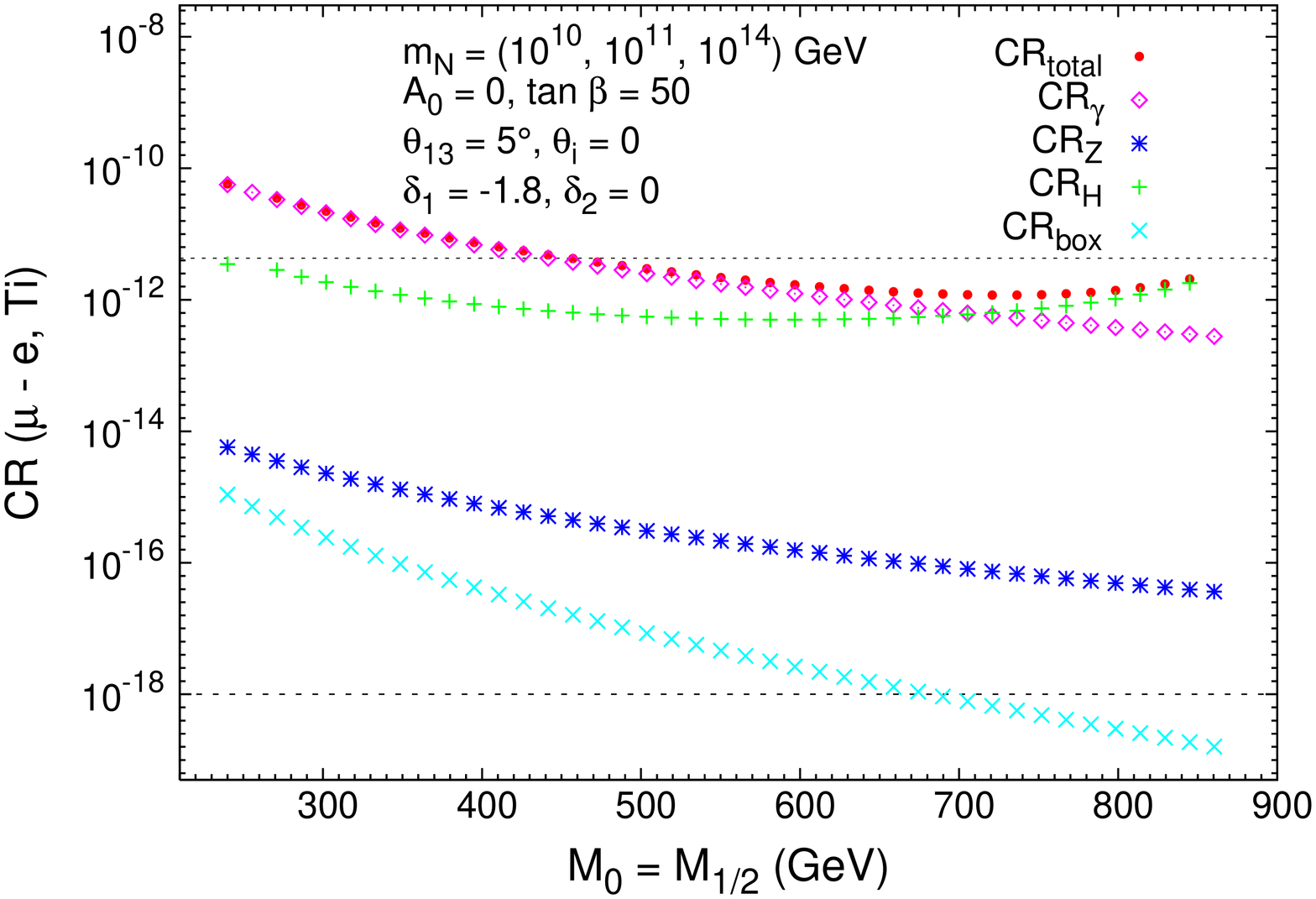,width=60mm,angle=270,clip=} &
    \psfig{file=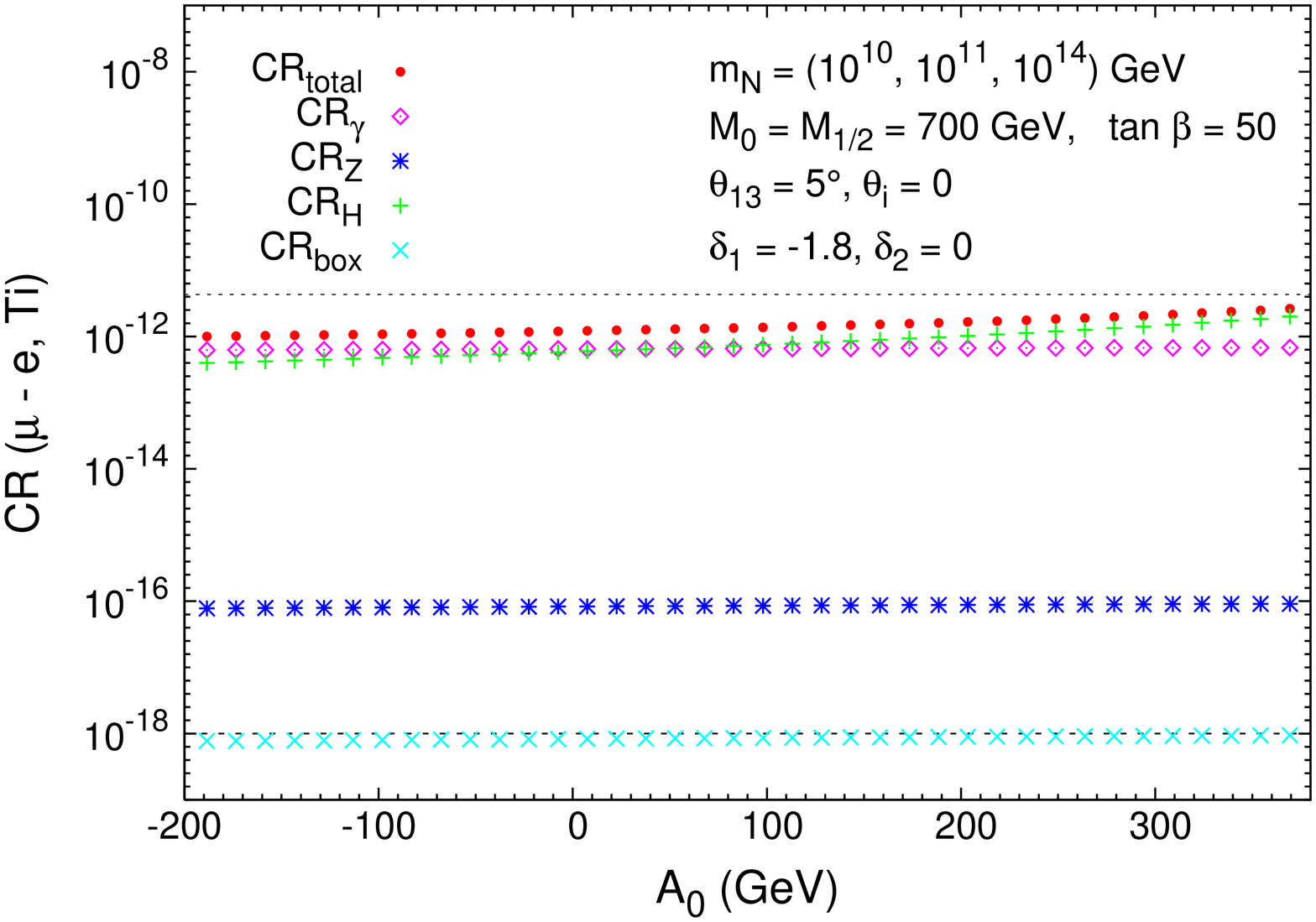,width=60mm,angle=270,clip=}
    \end{tabular}
    \caption{Contributions to CR($\mu -e$, Ti) 
      as a function of $M_0(=M_{1/2})$ (left) and $A_0$ (right): total (dots),
      $\gamma$-penguins ( diamonds),
      $Z$-penguins (asterisks), $H$-penguins (crosses) and
      box diagrams (times), for
      $\delta_1=-1.8$ and $\delta_2=0$. We set $\tan \beta =
      50$ and take $\theta_{13}=5^\circ$,
      $R=1$ ($\theta_i=0$) and $m_{N_i} = (10^{10},10^{11},10^{14})$
      GeV. On the left $A_0=0$, while on the right we choose 
      $M_0(=M_{1/2})=700$ GeV. In each case, a dashed (dotted)
      horizontal line denotes the present experimental bound (future
      sensitivity).
        }\label{fig:CR:NUHM:M0M12:A0} 
  \end{center}
\end{figure}

\begin{figure}[h]
  \begin{center} 
 \psfig{file=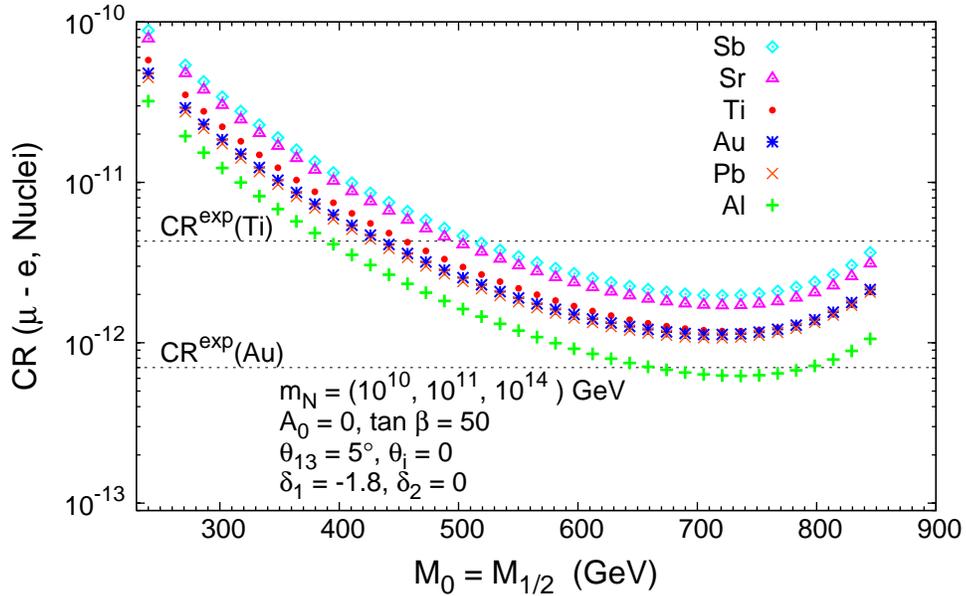,width=90mm,angle=270,clip=}
\caption{$\mu-e$ conversion rates for various nuclei as a function of
  $M_0=M_{1/2}$ in the NUHM-seesaw. We display the theoretical
  predictions for  Sb, Sr, Ti, Au, Pb and Al nuclei (diamonds,
  triangles, dots, asterisks, times and crosses, respectively).
  We have taken $m_{N_i} =
  (10^{10},10^{11},10^{14})$ GeV, $A_0=0$, $\tan \beta =
  50$,  $\theta_{13}=5^\circ$ and $R=1$ ($\theta_i=0$). The
  non-universality parameters are set to 
  $\delta_1=-1.8$ and $\delta_2=0$. From top to bottom, 
  the horizontal dashed lines denote the present experimental bounds
  for CR($\mu -e$, Ti) and CR($\mu -e$, Au).
}
\label{fig:variosnucleos} 
  \end{center}
\end{figure}

\begin{figure}[t]
  \begin{center}
    \begin{tabular}{cc}\hspace*{-10mm}
    \psfig{file=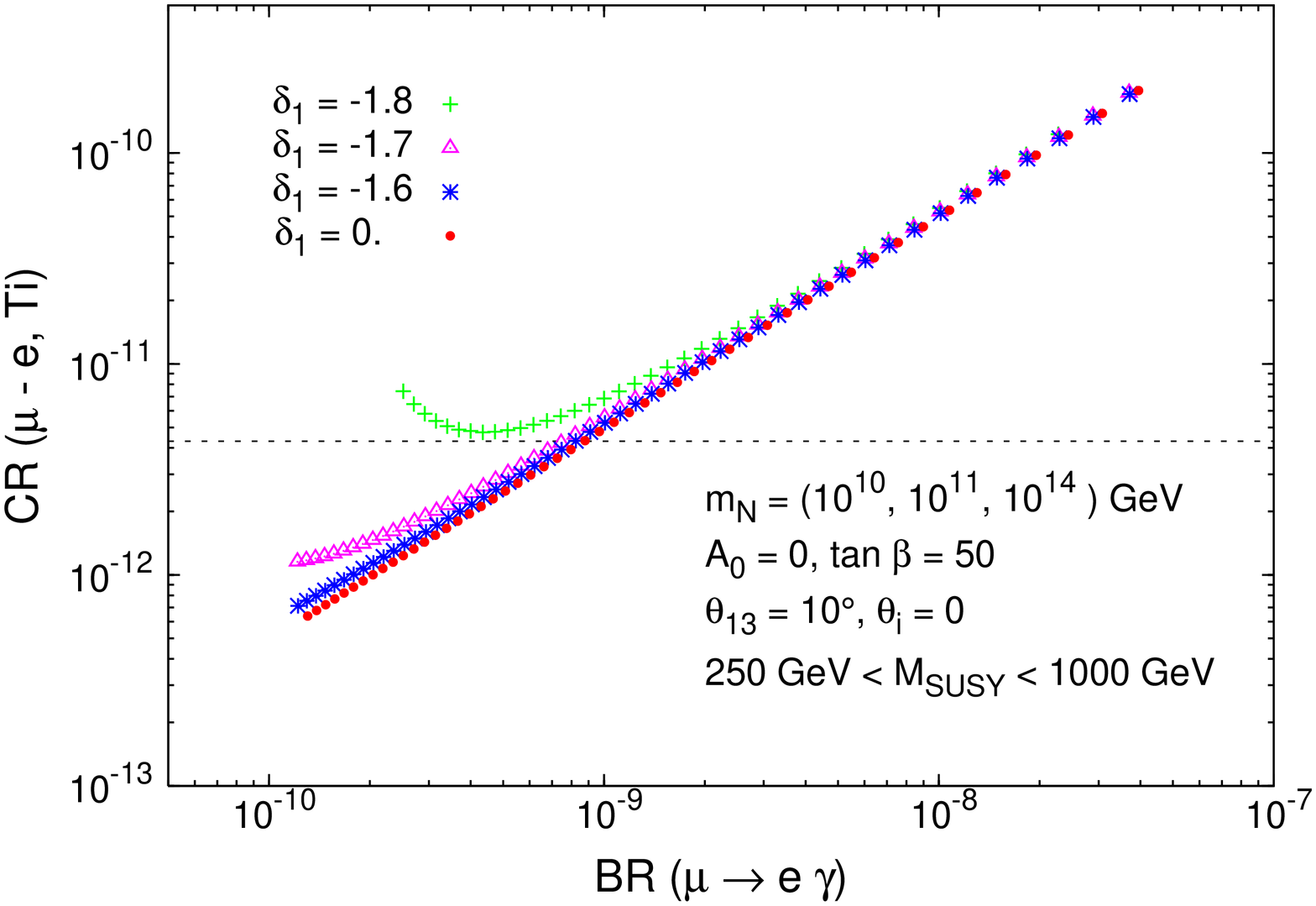,width=60mm,angle=270,clip=} &
    \psfig{file=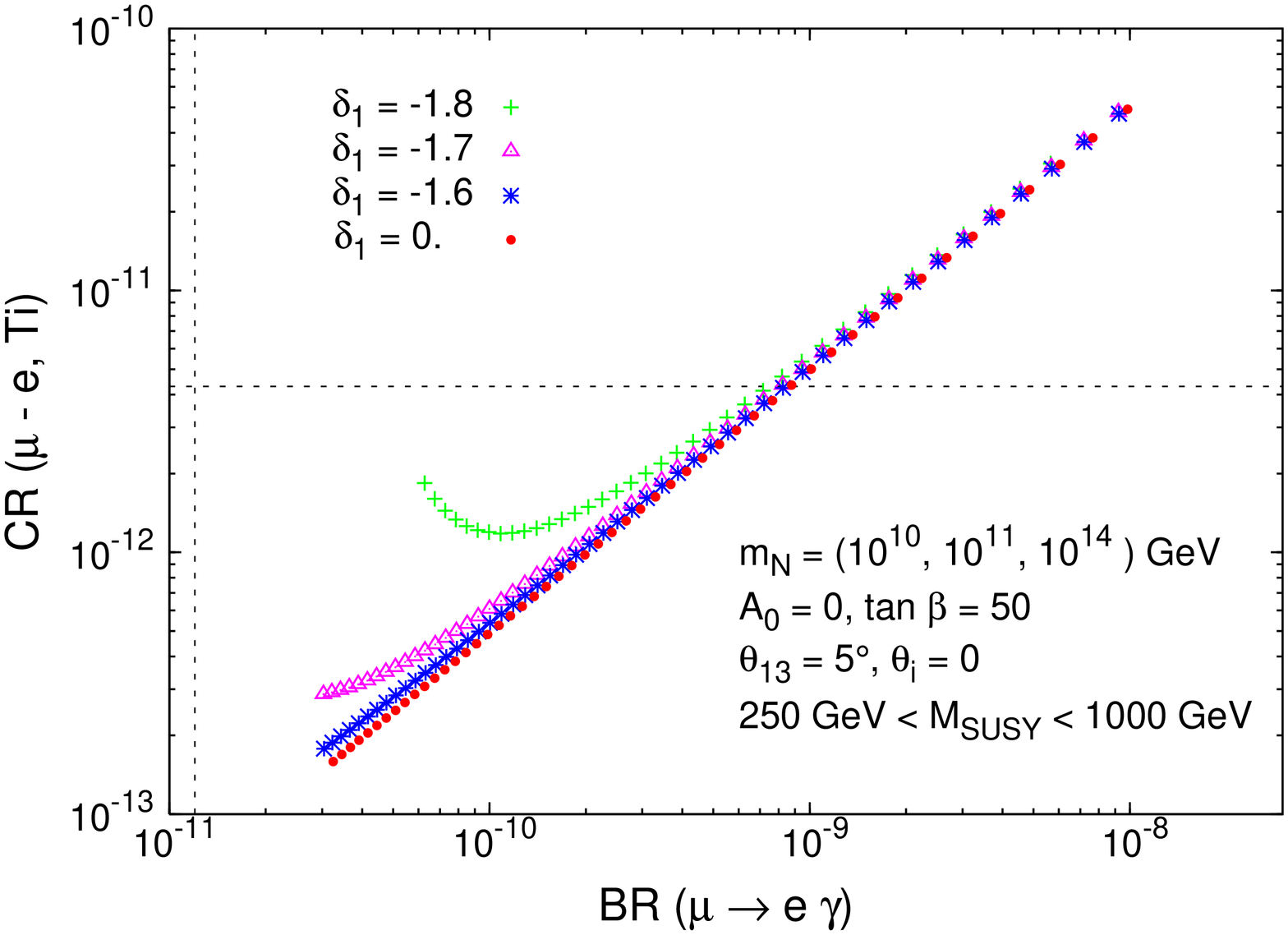,width=60mm,angle=270,clip=} \\\hspace*{-10mm}
    \psfig{file=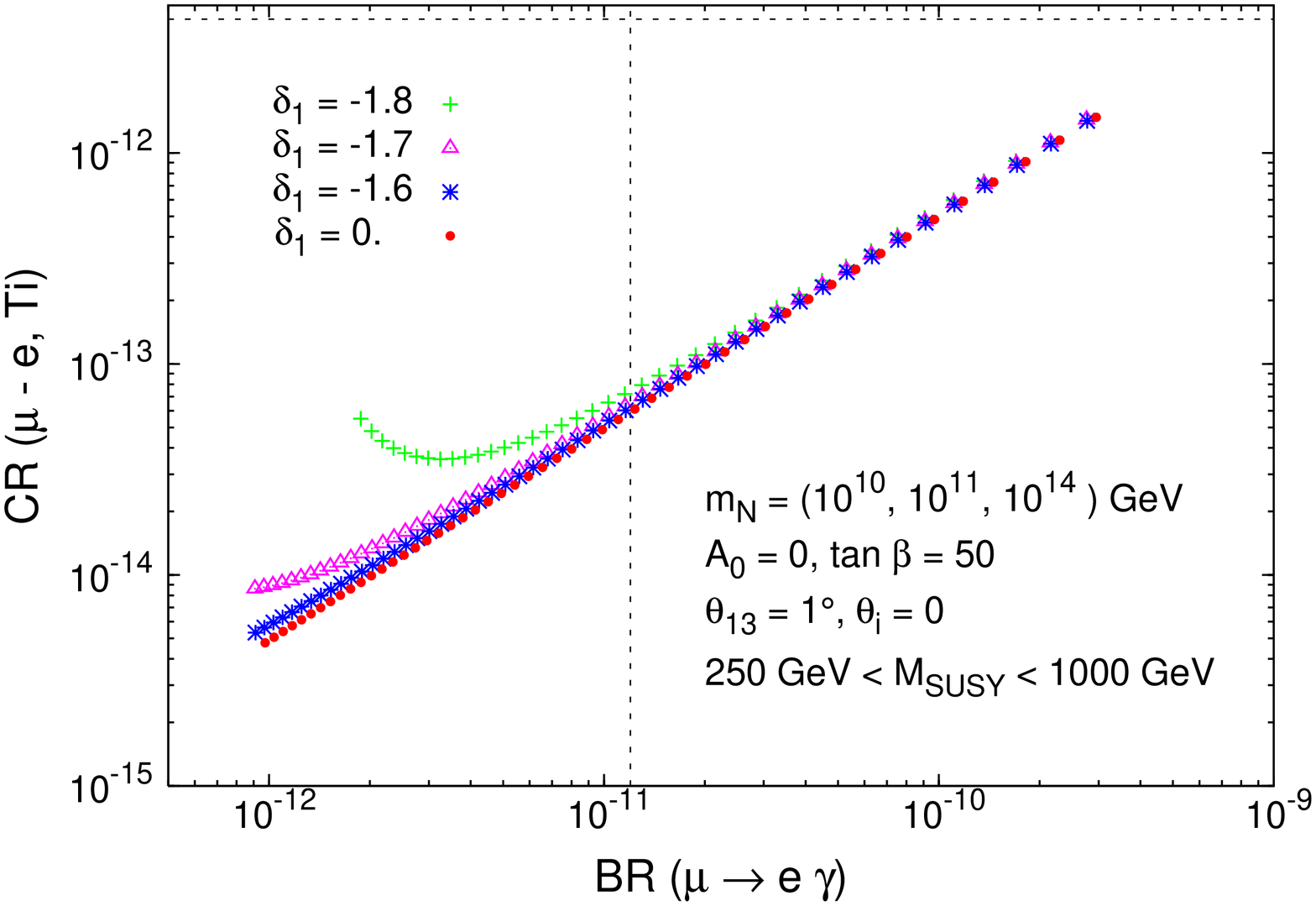,width=60mm,angle=270,clip=} &
   \psfig{file=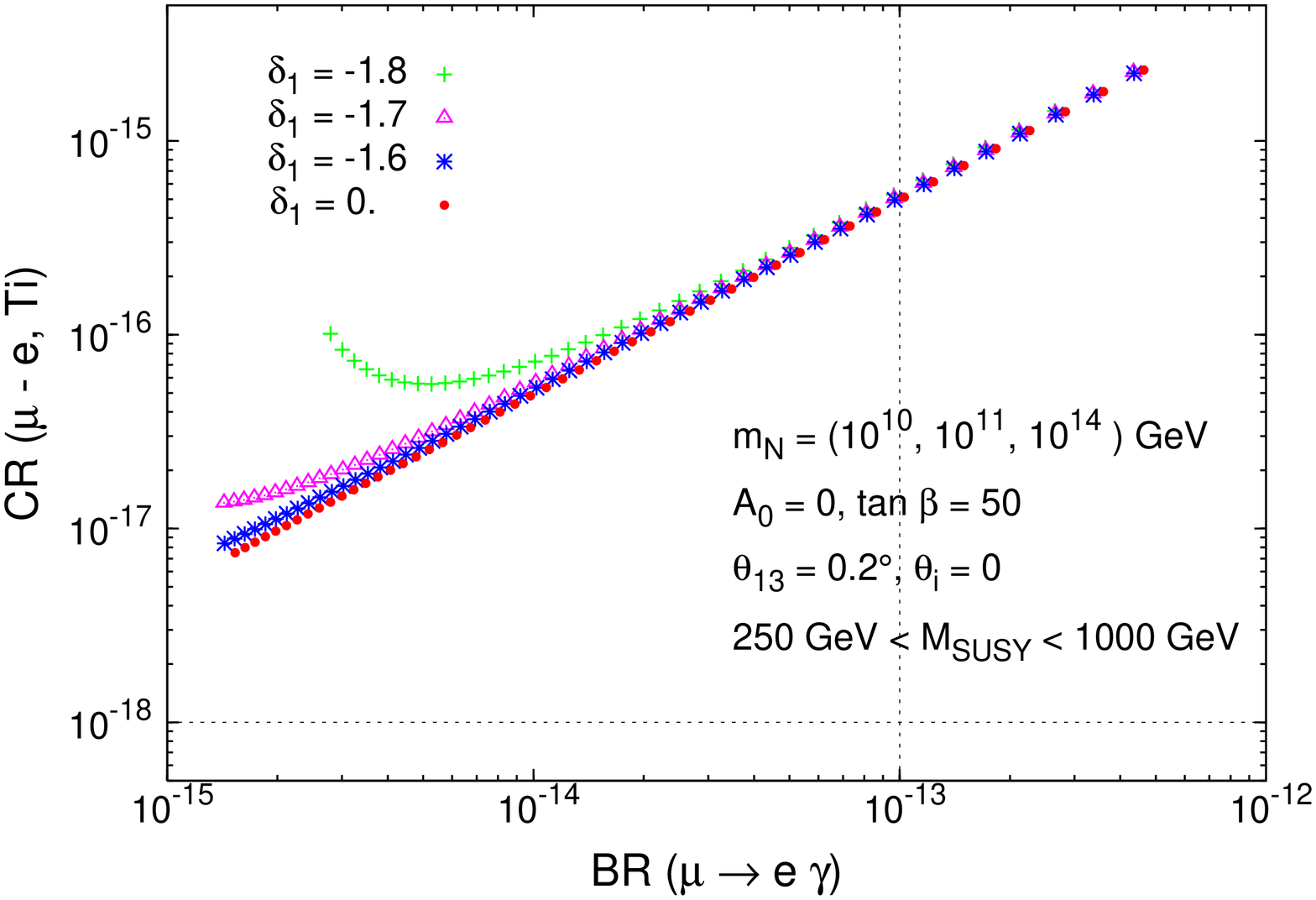,width=60mm,angle=270,clip=}
    \end{tabular}
    \caption{CR($\mu -e$, Ti) versus BR($\mu \to e \gamma$) for $250
    \text{ GeV} \leq M_\text{SUSY}\leq 1000 \text{ GeV}$, and
    $\delta_1= -1.8,\,-1.7,\,-1.6,\,0$ (crosses, triangles, asterisks,
    dots, respectively). We set $\delta_2=0$, and take $m_{N_i} =
    (10^{10},10^{11},10^{14})$ GeV, $A_0=0$, $\tan \beta =
    50$ and $R=1$ ($\theta_i=0$). From left to right and top to bottom,
    the panels are associated with $\theta_{13}=10^\circ, 5^\circ,
    1^\circ$ and $0.2^\circ$. In each case, the horizontal and vertical dashed
    (dotted) lines denote the present experimental bounds (future
    sensitivities) for  CR($\mu -e$, Ti) and BR($\mu \to e \gamma$),
    respectively.
    }\label{fig:CR_versus_muegamma} 
  \end{center}
\end{figure}

\begin{figure}[t]
  \begin{center} 
 \psfig{file=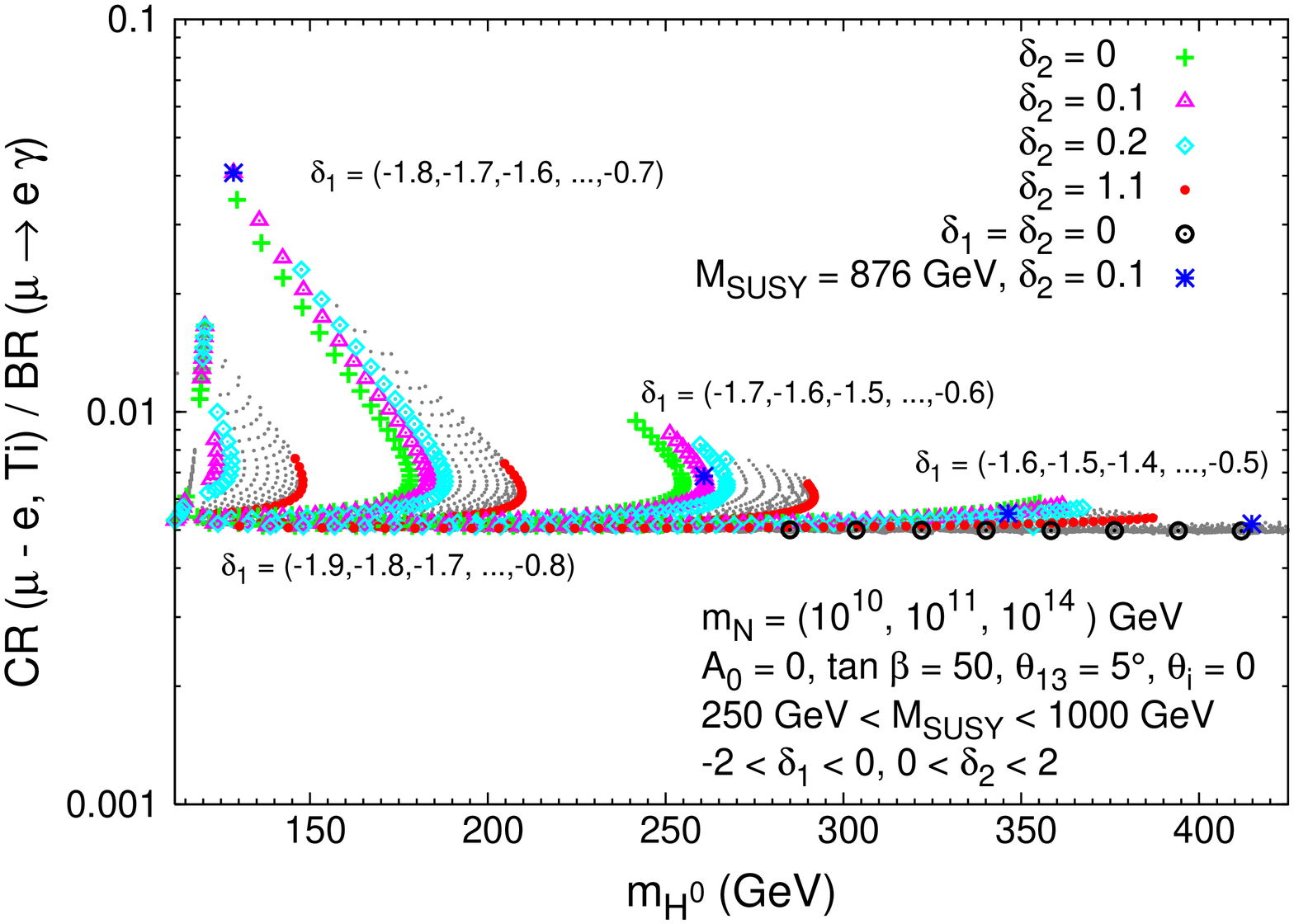,width=90mm,angle=270,clip=}  
 \caption{Ratio CR($\mu -e$, Ti)/BR($\mu \to e \gamma$) as a function
   of the Higgs mass, $m_{H^0}$. We take $m_{N_i} =
   (10^{10},10^{11},10^{14})$ GeV, $A_0=0$, $\tan \beta =
   50$, $\theta_{13}=5^\circ$ and $R=1$ ($\theta_i=0$), and scan over
   250 GeV $\leq$ $M_{\rm SUSY}$ $\leq$ 1000 GeV, $-2 \leq \delta_1
   \leq 0$, and  $0\leq \delta_2 \leq 2$ (grey dots). We have
   highlighted specific choices of $\delta_2=0,\,0.1,\,0.2,\,1.1$
   (crosses, triangles, diamonds, dots, respectively). In each tilted
   cluster, we have also indicated the values of $\delta_1$ associated
   with the $\delta_2$ coloured points. The universality limit
   ($\delta_1=\delta_2=0$) is denoted by a circle. Asterisks denote
   points with $M_{\text{SUSY}}=876$ GeV and $\delta_2=0.1$.}\label{fig:mu-e/muegamma} 
  \end{center}
\end{figure}

The numerical results for the NUHM-seesaw scenario are collected 
in figures~\ref{fig:MH0andmu:delta1delta2} through~\ref{fig:mu-e/muegamma}.

In order to study the influence of the hypothesis of non-universal Higgs
soft SUSY breaking masses, $M_{H_{1,2}}$, on the $\mu-e$ conversion rates, we
have first explored the impact of the non-universality parameters 
$\delta_1$ and $\delta_2$
on the predicted Higgs boson masses. 
The values for these parameters have been taken to lie 
within the interval $-2 \leq \delta_{1,2} \leq 2$.

The predictions for the relevant Higgs boson mass, $m_{H^0}$, as a function
of $\delta_1$ and $\delta_2$ are summarised in 
Fig.~\ref{fig:MH0andmu:delta1delta2}. We have chosen here the largest value of
$\tan \beta=50$ and three representative values of
$M_{\rm SUSY}=$ 250, 500 and 850 GeV for moderate, heavy
and very heavy SUSY spectra, respectively. The other parameters are set to our
``reference'' values of $m_{N_i} = (10^{10},10^{11},10^{14})$ GeV, 
$\theta_i=0$, $A_0=0$, $\theta_{13}=5^\circ$ and ${\rm sign} (\mu)=+1$.

First, it is important to mention that not all the considered values
of the $\delta_{1,2}$ parameters and $M_{\rm SUSY}$
allow for a correct $SU(2)\times U(1)$ breaking. In fact some particular
choices for $\delta_1$, $\delta_2$, and $M_{\rm SUSY}$ lead to unacceptable
negative values of $B\mu$ (and hence, negative $m^2_{A^0}$). 
For instance, this is the case when $\delta_{1,2}$ are simultaneously
positive or negative. Some other points, despite 
leading to a proper $SU(2)\times U(1)$ breaking, are nevertheless not 
acceptable, since they lead to a Higgs boson sector 
which is too light, with masses
below the present experimental lower limits. To ensure that our
results are indeed experimentally viable, 
we have included in this, and in the following figures, 
only the solutions where the three neutral 
Higgs boson masses are above the experimental bound for the lightest MSSM
Higgs boson, which at present is 110 GeV for $\tan{\beta} > 5$ ($99.7\%$ C.L.)~\cite{Yao:2006px}.
The most interesting solutions with important phenomenological
implications are found for negative $\delta_1$ and positive $\delta_2$,
the choice selected for Fig.~\ref{fig:MH0andmu:delta1delta2}.
In this figure, 
for all the explored values of $\delta_1$ and $\delta_2$,
we find a value of $m_{H^0}$ that is significantly 
smaller than what one would encounter in the universal
case (here represented by the choice $\delta_1=\delta_2=0$). 
This is truly remarkable in the case of large soft breaking masses,
as can be seen, for instance,
in the panel with $M_{\rm SUSY}=850$ GeV, where low values of $m_{H^0}$ are
still found, 
even close to the experimental limit. 
For completeness we have also shown in
Fig.~\ref{fig:MH0andmu:delta1delta2} the 
predictions for the $\mu$ parameter as a function of 
$\delta_1$ and $\delta_2$. 
This parameter turns out to be nearly independent of
$\delta_1$, and its largest values are obtained for $\delta_2= 0$.

The behaviour of the predicted $m_{H^0}$ and $\mu$ parameter as a function of 
$M_{\rm SUSY}=M_0=M_{1/2}$ is shown in
Fig.~\ref{fig:MH0andmu:M0M12}. Here the
specific values of $\delta_1=\{-1.8,\,-1.6,\,-1,\,0\}$ 
and $\delta_2=0$ have been considered. 
This figure again illustrates the interesting departure from the linear
behaviour of $m_{H^0}$ with $M_{\rm SUSY}$, which is 
generic in the universal case ($\delta_{1,2}=0$). 
In contrast, the $\mu$ parameter conserves 
a similar linear behaviour with $M_{\rm SUSY}$ in all the studied
scenarios (universal and non-universal).  

As a representative example of these interesting 
non-universal points, we explicitly refer to the case with $\delta_1=-1.8$
and $\delta_2=0$, where the predicted masses are $m_{H^0}=$ 113, 174 and 
127 GeV for $M_{\rm SUSY}=$ 250, 500 and 850 GeV, respectively. For
completeness, we have also collected the corresponding 
masses of the other relevant SUSY particles in Table~\ref{NUHMspectra}. 
Notice that, in the case of $M_{\rm SUSY}=$ 850 GeV, this 
table illustrates a very heavy SUSY spectrum, 
even with a considerably heavy lightest SUSY particle, 
$m_{\tilde{\chi}_1^0}=362$ GeV, but where the relevant Higgs boson
is still 
light, $m_{H^0}=127$ GeV.

\begin{center}
\begin{table}[h]\hspace*{25mm}
\hspace{2.5cm}
\begin{tabular}{|c|c|c|c|}
\hline
MSSM masses & \multicolumn{3}{c|}{${M_{\rm SUSY}}$ (GeV)} \\
\cline{2-4}
(GeV) & 250 & 500 & 850 \\\hline
$m_{\tilde{l}_1}$ & 175 & 415 & 734 \\
$m_{\tilde{l}_2}$ & 258 & 511 & 867 \\
$m_{\tilde{l}_3}$ & 258 & 511 & 867 \\
$m_{\tilde{l}_4}$ & 307 & 594 & 985 \\
$m_{\tilde{l}_5}$ & 309 & 607 & 1025 \\
$m_{\tilde{l}_6}$ & 323 & 609 & 1031 \\ \hline
$m_{\tilde{\nu}_1}$ & 281 & 571 & 971 \\
$m_{\tilde{\nu}_2}$ & 297 & 601 & 1022 \\
$m_{\tilde{\nu}_3}$ & 299 & 605 & 1028 \\ \hline
$m_{\tilde{\chi}_1^-}$ & 185 & 395 & 687 \\
$m_{\tilde{\chi}_2^-}$ & 379 & 679 & 1075 \\ \hline
$m_{\tilde{\chi}_1^0}$ & 99 & 207 & 362 \\
$m_{\tilde{\chi}_2^0}$ & 185 & 394 & 687 \\
$m_{\tilde{\chi}_3^0}$ & 363 & 668 & 1067 \\
$m_{\tilde{\chi}_4^0}$ & 377 & 678 & 1074 \\ \hline
$m_{h^0}$ & 110 & 119 & 123 \\
$m_{H^0}$ & 113 & 174 & 127 \\ \hline
\end{tabular} 
\caption{Relevant MSSM spectra for $M_0 = M_{1/2} = M_{\rm SUSY}$, 
$\tan{\beta} = 50$, $A_0 = 0$, $\theta_i = 0$, $\theta_{13} =
5^{\circ}$, $M_{N_i} = (10^{10}, 10^{11}, 10^{14})$ GeV, 
$\delta_1 = -1.8$ and $\delta_2 = 0$.}
\label{NUHMspectra}
\end{table}
\end{center}

In the following we present the predictions of the $\mu-e$ conversion rates in Titanium nuclei
within the  NUHM-seeesaw scenario.
First we display in Fig.~\ref{fig:CR:NUHM:M0M12:A0} the CR$(\mu-e, \text{Ti})$
as a function of $M_0 = M_{1/2} = M_{\text{SUSY}}$ and of $A_0$ for the
particular choice $\delta_1 = -1.8$ and $\delta_2 = 0$.
In order to illustrate the impact of the non-universality hypothesis 
on the conversion rates, we have separately displayed in this plot 
the various contributions from the
$\gamma$-, $Z$-, Higgs mediated penguins and box diagrams. We observe a very distinct behaviour with $M_{\text{SUSY}}$ of the Higgs-mediated 
contributions when compared to what was found for the CMSSM
(universal) case, shown in Fig.~\ref{fig:CR:several:tanb:M0}. 
In fact, for the choice 
of input parameters in this plot, the Higgs-mediated contribution can
equal, or even exceed that of the photon,
dominating the total conversion rate in the large $M_\text{SUSY}$
region. Both photon- and Higgs-mediated contributions are similar around
$M_\text{SUSY}=700$ GeV. 
These larger Higgs
contributions are the obvious consequence of  
the lighter Higgs boson mass values encountered in this region, 
as previously illustrated in Figs.~\ref{fig:MH0andmu:delta1delta2} 
and~\ref{fig:MH0andmu:M0M12}.
The non-decoupling behaviour of the SUSY
particles for the large $M_\text{SUSY}$ regime can be seen in the Higgs
contribution, and thus in the total rates for the Higgs-dominated case. 

For completeness, we have also explored other choices of $A_0$ and sign$(\mu)$.
The case of sign$(\mu) = -1$, whose numerical results are not presented here, 
does not evidence any interesting new feature. In fact, there 
is a much more reduced $\delta_1$, $\delta_2$ parameter space
allowing for the correct $SU(2)\times U(1)$ breaking. In addition, for 
sign$(\mu) = -1$ we have not found solutions 
displaying as small values of $m_{H^0}$ as in the case of
sign$(\mu) = +1$. The predicted Higgs contributions to the conversion
rates are correspondingly smaller, and therefore less interesting. 
Regarding $A_0$, the right panel in Fig.~\ref{fig:CR:NUHM:M0M12:A0} 
shows that all the contributions are essentially independent of the
value of the universal trilinear coupling, 
so that our selected value, $A_0= 0$, is in fact a good representative 
point.   

Within the  NUHM-seesaw scenario, we
have also studied the $\mu-e$ conversion rates 
for other nuclei. The case of Gold is particularly interesting 
since its present experimental bound of 
$7\times 10^{-13}$~\cite{Bertl:2001fu} is more stringent than the
present bound for Titanium ($4.3 \times
10^{-12}$~\cite{Dohmen:1993mp}). In Fig.~\ref{fig:variosnucleos} we
display the predicted $\mu-e$ conversion rates for
Al, Ti, Sr, Sb, Au and Pb, as a
function of $M_\text{SUSY}$. 
We have chosen two light, two moderate and two heavy nuclei 
and we have fixed the other parameters to those of the previously elected
non-universality reference point (with $\delta_1=-1.8$ and $\delta_2=0$).
For completeness, the values of the relevant parameters for
these nuclei, $Z_{\rm eff}$, $F_p$ and $\Gamma_{\rm capture}$, have
been collected in Table~\ref{datanuclei} and follow~\cite{Kitano:2002mt}. 
In this figure we clearly see that throughout most of the explored
$M_\text{SUSY}$ interval, 
the relative conversion rates obey the hierarchy 
CR($\mu -e$, Sb) $>$  CR($\mu -e$, Sr)  $>$  CR($\mu -e$, Ti)  $>$
CR($\mu -e$, Au) $>$  CR($\mu -e$, Pb)  $>$  CR($\mu -e$, Al), in agreement
with the generic results in~\cite{Kitano:2002mt}.
We do not find a significant difference in the large 
$M_{\rm SUSY}$ region, where 
the Higgs-contribution dominates the ratios. 
The predicted rates for Ti, Au and Pb tend to converge
whereas the corresponding curve for Al nuclei deviates slightly from the others
at large $M_{\text{SUSY}}$, but we do not consider these differences among the
predictions for the various nuclei to be relevant. 
The most important conclusion from Fig.~\ref{fig:variosnucleos}
concerns the fact that we have found predictions for Gold nuclei which, for the
input parameters in this plot, are clearly above its present experimental bound
throughout the explored $M_\text{SUSY}$ interval. 
However, it should be recalled that 
the formulae here used for these estimates come from approximations
that may not properly work for the case of very heavy nuclei.
These heavy nuclei deserve a more
dedicated and refined study.

Before proceeding with our analysis, let us briefly mention that for
the region investigated in Fig.~\ref{fig:variosnucleos}, the SUSY
contributions to the anomalous magnetic moment of the muon, 
$a_\mu=(g_\mu-2)$, range from $a_\mu^{\text{SUSY}} = 10^{-8}$ for $M_\text{SUSY}=250$ GeV to 
$a_\mu^{\text{SUSY}} = 10^{-9}$, in association with $M_\text{SUSY}=850$ GeV. The latter
values are in fair agreement with the observed excess in
$a_\mu^\text{exp}$ when compared to the SM prediction, which, at the 
3.8 $\sigma$ is given by
$a_\mu^\text{SUSY}=a_\mu^\text{exp}-a_\mu^\text{SM}= 3.32 \times 10^{-9}$ at
$3.8 \sigma$ (for
a review, see for instance~\cite{Passera:2007fk} and references therein).
\begin{center}
\begin{table}\hspace*{25mm}
\hspace{2cm}
\begin{tabular}{|c|c|c|c|}
\hline
$_Z^A$Nucleus & $Z_{\rm eff}$ 
& $F_p$ & $\Gamma_{\rm capt}{\rm (GeV)}$ \\ \hline
$_{13}^{27}$Al & 11.5 & 0.64 & $4.64079 \times 10^{-19}$ \\
$_{22}^{48}$Ti & 17.6 & 0.54 & $1.70422 \times 10^{-18}$ \\
$_{38}^{80}$Sr & 25.0 & 0.39 & $4.61842 \times 10^{-18}$ \\
$_{51}^{121}$Sb & 29.0 & 0.32 & $6.71711 \times 10^{-18}$ \\
$_{79}^{197}$Au & 33.5 & 0.16 & $8.59868 \times 10^{-18}$ \\
$_{82}^{207}$Pb & 34.0 & 0.15 & $8.84868 \times 10^{-18}$ \\ \hline
\end{tabular}
\caption{Values of $Z_{\rm eff}$, $F_p$ and $\Gamma_{\rm capt}$ for different
  nuclei, as taken from~\cite{Kitano:2002mt}.}
\label{datanuclei}
\end{table}
\end{center}

To complete our study of the $\mu-e$ conversion rates in the NUHM-seesaw 
scenario we have compared the theoretical predictions for the CR($\mu
-e$, Ti) with those for the BR($\mu \to e \gamma$). We recall that
both observables are sensitive to the same leptonic mixing given by the 
slepton mass matrix entries connecting the first and the second generation.
In the usual photon-penguin dominated case, the latter two quantities
are known to be highly correlated, and this is indeed what occurred for
the CMSSM-seesaw discussed in Sec.~\ref{CMSSM:seesaw}. In other seesaw scenarios, as for instance, SUSY-GUT seesaw~\cite{Calibbi:2006nq} or the inverse seesaw~\cite{Deppisch:2005zm}, this strong correlation still persists.
However, for some scenarios where the photon-mediated diagrams are no
longer the dominant contributions to the conversion
rates, the strong correlation between CR$(\mu-e, \text{Ti})$ and BR$(\mu \to e
\gamma)$ can be lost. For
instance, this loss of correlation has been found in the case of 
Littlest Higgs Models, as recently pointed out in~\cite{Blanke:2007db}.

We have also found an interesting loss of correlation in the present
case of the NUHM-scenario, where, as previously discussed, the
Higgs-contributions can be the dominant ones. The departure from the
strongly correlated regime for (CR($\mu -e$, Ti), BR($\mu \to e
\gamma$)) is illustrated in Fig.~\ref{fig:CR_versus_muegamma},
considering several choices of the neutrino mixing angle 
$\theta_{13}=10^\circ, 5^\circ, 1^\circ, 0.2^\circ$. 
For all plots the predictions for (CR($\mu -e$, Ti), BR($\mu
\to e \gamma$)) have been derived for several choices of the
non-universality parameter $\delta_1$, scanning over the following
interval $250 \text{ GeV} \leq M_\text{SUSY}\leq 1000 \text{ GeV}$.
In each of the panels, the predictions for 
(CR($\mu -e$, Ti), BR($\mu \to e \gamma$)) that correspond to $\delta_1 =
\delta_2 = 0$ fall upon a straight
line, which strongly supports the correlated behaviour of the two
observables in this case. As $M_\text{SUSY}$ increases within the considered
interval, (CR($\mu -e$, Ti), BR($\mu \to e \gamma$)) moves left and downwards
along the straight line due to the obvious decrease of the rate with
$M_{\text{SUSY}}$.

However a clear departure from the previous strongly correlated predictions
is found for other values of $\delta_1$, $\delta_2$. In particular, for the specific 
$\delta_1$ and 
$\delta_2$ values where, as previously shown, the Higgs contributions
dominate the $\mu-e$ conversion rates, 
the predicted (CR($\mu -e$, Ti), BR($\mu \to e \gamma$)) 
points exhibit a different behaviour, deviating from the straight line
associated with the universal case.
The separation between the correlated and uncorrelated regimes is
maximal for the $\delta_1=-1.8$, $\delta_2=0$ non-universal case, 
as can be clearly understood from our previous results. 
We find this loss of correlation  
a very promising phenomenon that could be fully explored if
future sensitivities 
of $10^{-18}$ are reached. 

Secondly, it is clear from
Fig.~\ref{fig:CR_versus_muegamma} that 
even in the most pessimistic situation of very small $\theta_{13}$, the
theoretical predictions for CR($\mu -e$, Ti), and in particular the
corresponding curved line, are well  
above the horizontal line at $10^{-18}$. 
This is quite a challenging possibility,  
since for those high values of $M_\text{SUSY}\sim 850$ GeV, 
whose predictions lie at the left end of
the curved and straight lines, 
the predicted BR($\mu \to e \gamma$) is far below the planned
$10^{-13}$ sensitivity.          
This clearly reflects that $\mu -e$ in nuclei can be a very
competitive process to study LFV within the SUSY seesaw.
 
Finally, and to summarise the most striking results 
for these NUHM-seesaw scenarios, we
plot in Fig.~\ref{fig:mu-e/muegamma} the ratio of the two predicted rates, 
CR($\mu -e$, Ti)/BR($\mu \to e \gamma$) 
as a function of $m_{H^0}$. 
Since both
observables exhibit the same dependence on $m_{N_i}$, $\theta_i$ and 
$\theta_{13}$, the consideration of this ratio of rates allows to
reduce the number of relevant parameters to $\tan \beta$, $M_{\rm SUSY}$ and
$\delta_{1,2}$. 
These last two are clearly the leading ones given that they drive the
solutions to the interesting low $m_{H^0}$ values.
In this figure, and in order to maximise the Higgs-contribution to the total
$\mu-e$ conversion rates we have again considered
the extreme $\tan \beta=50$ value. 
For consistency, we have set the remaining parameters to their 
reference values, but as we have
said, they will not play a relevant role in this study. 

Leading to this scatter plot, we have scanned in the intervals $-2 <
\delta_1 < 0$, $0<\delta_2<2$ and 250 GeV $<$ $M_{\rm SUSY}$ $<$ 1000 GeV.
The most important conclusion from this plot is that the ratio  
CR($\mu -e$, Ti)/BR($\mu \to e \gamma$) 
can deviate from the constant
prediction of $5 \times 10^{-3}$ of the universality case by 
as much as a factor of almost 10. For the scan here conducted, 
the maximum value 
of this ratio of rates is found for $\delta_1=-1.7$, $\delta_2=0.1$ 
and $M_{\text{SUSY}}=876$ GeV, and 
its size is 0.04. 

Considering larger values of $M_\text{SUSY}$ and identical intervals
for $\delta_{1,2}$ leads to somewhat similar results: one finds the same
pattern of clusters departing from the constant value of the universal
case, but the maximum value of CR($\mu -e$, Ti)/BR($\mu \to e \gamma$)
is in general smaller than the 0.04 obtained in the scan of
Fig.~\ref{fig:mu-e/muegamma}. The reason why this ratio is not improved at
larger values of $M_{\text{SUSY}}$ than 1 TeV is because the acceptable
solutions producing the proper $SU(2) \times U(1)$ breaking do not lead to
sufficiently light Higgs bosons.

Even without the knowledge
of the seesaw parameters, a measurement of CR($\mu -e$,
Ti) and BR($\mu \to e \gamma$), together with information on $\tan{\beta}$ and
the SUSY scale, may allow to shed some light into the Higgs sector.

\section{Conclusions}\label{concs}
In this work we have extensively studied the lepton flavour violating
process of $\mu - e$ conversion in nuclei, within the context of the
SUSY-seesaw. In particular, we considered two distinct scenarios, the
CMSSM-seesaw, and the NUHM-seesaw, obtained by partially relaxing the
universality conditions of the Higgs boson masses. 
Throughout our analysis, we compared our theoretical predictions
with the present experimental bounds, and with the challenging future
sensitivities. In fact, the latter may convert 
processes like CR($\mu -e$, Ti) into one of the most sensitive probes to
new physics.

We have presented here the first full one-loop computation of the $\mu-e$
conversion in nuclei, including the complete set of SUSY-loop diagrams: 
$\gamma$-mediated, $Z$- and Higgs-boson mediated penguins and 
box diagrams. We have also provided the full analytical results
working in terms of physical eigenstates (for all intervening SUSY and
Higgs particles).

For the CMSSM-seesaw, we have considered the dependence of the
conversion rates on the several parameters defining the
scenario. Choosing the well known SPS benchmark points to specify the
CMSSM parameters, we focused on the most relevant
parameters in the neutrino sector, namely on the heavy neutrino masses
($m_{N_i}$), the complex $\theta_i$ mixing angles and the still
undetermined angle of the $U_\text{MNS}$ matrix, $\theta_{13}$. 
As discussed here, the CRs exhibit a very pronounced dependence on the
previous parameters, with variations that can reach up to ten orders of
magnitude in the case of $m_{N_i}$ and up to five orders of magnitude in the cases of
$\theta_i$ and $\theta_{13}$,
for the investigated ranges. 
In turn, this strong
dependence implies that a comparison of the theoretical predictions
with the present
experimental bound allows to derive indirect upper bounds for the
unknown seesaw parameters.

We have pointed out that the highest sensitivity is found for the case
of hierarchical heavy neutrinos. In this case, the conversion rates
are essentially dependent on $m_{N_3}$ and $\theta_{1,2}$, manifesting
an extreme sensitivity to $\theta_{13}$ (for the case of vanishing 
$\theta_i$). 
In fact, the values of these parameters in the upper part
of their studied intervals, 
$10^{12}\, {\rm GeV} \leq  m_{N_3} \leq 10^{15} \, {\rm GeV}$, 
$0 \leq |\theta_{1,2}|<\pi$, $ 0\leq {\rm arg}(\theta_{1,2}) \leq \pi/2$ and 
$0^\circ \leq \theta_{13} \leq 10^\circ$
are already in conflict with the 
present upper bounds on CR($\mu -e$, Ti) and CR($\mu -e$, Au).
 
We have put special emphasis on the sensitivity of the CR($\mu -e$,Ti)
to $\theta_{13}$, given that either a measurement, or a more stringent
bound on this parameter is expected in the near
future~\cite{theta13_future}. Therefore, and once  $\theta_{13}$ is
measured, a dedicated study of the $\mu-e$ conversion
rates could provide some insight into the potentially unreachable 
heavy neutrino parameters.

In all the studied examples of the CMSSM-seesaw, the dominant
contribution to the $\mu-e$ conversion rates clearly arises from the
photon-penguins. Even though we have verified that the
Higgs contributions do indeed grow with $\tan^6 \beta$, they induce
contributions which are several orders of magnitude below those of the
photon (which grow as $\tan^2 \beta$) for all the studied interval 
$5 \leq \tan \beta \leq 50$.
A very interesting departure from this situation occurs when one
relaxes the universality condition for the Higgs soft breaking masses,
and this fuelled our interest to consider the NUHM-seesaw.

In the case of the NUHM-seesaw, we explored the
influence of the non-universality hypothesis of the
soft SUSY breaking masses $M_{H_{1,2}}$ on the $\mu-e$ conversion rates.
The $\delta_1$ and $\delta_2$ parameters that describe the departure
from universality in the Higgs sector have an important impact on the
predicted Higgs boson masses.  In particular, we have found regimes
for $\delta_{1,2}$ with very interesting phenomenological
implications, namely the possibility of a light Higgs spectrum, even
in the limit of large soft SUSY masses. As a concrete example, we
recall that for the reference choice of $\delta_1=-1.8$, $\delta_2=0$,
we find $m_{H^0}=$ 113, 174 and 127 GeV for $M_{\rm SUSY}=250, 500,
850$ GeV respectively (in turn associated with moderate, heavy and
very heavy sparticle spectra).

The distinctive NUHM-seesaw scenarios associated with light $H^0$
bosons and a relatively heavy SUSY spectra induce very interesting and
unique predictions for the $\mu-e$ conversion rates. Specifically, we
have shown that in the large $M_{\rm SUSY}=M_0=M_{1/2}$ region
(e.g. above 700 GeV), there
is a strong enhancement in the Higgs-dominated rates, leading
to a remarkable loss of correlation between the CRs 
in nuclei and the BRs of $\mu \to e \gamma$ decays.
As we aimed at illustrating in Fig.~\ref{fig:CR_versus_muegamma}, the  
departure from the linear correlation of these two observables can be
sizable.
It is worth stressing that if both these rates and $\theta_{13}$ are
measured, values of BR($\mu \to e \gamma$) and CR($\mu -e$, Ti) 
that clearly deviate from the expected
SUSY-seesaw ratio in the photon-dominated case, can provide indirect
information into the structure of the Higgs sector. 

It is also important to remark that with the expected future sensitivities, 
$\mu-e$ conversion in nuclei maybe sensitive to LFV signals that lie
beyond the reach of the future sensitivities to $\mu \to e \gamma$
decays. For example, this can occur for a heavy SUSY spectrum, and very
small values of $\theta_{13}$. 

Finally, we considered the predictions for the 
ratio CR($\mu -e$, Ti)/BR($\mu \to e \gamma$) as a 
a function of $m_{H^0}$ in NUHM-seesaw scenarios, comparing the
results with those obtained for the CMSSM-seesaw case. 
The most important conclusion to be drawn from this
study (which is presented in Fig.~\ref{fig:mu-e/muegamma}) 
is that in the NUHM-seesaw
one can observe a clear deviation from the constant prediction
of the CMSSM-seesaw by as much as a factor close to 10.
If such deviation is indeed observed, we can obtain some indirect hints
regarding the SUSY Higgs sector.

In summary, with the expected 
future sensitivities, $\mu-e$ conversion in nuclei can clearly be 
more competitive for the study of LFV in SUSY-seesaw than $\mu \to e \gamma$, and
certainly provide an important tool for the study of the Higgs sector. 

\section*{Acknowledgements}
We acknowledge S.~Antusch for his participation in the initial stages of this 
work. We are grateful to S.~Ritt for providing important information on the
experimental status of present bounds and future sensitivities 
of LFV muon decays and $\mu-e$ conversion rates in nuclei. We are also 
indebted to D.G.~Cerde\~no for valuable discussions on the NUHM scenarios.
E.~Arganda acknowledges the Spanish MEC for financial support under the grant
AP2003-3776. 
The work of A.~M.~Teixeira has been supported by the French ANR
project PHYS@COL\&COS.
This work was partially supported by the Spanish MEC under project
FPA2006-05423 and by the regional ``Comunidad de Madrid'' HEPHACOS
project. 

\appendix
\label{apendices}

\section{One-loop formulae for $\mu-e$ conversion in nuclei }
\label{apendice1}
\begin{figure}[hbtp]
  \begin{center} 
        \psfig{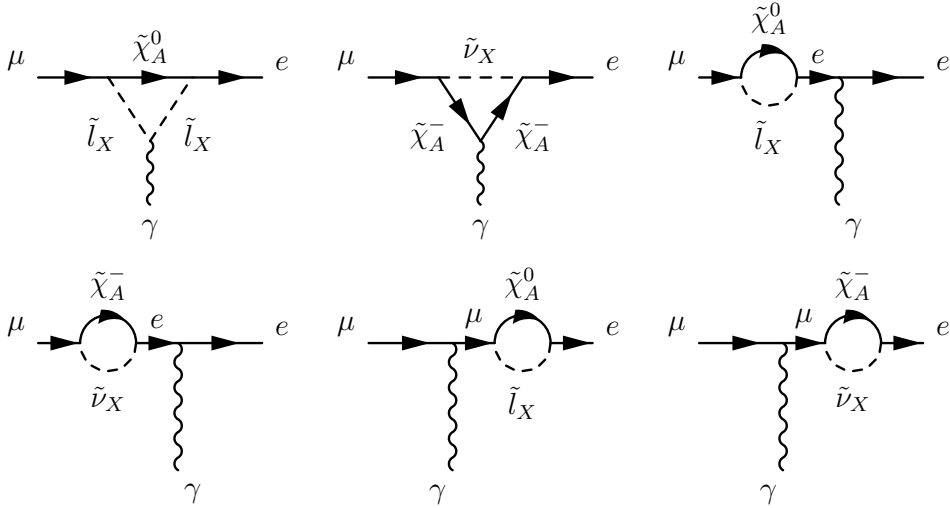}
    \caption{Relevant SUSY one-loop diagrams for the photon-mediated contributions to 
    $\mu-e$ conversion in nuclei.}
    \label{Photon_diagrams} 
  \end{center}
\end{figure}

\begin{figure}[hbtp]
  \begin{center} 
        \psfig{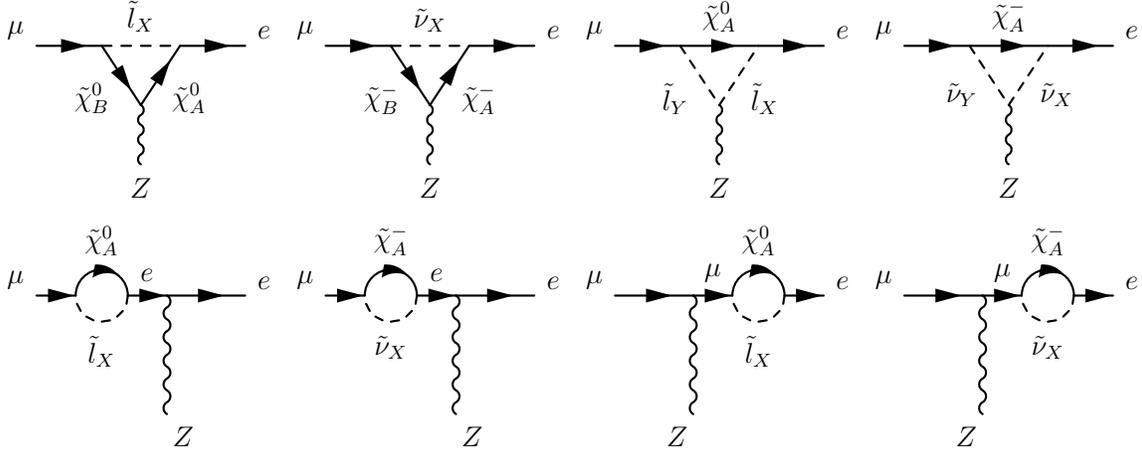}
    \caption{Relevant SUSY one-loop diagrams for the $Z$-mediated contributions to 
    $\mu-e$ conversion in nuclei
    }\label{Z_diagrams} 
  \end{center}
\end{figure}

\begin{figure}[hbtp]
  \begin{center} 
        \psfig{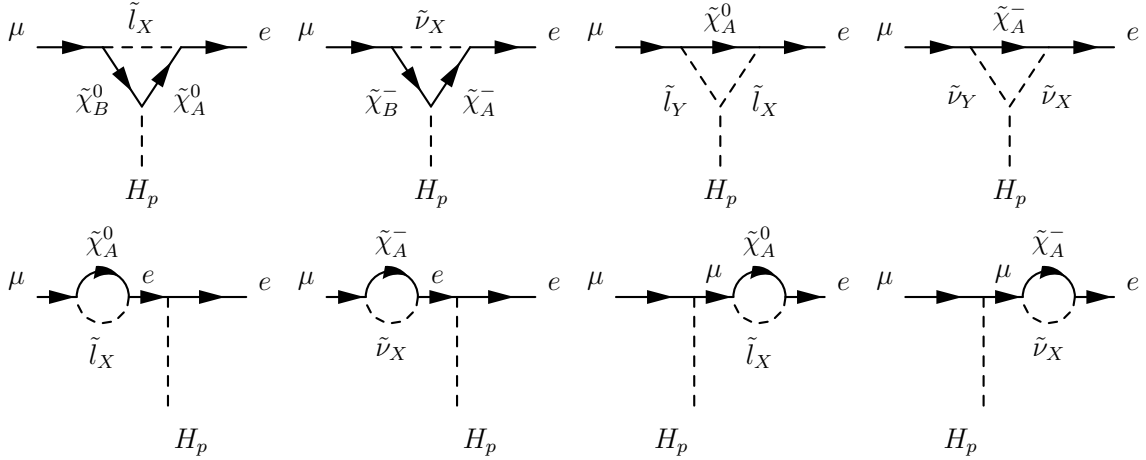}
    \caption{Relevant SUSY one-loop diagrams for the Higgs-mediated contributions to 
    $\mu-e$ conversion in nuclei.}\label{H_diagrams} 
  \end{center}
\end{figure}

\begin{figure}[hbtp]
  \begin{center} 
        \psfig{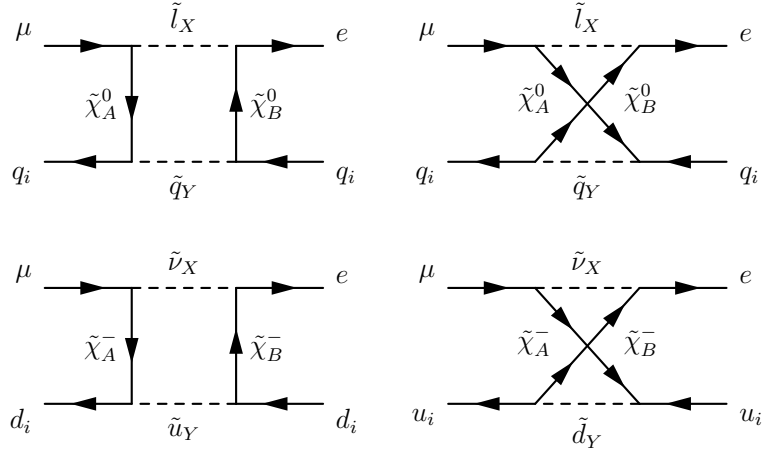}
    \caption{Box diagrams contributing to $\mu-e$ conversion in nuclei.
    }\label{Boxes} 
  \end{center}
\end{figure}

In this appendix we collect all the analytical results of the SUSY one-loop 
diagrams 
that contribute to the $\mu - e$ conversion rates in nuclei. These are 
summarised by the photon-, $Z$-boson- and Higgs-boson- penguins and box diagrams in
Figs.~\ref{Photon_diagrams},~\ref{Z_diagrams},~\ref{H_diagrams} and~\ref{Boxes},
respectively.  In the following subsections we present the relevant formulae for 
each separate contribution. All the loop functions in the formulae are taken from~\cite{Arganda:2005ji,Hollik}.

\subsection{Form factors for the $\gamma \mu e$ vertex}

Our convention for the form factors $A_{1,2}^{L,R}$ defining the $\gamma \mu e$ 
vertex is as follows:
\begin{equation}
i e \left[q^2 \gamma_\alpha (A_1^LP_L+A_1^RP_R)+
im_\mu\sigma_{\alpha\beta}q^\beta(A_a^LP_L+A_2^RP_R)\right],
\end{equation}
where $q$ is the off-shell photon momentum, $P_{L,R}=(1\mp\gamma_5)/2$, $e$ is the
electromagnetic charge and $m_\mu$ is the muon mass. 

In the SUSY-seesaw context there are one-loop contributions to these form factors that come 
from the chargino and neutralino sectors respectively,
\begin{equation}
A_a^{L,R} = A_a^{(n)L.R} + A_a^{(c)L,R}, \quad a = 1, 2 \, .
\end{equation}
The neutralino contributions are given by,
\begin{eqnarray}
A_1^{(n)L} &=& \frac{1}{576 \pi^2} N_{eAX}^R N_{\mu AX}^{R \ast} \frac{1}{m_{\tilde{l}_X}^2} \frac{2 - 9 x_{AX} + 18 x_{AX}^2 - 11 x_{AX}^3 + 6 x_{AX}^3 \log{x_{AX}}}{\left( 1 - x_{AX} \right)^4} \nonumber \\
\\
A_2^{(n)L} &=& \frac{1}{32 \pi^2} \frac{1}{m_{\tilde{l}_X}^2} \left[ N_{eAX}^L N_{\mu AX}^{L \ast} \frac{1 - 6 x_{AX} + 3 x_{AX}^2 + 2 x_{AX}^3 - 6 x_{AX}^2 \log{x_{AX}}}{6 \left( 1 - x_{AX} \right)^4} \right. \nonumber \\
&+& N_{eAX}^R N_{\mu AX}^{R \ast} \frac{m_{e}}{m_{\mu}} \frac{1 - 6 x_{AX} + 3 x_{AX}^2 + 2 x_{AX}^3 - 6 x_{AX}^2 \log{x_{AX}}}{6 \left( 1 - x_{AX} \right)^4} \nonumber \\
&+& \left. N_{eAX}^L N_{\mu AX}^{R \ast} \frac{m_{\tilde{\chi}_A^0}}{m_{\mu}} \frac{1 -
x_{AX}^2 +2 x_{AX} \log{x_{AX}}}{\left( 1 - x_{AX} \right)^3} \right], \label{A2Lneut}\\
A_a^{(n)R} &=& \left. A_a^{(n)L} \right|_{L \leftrightarrow R},\label{ARneut}
\end{eqnarray}
where $x_{AX} = m_{\tilde{\chi}_A^0}^2/m_{\tilde{l}_X}^2$ and the indices are 
$A=1,..,4$, $X=1,..,6$.

The chargino contributions are given by
\begin{eqnarray}
A_1^{(c)L} &=& -\frac{1}{576 \pi^2} C_{eAX}^R C_{\mu AX}^{R \ast} \frac{1}{m_{\tilde{\nu}_X}^2} \frac{16 - 45 x_{AX} + 36 x_{AX}^2 - 7 x_{AX}^3 + 6 (2 - 3 x_{AX}) \log{x_{AX}}}{\left( 1 - x_{AX} \right)^4}, \nonumber \\
& & \\ \cr
A_2^{(c)L} &=& -\frac{1}{32 \pi^2} \frac{1}{m_{\tilde{\nu}_X}^2} \left[ C_{eAX}^L C_{\mu AX}^{L \ast} \frac{2 + 3 x_{AX} - 6 x_{AX}^2 + x_{AX}^3 + 6 x_{AX} \log{x_{AX}}}{6 \left( 1 - x_{AX} \right)^4} \right. \nonumber \\
&+& C_{eAX}^R C_{\mu AX}^{R \ast} \frac{m_{e}}{m_{\mu}} \frac{2 + 3 x_{AX} - 6 x_{AX}^2 + x_{AX}^3 + 6 x_{AX} \log{x_{AX}}}{6 \left( 1 - x_{AX} \right)^4} \nonumber \\
&+& \left. C_{eAX}^L C_{\mu AX}^{R \ast} \frac{m_{\tilde{\chi}_A^-}}{m_{\mu}} \frac{-3 +
4 x_{AX} - x_{AX}^2 - 2 \log{x_{AX}}}{\left( 1 - x_{AX} \right)^3} \right],
\label{A2Lchar} \\
A_a^{(c)R} &=& \left. A_a^{(c)L} \right|_{L \leftrightarrow R},\label{ARchar}
\end{eqnarray}
where in this case $x_{AX} = m_{\tilde{\chi}_A^-}^2/m_{\tilde{\nu}_X}^2$ and the indices are 
$A=1,2$, $X=1,2,3$. Notice that in 
both neutralino and chargino contributions a summation over the indices 
$A$ and $X$ is understood.

\subsection{Form factors for the $Z \mu e$ vertex}
Our convention for the form factors $F_{L,R}$ defining the $Z \mu e$ 
vertex is as follows:
\begin{equation}
-i\gamma_\mu\left[F_LP_L+F_RP_R\right].
\end{equation}

The $Z$-boson form factors have also the two kinds of contributions, from neutralinos
$(n)$ and charginos $(c)$, 
\begin{equation}\label{Zformfactors}
F_{L(R)} = F_{L(R)}^{(n)} + F_{L(R)}^{(c)} \, .
\end{equation}
The results for the corresponding form factors are the following:
\begin{eqnarray}\label{Zformafactor_expressions}
F_L^{(n)} &=& -\frac{1}{16 \pi^2} \left\{ N_{eBX}^R N_{\mu AX}^{R \ast} \left[ 2 E_{BA}^{R(n)} C_{24}(m_{\tilde{l}_X}^2, m_{\tilde{\chi}_A^0}^2, m_{\tilde{\chi}_B^0}^2) - E_{BA}^{L(n)} m_{\tilde{\chi}_A^0} m_{\tilde{\chi}_B^0} C_0(m_{\tilde{l}_X}^2, m_{\tilde{\chi}_A^0}^2, m_{\tilde{\chi}_B^0}^2) \right] \right. \nonumber \\
&+& \left. N_{eAX}^R N_{\mu AY}^{R \ast} \left[ 2 Q_{XY}^{\tilde{l}} C_{24}(m_{\tilde{\chi}_A^0}^2, m_{\tilde{l}_X}^2, m_{\tilde{l}_Y}^2) \right] + N_{eAX}^R N_{\mu AX}^{R \ast} \left[ Z_L^{(l)} B_1(m_{\tilde{\chi}_A^0}^2, m_{\tilde{l}_X}^2) \right] \right\}, \\
F_R^{(n)} &=& \left. F_L^{(n)} \right|_{L \leftrightarrow R}, \\
F_L^{(c)} &=& -\frac{1}{16 \pi^2} \left\{ C_{eBX}^R C_{\mu AX}^{R \ast} \left[ 2 E_{BA}^{R(c)} C_{24}(m_{\tilde{\nu}_X}^2, m_{\tilde{\chi}_A^-}^2, m_{\tilde{\chi}_B^-}^2) - E_{BA}^{L(c)} m_{\tilde{\chi}_A^-} m_{\tilde{\chi}_B^-} C_0(m_{\tilde{\nu}_X}^2, m_{\tilde{\chi}_A^-}^2, m_{\tilde{\chi}_B^-}^2) \right] \right. \nonumber \\
&+& \left. C_{eAX}^R C_{\mu AY}^{R \ast} \left[ 2 Q_{XY}^{\tilde{\nu}} C_{24}(m_{\tilde{\chi}_A^-}^2, m_{\tilde{\nu}_X}^2, m_{\tilde{\nu}_Y}^2) \right] + C_{eAX}^R C_{\mu AX}^{R \ast} \left[ Z_L^{(l)} B_1(m_{\tilde{\chi}_A^-}^2, m_{\tilde{\nu}_X}^2) \right] \right\}, \\
F_R^{(c)} &=& \left. F_L^{(c)} \right|_{L \leftrightarrow R},
\end{eqnarray}
where again the indices are $A,B=1,..,4$, $X,Y=1,..,6$ in the contributions from the
neutralino sector and $A,B=1,2$, $X,Y=1,2,3$ in the contributions from the chargino
sector, and a summation over the various indices is understood.

\subsection{Form factors for the $H \mu e$ vertex}
Our convention for the form factors $H_{L,R}$ defining the $H \mu e$ 
vertex is as follows:
\begin{equation}
i\gamma_\mu\left[H_LP_L+H_RP_R\right].
\end{equation}

As in the previous cases, we separate the contributions from the neutralino and
chargino sectors,
\begin{equation}
H_{L(R)} = H_{L(R),n}^{(p)} + H_{L(R),c}^{(p)}.
\end{equation}

The results for the form factors are the following,
\begin{eqnarray}
H_{L, n}^{(p)} &=& -\frac{1}{16 \pi^2} \left\{ \left[ B_0(m_{\tilde{\chi}_A^0}^2, m_{\tilde{\chi}_B^0}^2) + m_{\tilde{l}_X}^2 C_0(m_{\tilde{l}_X}^2, m_{\tilde{\chi}_A^0}^2, m_{\tilde{\chi}_B^0}^2) + m_{\mu}^2 C_{12}(m_{\tilde{l}_X}^2, m_{\tilde{\chi}_A^0}^2, m_{\tilde{\chi}_B^0}^2) \right. \right. \nonumber \\
&+& \left. m_{e}^2 (C_{11} - C_{12})(m_{\tilde{l}_X}^2, m_{\tilde{\chi}_A^0}^2, m_{\tilde{\chi}_B^0}^2) \right] N_{eAX}^L D_{R, AB}^{(p)} N_{\mu BX}^{R \ast} \nonumber \\
&+& m_{e} m_{\mu} (C_{11} + C_0)(m_{\tilde{l}_X}^2, m_{\tilde{\chi}_A^0}^2, m_{\tilde{\chi}_B^0}^2) N_{eAX}^R D_{L, AB}^{(p)} N_{\mu BX}^{L \ast} \nonumber \\
&+& m_{e} m_{\tilde{\chi}_B^0} (C_{11} - C_{12} + C_0)(m_{\tilde{l}_X}^2, m_{\tilde{\chi}_A^0}^2, m_{\tilde{\chi}_B^0}^2) N_{eAX}^R D_{L, AB}^{(p)} N_{\mu BX}^{R \ast} \nonumber \\
&+& m_{\mu} m_{\tilde{\chi}_B^0} C_{12}(m_{\tilde{l}_X}^2, m_{\tilde{\chi}_A^0}^2, m_{\tilde{\chi}_B^0}^2) N_{eAX}^L D_{R, AB}^{(p)} N_{\mu BX}^{L \ast} \nonumber \\
&+& m_{e} m_{\tilde{\chi}_A^0} (C_{11} - C_{12})(m_{\tilde{l}_X}^2, m_{\tilde{\chi}_A^0}^2, m_{\tilde{\chi}_B^0}^2) N_{eAX}^R D_{R, AB}^{(p)} N_{\mu BX}^{R \ast} \nonumber \\
&+& m_{\mu} m_{\tilde{\chi}_A^0} (C_{12} + C_0)(m_{\tilde{l}_X}^2, m_{\tilde{\chi}_A^0}^2, m_{\tilde{\chi}_B^0}^2) N_{eAX}^L D_{L, AB}^{(p)} N_{\mu BX}^{L \ast} \nonumber \\
&+& m_{\tilde{\chi}_A^0} m_{\tilde{\chi}_B^0} C_0(m_{\tilde{l}_X}^2, m_{\tilde{\chi}_A^0}^2, m_{\tilde{\chi}_B^0}^2) N_{eAX}^L D_{L, AB}^{(p)} N_{\mu BX}^{R \ast} \nonumber \\
&+& G_{XY}^{(p) \tilde{l}} \left[ - m_{e} (C_{11} - C_{12})(m_{\tilde{\chi}_A^0}^2, m_{\tilde{l}_X}^2, m_{\tilde{l}_Y}^2) N_{eAX}^R N_{\mu AY}^{R \ast} \right. \nonumber \\
&-& \left. m_{\mu} C_{12}(m_{\tilde{\chi}_A^0}^2, m_{\tilde{l}_X}^2, m_{\tilde{l}_Y}^2) N_{eAX}^L N_{\mu AY}^{L \ast} + m_{\tilde{\chi}_A^0} C_0(m_{\tilde{\chi}_A^0}^2, m_{\tilde{l}_X}^2, m_{\tilde{l}_Y}^2) N_{eAX}^L N_{\mu AY}^{R \ast} \right] \nonumber \\
&+& \frac{S_{L, j}^{(p)}}{m_{e}^2 - m_{\mu}^2} \left[ - m_{e}^2 B_1(m_{\tilde{\chi}_A^0}^2, m_{\tilde{l}_X}^2) N_{eAX}^L N_{\mu AX}^{L \ast} + m_{e} m_{\tilde{\chi}_A^0} B_0(m_{\tilde{\chi}_A^0}^2, m_{\tilde{l}_X}^2) N_{eAX}^R N_{\mu AX}^{L \ast} \right. \nonumber \\
&-& \left. m_{e} m_{\mu} B_1(m_{\tilde{\chi}_A^0}^2, m_{\tilde{l}_X}^2) N_{eAX}^R N_{\mu AX}^{R \ast} + m_{\mu} m_{\tilde{\chi}_A^0} B_0(m_{\tilde{\chi}_A^0}^2, m_{\tilde{l}_X}^2) N_{eAX}^L N_{\mu AX}^{R \ast} \right] \nonumber \\
&+& \frac{S_{L, i}^{(p)}}{m_{\mu}^2 - m_{e}^2} \left[ - m_{\mu}^2 B_1(m_{\tilde{\chi}_A^0}^2, m_{\tilde{l}_X}^2) N_{eAX}^R N_{\mu AX}^{R \ast} + m_{\mu} m_{\tilde{\chi}_A^0} B_0(m_{\tilde{\chi}_A^0}^2, m_{\tilde{l}_X}^2) N_{eAX}^R N_{\mu AX}^{L \ast} \right. \nonumber \\
&-& \left. \left. m_{e} m_{\mu} B_1(m_{\tilde{\chi}_A^0}^2, m_{\tilde{l}_X}^2) N_{eAX}^L N_{\mu AX}^{L \ast} + m_{e} m_{\tilde{\chi}_A^0} B_0(m_{\tilde{\chi}_A^0}^2, m_{\tilde{l}_X}^2) N_{eAX}^L N_{\mu AX}^{R \ast} \right] \right\}, \\
H_{R, n}^{(p)} &=& \left. H_{L, n}^{(p)} \right|_{L \leftrightarrow R} \quad p = 1, 2, 3.
\end{eqnarray}
Correspondingly, the result for the chargino contribution $H_{L (R), c}^{(p)}$ can be obtained from the previous $H_{L (R), n}^{(p)}$  by replacing everywhere,
\begin{eqnarray}
\tilde{l} &\to& \tilde{\nu} \nonumber \\
\tilde{\chi}^0 &\to& \tilde{\chi}^- \nonumber \\
N^{L(R)} &\to& C^{L(R)} \nonumber \\
D_{L(R)} &\to& W_{L(R)} \nonumber
\end{eqnarray}
In the previous formulae, the index $p$ refers to the each of the Higgs bosons.
Concretely,  $H_p = h^0, H^0, A^0$ for $p = 1, 2, 3$, respectively. The other indices
are again $A,B=1,..,4$, $X,Y=1,..,6$ in the contributions from the
neutralino sector and $A,B=1,2$ and $X,Y=1,2,3$ in the contributions from the chargino
sector. A summation over all the indices is also understood.

\subsection{Contributions from box diagrams}
We follow here a simmilar notation as in the previous formulae 
for the separate contributions from the neutralino and
the chargino sectors. Our convention for the box diagrams at the quark level
is $iB_q$.    

The results for the vector contributions are the following:
\begin{eqnarray}
B_q^{(n)LV} &=& \frac{1}{16 \pi^2} \left\{ -\frac{1}{8} \tilde{D}_0(m_{\tilde{\chi}_A^0}^2, m_{\tilde{\chi}_B^0}^2, m_{\tilde{l}_X}^2, m_{\tilde{q}_Y}^2) \times \left[ N_{\mu AX}^{R(l) \ast} N_{e BX}^{R(l)} N_{q AY}^{R(q) \ast} N_{q BY}^{R(q)} \right. \right. \nonumber \\
&-& \left. N_{\mu AX}^{R(l) \ast} N_{e BX}^{R(l)} N_{q AY}^{L(q)} N_{q BY}^{L(q) \ast} \right] + \frac{1}{4} m_{\tilde{\chi}_A^0} m_{\tilde{\chi}_B^0} D_0(m_{\tilde{\chi}_A^0}^2, m_{\tilde{\chi}_B^0}^2, m_{\tilde{l}_X}^2, m_{\tilde{q}_Y}^2) \times \nonumber \\
&& \left. \left[ N_{\mu AX}^{R(l) \ast} N_{e BX}^{R(l)} N_{q AY}^{L(q)} N_{q BY}^{L(q) \ast} - N_{\mu AX}^{R(l) \ast} N_{e BX}^{R(l)} N_{q AY}^{L(q) \ast} N_{q BY}^{R(q)} \right] \right\}, \\
B_q^{(n)RV} &=& \left. B_q^{(n)LV} \right|_{L \leftrightarrow R}, \\
B_d^{(c)LV} &=& \frac{1}{16 \pi^2} \left\{ -\frac{1}{8} \tilde{D}_0(m_{\tilde{\chi}_A^-}^2, m_{\tilde{\chi}_B^-}^2, m_{\tilde{\nu}_X}^2, m_{\tilde{u}_Y}^2) C_{\mu AX}^{R(l) \ast} C_{e BX}^{R(l)} C_{d AY}^{R(d)} C_{d BY}^{R(d) \ast} \right. \nonumber \\
&+& \left. \frac{1}{4} m_{\tilde{\chi}_A^-} m_{\tilde{\chi}_B^-} D_0(m_{\tilde{\chi}_A^-}^2, m_{\tilde{\chi}_B^-}^2, m_{\tilde{\nu}_X}^2, m_{\tilde{u}_Y}^2) C_{\mu AX}^{R(l) \ast} C_{e BX}^{R(l)} C_{d AY}^{L(d)} C_{d BY}^{L(d) \ast} \right\}, \\
B_u^{(c)LV} &=& \frac{1}{16 \pi^2} \left\{ \frac{1}{8} \tilde{D}_0(m_{\tilde{\chi}_A^-}^2, m_{\tilde{\chi}_B^-}^2, m_{\tilde{\nu}_X}^2, m_{\tilde{d}_Y}^2) C_{\mu AX}^{R(l) \ast} C_{e BX}^{R(l)} C_{u AY}^{L(u) \ast} C_{u BY}^{L(u)} \right. \nonumber \\
&+& \left. \frac{1}{4} m_{\tilde{\chi}_A^-} m_{\tilde{\chi}_B^-} D_0(m_{\tilde{\chi}_A^-}^2, m_{\tilde{\chi}_B^-}^2, m_{\tilde{\nu}_X}^2, m_{\tilde{d}_Y}^2) C_{\mu AX}^{R(l) \ast} C_{e BX}^{R(l)} C_{u AY}^{R(u) \ast} C_{u BY}^{R(u)} \right\}, \\
B_{u,d}^{(c)RV} &=& \left. B_{u,d}^{(c)LV} \right|_{L \leftrightarrow R}.
\end{eqnarray}

The results for the scalar contributions are given by:
\begin{eqnarray}
B_q^{(n)LS} &=& \frac{1}{16 \pi^2} \left\{ \frac{1}{4} \tilde{D}_0(m_{\tilde{\chi}_A^0}^2, m_{\tilde{\chi}_B^0}^2, m_{\tilde{l}_X}^2, m_{\tilde{q}_Y}^2) \times \left[ N_{\mu AX}^{R(l) \ast} N_{e BX}^{L(l)} N_{q AY}^{R(q)} N_{q BY}^{L(q) \ast} \right. \right. \nonumber \\
&+& \left. N_{\mu AX}^{R(l) \ast} N_{e BX}^{L(l)} N_{q AY}^{L(q) \ast} N_{q BY}^{R(q)} \right] + \frac{1}{4} m_{\tilde{\chi}_A^0} m_{\tilde{\chi}_B^0} D_0(m_{\tilde{\chi}_A^0}^2, m_{\tilde{\chi}_B^0}^2, m_{\tilde{l}_X}^2, m_{\tilde{q}_Y}^2) \times \nonumber \\
&& \left. \left[ N_{\mu AX}^{R(l) \ast} N_{e BX}^{L(l)} N_{q AY}^{L(q)} N_{q BY}^{R(q) \ast} + N_{\mu AX}^{R(l) \ast} N_{e BX}^{L(l)} N_{q AY}^{R(q) \ast} N_{q BY}^{L(q)} \right] \right\}, \\
B_q^{(n)RS} &=& \left. B_q^{(n)LS} \right|_{L \leftrightarrow R}, \\
B_d^{(c)LS} &=& \frac{1}{16 \pi^2} \left\{ \frac{1}{4} \tilde{D}_0(m_{\tilde{\chi}_A^-}^2, m_{\tilde{\chi}_B^-}^2, m_{\tilde{\nu}_X}^2, m_{\tilde{u}_Y}^2) C_{\mu AX}^{R(l) \ast} C_{e BX}^{L(l)} C_{d AY}^{R(d)} C_{d BY}^{L(d) \ast} \right. \nonumber \\
&+& \left. \frac{1}{4} m_{\tilde{\chi}_A^-} m_{\tilde{\chi}_B^-} D_0(m_{\tilde{\chi}_A^-}^2, m_{\tilde{\chi}_B^-}^2, m_{\tilde{\nu}_X}^2, m_{\tilde{u}_Y}^2) C_{\mu AX}^{R(l) \ast} C_{e BX}^{L(l)} C_{d AY}^{L(d)} C_{d BY}^{R(d) \ast} \right\}, \\
B_u^{(c)LS} &=& \frac{1}{16 \pi^2} \left\{ \frac{1}{4} \tilde{D}_0(m_{\tilde{\chi}_A^-}^2, m_{\tilde{\chi}_B^-}^2, m_{\tilde{\nu}_X}^2, m_{\tilde{d}_Y}^2) C_{\mu AX}^{R(l) \ast} C_{e BX}^{L(l)} C_{u AY}^{L(u) \ast} C_{u BY}^{R(u)} \right. \nonumber \\
&+& \left. \frac{1}{4} m_{\tilde{\chi}_A^-} m_{\tilde{\chi}_B^-} D_0(m_{\tilde{\chi}_A^-}^2, m_{\tilde{\chi}_B^-}^2, m_{\tilde{\nu}_X}^2, m_{\tilde{d}_Y}^2) C_{\mu AX}^{R(l) \ast} C_{e BX}^{L(l)} C_{u AY}^{R(u) \ast} C_{u BY}^{L(u)} \right\}, \\
B_{u,d}^{(c)RS} &=& \left. B_{u,d}^{(c)LS} \right|_{L \leftrightarrow R}.
\end{eqnarray}
The indices in the previous formulae are again, $A,B=1,..,4$, $X,Y=1,..,6$ in the contributions from the
neutralino sector and $A,B=1,2$, $X,Y=1,2,3$ in the contributions from the chargino
sector. A summation over all the indices is also understood.

\section{Relevant couplings for $\mu-e$ conversion in nuclei}
\label{apendice2}

In this appendix we collect the formulae for the couplings 
that are relevant in this work. We follow the same notation for the couplings as 
in~\cite{Arganda:2005ji} and include here some of the formulae presented there, 
for completeness. 
The couplings are expressed in the physical 
eigenstate basis, for all the MSSM sectors involved: sleptons $\tilde{l}_X$ 
$(X=1,..,6)$, sneutrinos $\tilde{\nu}_X$ $(X=1,2,3)$, neutralinos 
$\tilde{\chi}^0_A$ $(A=1,..,4)$, charginos $\tilde{\chi}^-_A$ $(A=1,2)$ and 
the neutral Higgs bosons $H_p \,(p=1,2,3)\,=h^0, H^0, A^0$. Notice that the case 
$H_p=A^0$ is
given for completeness but, as explained in the text, it does not contribute
to the $\mu-e$ conversion in nuclei in the coherent approximation assumed in
the present work.

The notation for the SM parameters that appear in the following couplings is
as follows: 
$g$ is the $SU(2)$ gauge coupling, $m_f$ is the fermion mass, 
$m_W$, $m_Z$ are the $W$-boson and $Z$-boson masses,
respectively, and $\theta_W$ is the 
weak angle.

\subsection{Neutralino couplings}

The couplings for neutralinos that enter in the one-loop diagrams computed here are the following:
\begin{eqnarray}
N_{iAX}^{L(l)} &=& -g \sqrt{2} \left\{ \frac{m_{l_i}}{2 m_W \cos{\beta}} N_{A3}^{\ast} R_{(1, 3, 5) X}^{(l)} + \tan{\theta_W} N_{A1}^{\ast} R_{(2, 4, 6) X}^{(l)} \right\}, \\
N_{iAX}^{R(l)} &=& -g \sqrt{2} \left\{ -\frac{1}{2} \left( \tan{\theta_W} N_{A1} + N_{A2} \right) R_{(1, 3, 5) X}^{(l)} + \frac{m_{l_i}}{2 m_W \cos{\beta}} N_{A3} R_{(2, 4, 6) X}^{(l)} \right\}, \\
N_{iAX}^{L(d)} &=& -g \sqrt{2} \left\{ \frac{m_{d_i}}{2 m_W \cos{\beta}} N_{A3}^{\ast} R_{(1, 3, 5) X}^{(d)} + \frac{1}{3} \tan{\theta_W} N_{A1}^{\ast} R_{(2, 4, 6) X}^{(d)} \right\}, \\
N_{iAX}^{R(d)} &=& -g \sqrt{2} \left\{ -\frac{1}{2} \left( -\frac{1}{3} \tan{\theta_W} N_{A1} + N_{A2} \right) R_{(1, 3, 5) X}^{(d)} + \frac{m_{d_i}}{2 m_W \cos{\beta}} N_{A3} R_{(2, 4, 6) X}^{(d)} \right\}, \\
N_{iAX}^{L(u)} &=& -g \sqrt{2} \left\{ \frac{m_{u_i}}{2 m_W \sin{\beta}} N_{A4}^{\ast} R_{(1, 3, 5) X}^{(u)} - \frac{2}{3} \tan{\theta_W} N_{A1}^{\ast} R_{(2, 4, 6) X}^{(u)} \right\}, \\
N_{iAX}^{R(u)} &=& -g \sqrt{2} \left\{ -\frac{1}{2} \left( \frac{2}{3} \tan{\theta_W} N_{A1} - N_{A2} \right) R_{(1, 3, 5) X}^{(u)} + \frac{m_{u_i}}{2 m_W \sin{\beta}} N_{A4} R_{(2, 4, 6) X}^{(u)} \right\}. 
\end{eqnarray}
Here, $R^{(l)}, R^{(d)}, R^{(u)}$ are the $6\times 6$ rotation matrices for the 
charged slepton, down squark and up squark sectors, respectively, and 
$N$ is the 
$4 \times 4$ rotation matrix for the neutralino sector. 
For completeness,
we have written the
full set of couplings, including the three fermion generations. 
The displayed notation for the sfermion
rotation matrices with three entries $R_{(\,\,,\,\,,\,\,)}$ correspond with the
three generic possibilities to fermion index $i$. The fermion masses are
correspondingly, $m_{l_i}=m_e, m_\mu, m_\tau$; $m_{d_i}=m_d,m_s,m_b$ and 
$m_{u_i}=m_u,m_c,m_t$. Notice also that, although we use the same notation 
for the squark and slepton sectors, and since we have not included mixing in
the quark sector, the $6\times 6$ rotation matrices $R^{(d)}$ and  
$R^{(u)}$are block diagonal in flavour space and only $L-R$ mixing occurs in that
case.   

\subsection{Chargino couplings}
The couplings for charginos that are present in the one-loop diagrams computed here are the following:
\begin{eqnarray}
C_{iAX}^{L(l)} &=& g \frac{m_{l_i}}{\sqrt{2} m_W \cos{\beta}} U_{A2}^{\ast} R_{(1, 2, 3) X}^{(\nu)}, \\
C_{iAX}^{R(l)} &=& -g V_{A1} R_{(1, 2, 3) X}^{(\nu)}, \\
C_{iAX}^{L(d)} &=& g \frac{m_{d_i}}{\sqrt{2} m_W \cos{\beta}} U_{A2}^{\ast} R_{(1, 3, 5) X}^{(u)}, \\
C_{iAX}^{R(d)} &=& -g V_{A1} R_{(1, 3, 5) X}^{(u)} + g \frac{m_{u_i}}{\sqrt{2} m_W \sin{\beta}} V_{A2} R_{(2, 4, 6) X}^{(u)}, \\
C_{iAX}^{L(u)} &=& g \frac{m_{u_i}}{\sqrt{2} m_W \sin{\beta}} V_{A2}^{\ast} R_{(1, 3, 5) X}^{(d)}, \\
C_{iAX}^{R(u)} &=& -g U_{A1} R_{(1, 3, 5) X}^{(d)} + g \frac{m_{d_i}}{\sqrt{2} m_W \cos{\beta}} U_{A2} R_{(2, 4, 6) X}^{(d)}.
\end{eqnarray}
Here $R^{(\nu)}$ is the $3\times 3$ rotation matrix for the sneutrino sector, and 
$U$ and $V$ are the $2 \times 2$ rotation matrices in the chargino sector.  
The displayed notation for the three entries in the sfermion rotation matrices is as in
the previous neutralino couplings. The rotation matrices for neutralinos and charginos can
be found in~\cite{Haber:1984rc} and~\cite{Gunion:1984yn}.

\subsection{$Z$ boson couplings}

$Z \tilde{\chi}_A^0 \tilde{\chi}_B^0$ coupling:
\begin{eqnarray}
E_{AB}^{L(n)} &=& \frac{g}{\cos{\theta_W}} O_{AB}^{\prime \prime L} = \frac{g}{c_W} \left( -\frac{1}{2} N_{A3} N_{B3}^{\ast} + \frac{1}{2} N_{A4} N_{B4}^{\ast} \right), \\
E_{AB}^{R(n)} &=& \frac{g}{\cos{\theta_W}} O_{AB}^{\prime \prime R} = -\frac{g}{c_W} \left( -\frac{1}{2} N_{A3}^{\ast} N_{B3} + \frac{1}{2} N_{A4}^{\ast} N_{B4} \right).
\end{eqnarray}

$Z \tilde{\chi}_A^+ \tilde{\chi}_B^-$ coupling:
\begin{eqnarray}
E_{AB}^{L(c)} &=& -\frac{g}{\cos{\theta_W}} O_{AB}^{\prime R} = -\frac{g}{c_W} \left[ -\left( \frac{1}{2} - s_W^2 \right) U_{A2}^{\ast} U_{B2} - c_W^2 U_{A1}^{\ast} U_{B1} \right], \\
E_{AB}^{R(c)} &=& -\frac{g}{\cos{\theta_W}} O_{AB}^{\prime L} = -\frac{g}{c_W} \left[ -\left( \frac{1}{2} - s_W^2 \right) V_{A2} V_{B2}^{\ast} - c_W^2 V_{A1} V_{B1}^{\ast} \right].
\end{eqnarray}

$Z \tilde{l}_X \tilde{l}_Y$ coupling:
\begin{equation}
Q_{XY}^{(\tilde{l})} = -\frac{g}{c_W} \sum_{k=1}^3 \left[ \left( -\frac{1}{2} + s_W^2 \right) R_{2k-1, X}^{(l) \ast} R_{2k-1, Y}^{(l)} + s_W^2 R_{2k, X}^{(l) \ast} R_{2k, Y}^{(l)}  \right].
\end{equation}

$Z \tilde{\nu}_X \tilde{\nu}_Y$ coupling:
\begin{equation}
Q_{XY}^{(\tilde{\nu})} = -\frac{g}{2 c_W} \delta_{XY}.
\end{equation}

$Z \bar{l} l$ coupling:
\begin{eqnarray}
Z_L^{(l)} &=& -\frac{g}{c_W} \left[ -\frac{1}{2} + s_W^2 \right], \\
Z_R^{(l)} &=& -\frac{g}{c_W} s_W^2.
\end{eqnarray}

$Z \bar{q} q$ coupling:
\begin{eqnarray}
Z_L^{(q)} &=& -\frac{g}{c_W} \left[ T_3^q - Q_q s_W^2 \right], \\
Z_R^{(q)} &=& \frac{g}{c_W} Q_q s_W^2.
\end{eqnarray}
We have used here the short notation $s_W = \sin{\theta_W}$ and $c_W = \cos{\theta_W}$.

\subsection{Higgs boson couplings}

$H_p \tilde{\chi}_A^0 \tilde{\chi}_B^0$ coupling:
\begin{eqnarray}
D_{L, AB}^{(p)} &=& -\frac{g}{2 \cos{\theta_W}} \left[ \left( s_W N_{B1}^{\ast} - c_W N_{B2}^{\ast} \right) \left( \sigma_1^{(p)} N_{A3}^{\ast} + \sigma_2^{(p)} N_{A4}^{\ast} \right) \right. \nonumber \\
&+& \left. \left( s_W N_{A1}^{\ast} - c_W N_{A2}^{\ast} \right) \left( \sigma_1^{(p)} N_{B3}^{\ast} + \sigma_2^{(p)} N_{B4}^{\ast} \right) \right], \\
D_{R, AB}^{(p)} &=& D_{L, AB}^{(p) \ast}.
\end{eqnarray}

$H_p \tilde{\chi}_A^+ \tilde{\chi}_B^-$ coupling:
\begin{eqnarray}
W_{L, AB}^{(p)} &=& -\frac{g}{\sqrt{2}} \left[ -\sigma_1^{(p)} U_{B2}^\ast V_{A1}^\ast + \sigma_2^{(p)} U_{B1}^\ast V_{A2}^\ast \right], \\
W_{R, AB}^{(p)} &=& -\frac{g}{\sqrt{2}} \left[ -\sigma_1^{(p) \ast} U_{A2} V_{B1} + \sigma_2^{(p) \ast} U_{A1} V_{B2} \right].
\end{eqnarray}

$H_p \tilde{l}_X \tilde{l}_Y$ coupling:
\begin{eqnarray}
G_{XY}^{p (\tilde{l})} &=& -g \left[ g_{LL, e}^{(p)} R_{1X}^{*(l)} R_{1Y}^{(l)} + g_{RR, e}^{(p)} R_{2X}^{*(l)} R_{2Y}^{(l)} + g_{LR, e}^{(p)} R_{1X}^{*(l)} R_{2Y}^{(l)} + g_{RL, e}^{(p)} R_{2X}^{*(l)} R_{1Y}^{(l)}\right. \nonumber \\
&+& g_{LL, \mu}^{(p)} R_{3X}^{*(l)} R_{3Y}^{(l)} + g_{RR, \mu}^{(p)} R_{4X}^{*(l)} R_{4Y}^{(l)} + g_{LR, \mu}^{(p)} R_{3X}^{*(l)} R_{4Y}^{(l)} + g_{RL, \mu}^{(p)} R_{4X}^{*(l)} R_{3Y}^{(l)} \nonumber \\
&+& \left. g_{LL, \tau}^{(p)} R_{5X}^{*(l)} R_{5Y}^{(l)} + g_{RR, \tau}^{(p)} R_{6X}^{*(l)} R_{6Y}^{(l)} + g_{LR, \tau}^{(p)} R_{5X}^{*(l)} R_{6Y}^{(l)} + g_{RL, \tau}^{(p)} R_{6X}^{*(l)} R_{5Y}^{(l)} \right],
\end{eqnarray}
with
\begin{eqnarray}
g_{LL, l}^{(p)} &=&  \frac{m_Z}{\cos{\theta_W}} \sigma_3^{(p)} \left(
  \frac{1}{2}- \sin^2{\theta_W} \right) + \frac{m_{l}^2}{m_W \cos{\beta}}
\sigma_4^{(p)} \,, \\
g_{RR, l}^{(p)} &=&  \frac{m_Z}{\cos{\theta_W}} \sigma_3^{(p)} \left(
  \sin^2{\theta_W} \right) + \frac{m_{l}^2}{m_W \cos{\beta}}  \sigma_4^{(p)} \,, \\
g_{LR, l}^{(p)} &=& \left(-\sigma_1^{(p)}A_l-\sigma_2^{(p)*}\mu\right)
\frac{m_{l}}{2 m_W \cos{\beta}} \,, \\
g_{RL, l}^{(p)} &=& g_{LR, l}^{(p)*} \,,
\end{eqnarray}
with $A_l = (A_l)^{ii}/(Y_l)^{ii}$ (at the EW scale), $i = 1, 2, 3$ for $l = e, \mu, \tau$, respectively.

$H_p \tilde{\nu}_X \tilde{\nu}_Y$ coupling:
\begin{equation}
G_{XY}^{p (\tilde{\nu})} = -g \left[ g_{LL, \nu}^{(p)} R_{1X}^{*(\nu)} R_{1Y}^{(\nu)} + g_{LL, \nu}^{(p)} R_{2X}^{*(\nu)} R_{2Y}^{(\nu)} + g_{LL, \nu}^{(p)} R_{3X}^{*(\nu)} R_{3Y}^{(\nu)} \right],
\end{equation}
with
\begin{equation}
g_{LL, \nu}^{(p)} = -\frac{m_Z}{2\cos{\theta_W}} \sigma_3^{(p)}.
\end{equation}

$H_p \bar{l} l$ coupling:
\begin{eqnarray}
S_{L, l}^{(p)} &=& g \frac{m_{l_i}}{2 m_W \cos{\beta}} \sigma_1^{(p) \ast}, \\
S_{R, l}^{(p)} &=& S_{L, l}^{(p) \ast}.
\end{eqnarray}

$H_p \bar{d} d$ coupling:
\begin{eqnarray}
S_{L, d}^{(p)} &=& g \frac{m_{d_i}}{2 m_W \cos{\beta}} \sigma_1^{(p) \ast}, \\
S_{R, d}^{(p)} &=& S_{L, d}^{(p) \ast}.
\end{eqnarray}

$H_p \bar{u} u$ coupling:
\begin{eqnarray}
S_{L, u}^{(p)} &=& -g \frac{m_{u_i}}{2 m_W \sin{\beta}} \sigma_2^{(p) \ast}, \\
S_{R, u}^{(p)} &=& S_{L, u}^{(p) \ast}.
\end{eqnarray}

In all the above equations,
\begin{eqnarray}
\sigma_1^{(p)} &=& \left( \begin{array}{c} \sin{\alpha} \\ -\cos{\alpha} \\ i
    \sin{\beta} \end{array} \right) \,, \\ 
\sigma_2^{(p)} &=& \left( \begin{array}{c} \cos{\alpha} \\ \sin{\alpha} \\ -i
    \cos{\beta} \end{array} \right) \,, \\
\sigma_3^{(p)} &=& \left( \begin{array}{c} \sin{(\alpha + \beta)} \\
    -\cos{(\alpha + \beta)} \\ 0 \end{array} \right) \,, \\
\sigma_4^{(p)} &=& \left( \begin{array}{c} -\sin{\alpha} \\ \cos{\alpha} \\ 0
  \end{array} \right) \,, \\
\sigma_5^{(p)} &=& \left( \begin{array}{c} -\cos{(\beta - \alpha)} \\
    \sin{(\beta - \alpha)} \\ i \cos{2\beta} \end{array} \right) \,.
\end{eqnarray}
We have also used here the standard notation for the low-energy MSSM
soft-gaugino-mass parameters $M_{1,2}$ and the $\mu$ parameter.

%


\begin{thebibliography}{99}

\bibitem{neutrinodata}
B.~T.~Cleveland {\it et al.},
Astrophys.\ J.\  {\bf 496} (1998) 505;
W.~Hampel {\it et al.},
Phys. \ Lett.\ B {\bf 447} (1999) 127;
Q.~R.~Ahmad {\it et al.}  [SNO Collaboration],
Phys.\ Rev.\ Lett.\  {\bf 87} (2001) 071301
[arXiv:nucl-ex/0106015];
Q.~R.~Ahmad {\it et al.}  [SNO Collaboration],
Phys.\ Rev.\ Lett.\  {\bf 89} (2002) 011302
[arXiv:nucl-ex/0204009];
R.~Becker-Szendy {\it et al.},
  Nucl.\ Phys.\ Proc.\ Suppl.\  {\bf 38} (1995) 331;
Y.~Fukuda {\it et al.}  [Kamiokande Collaboration],
  Phys.\ Lett.\ B {\bf 335} (1994) 237;
Y.~Ashie {\it et al.}  [Super-Kamiokande Collaboration],
  Phys.\ Rev.\ Lett.\  {\bf 93} (2004) 101801 
  [arXiv:hep-ex/0404034];
T.~Araki {\it et al.}  [KamLAND Collaboration],
  Phys.\ Rev.\ Lett.\  {\bf 94} (2005) 081801 
  [arXiv:hep-ex/0406035];
E.~Aliu {\it et al.}  [K2K Collaboration],
  Phys.\ Rev.\ Lett.\  {\bf 94} (2005) 081802
  [arXiv:hep-ex/0411038];
T.~Araki {\it et al.}  [KamLAND Collaboration],
  Phys.\ Rev.\ Lett.\  {\bf 94}  (2005) 081801 
  [arXiv:hep-ex/0406035].

\bibitem{Yao:2006px}
  W.~M.~Yao {\it et al.}  [Particle Data Group],
  J.\ Phys.\ G {\bf 33} (2006) 1.

\bibitem{Kuno:1999jp}
  Y.~Kuno and Y.~Okada,
  Rev.\ Mod.\ Phys.\  {\bf 73} (2001) 151
  [arXiv:hep-ph/9909265].

\bibitem{seesaw:I}
P.~Minkowski,
Phys.\ Lett.\ B {\bf 67} (1977) 421;
%
M.~Gell-Mann, P.~Ramond and R.~Slansky, in {\it Complex Spinors and
  Unified Theories} eds. P.~Van.~Nieuwenhuizen and D.~Z.~Freedman,
  {\it Supergravity} (North-Holland, Amsterdam, 1979), 
  p.315 [Print-80-0576 (CERN)];
%
T.~Yanagida, in {\it Proceedings of the Workshop on the Unified Theory
and the Baryon Numebr in the Universe}, eds. O.~Sawada and
A.~Sugamoto (KEK, Tsukuba, 1979), p.95;
%
S.~L.~Glashow, in {\it Quarks and Leptons}, eds. M.~L\'evy et
al. (Plenum Press, New York, 1980), p.687;
%
R.~N.~Mohapatra and G.~Senjanovi\'c,
Phys.\ Rev.\ Lett.\  {\bf 44} (1980) 912.

\bibitem{seesaw:II}
%
R.~Barbieri, D.~V.~Nanopolous, G.~Morchio and F.~Strocchi, 
Phys.\ Lett.\ B {\bf 90} (1980) 91;
%
R.~E.~Marshak and R.~N.~Mohapatra, {\it Invited talk given at Orbis
  Scientiae, Coral Gables, Fla., Jan. 14-17, 1980}, VPI-HEP-80/02;
%
T.~P.~Cheng and L.~F.~Li,
Phys.\ Rev.\ D {\bf 22} (1980) 2860;
%
M.~Magg and C.~Wetterich,
Phys.\ Lett.\ B {\bf 94} (1980) 61;
%
  G.~Lazarides, Q.~Shafi and C.~Wetterich,
  Nucl.\ Phys.\  B {\bf 181} (1981) 287;
%
J.~Schechter and J.~W.~F.~Valle,
Phys.\ Rev.\ D {\bf 22} (1980) 2227;
%
R.~N.~Mohapatra and G.~Senjanovic,
Phys.\ Rev.\ D {\bf 23} (1981) 165.
%
E.~Ma and U.~Sarkar,
Phys.\ Rev.\ Lett.\  {\bf 80} (1998) 5716 [arXiv:hep-ph/9802445].

\bibitem{Borzumati:1986qx}
F.~Borzumati and A.~Masiero,
Phys.\ Rev.\ Lett.\  {\bf 57} (1986) 961.

\bibitem{Brooks:1999pu}
M.~L.~Brooks {\it et al.}  [MEGA Collaboration],
Phys.\ Rev.\ Lett.\  {\bf 83} (1999) 1521
[arXiv:hep-ex/9905013].

\bibitem{Bellgardt:1987du}
  U.~Bellgardt {\it et al.}  [SINDRUM Collaboration],
  Nucl.\ Phys.\ B {\bf 299} (1988) 1.

\bibitem{Dohmen:1993mp}
  C.~Dohmen {\it et al.}  [SINDRUM II Collaboration.],
  Phys.\ Lett.\  B {\bf 317} (1993) 631.

\bibitem{Bertl:2001fu}
  W.~Bertl {\it et al.},
  Eur.\ Phys.\ J.\  C {\bf 47} (2006) 337.

\bibitem{Ritt:2006cg}
  S.~Ritt  [MEG Collaboration],
  Nucl.\ Phys.\ Proc.\ Suppl.\  {\bf 162} (2006) 279.

\bibitem{Ritt:private}
  S.~Ritt, {\it private communication}.

\bibitem{PRIME}
The PRIME working group,
``Search for the $\mu-e$ Conversion Process at an Ultimate
Sensitivity of the Order of $10^{18}$ with PRISM",
unpublished;
LOI to J-PARC 50-GeV PS, LOI-25,
{\tt http://psux1.kek.jp/jhf-np/LOIlist/LOIlist.html}

\bibitem{Kane:1993td}
  G.~L.~Kane, C.~F.~Kolda, L.~Roszkowski and J.~D.~Wells,
  Phys.\ Rev.\  D {\bf 49} (1994) 6173
  [arXiv:hep-ph/9312272].


\bibitem{NUHMrefs}

  J.~R.~Ellis, K.~A.~Olive and Y.~Santoso,
  Phys.\ Lett.\  B {\bf 539} (2002) 107
  [arXiv:hep-ph/0204192];
  J.~R.~Ellis, T.~Falk, K.~A.~Olive and Y.~Santoso,
  Nucl.\ Phys.\  B {\bf 652} (2003) 259
  [arXiv:hep-ph/0210205];
  M.~Olechowski and S.~Pokorski,
  Phys.\ Lett.\  B {\bf 344} (1995) 201
  [arXiv:hep-ph/9407404];
  V.~Berezinsky, A.~Bottino, J.~R.~Ellis, N.~Fornengo, G.~Mignola and S.~Scopel,
  Astropart.\ Phys.\  {\bf 5} (1996) 1
  [arXiv:hep-ph/9508249];
  M.~Drees, M.~M.~Nojiri, D.~P.~Roy and Y.~Yamada,
  Phys.\ Rev.\  D {\bf 56} (1997) 276
  [Erratum-ibid.\  D {\bf 64} (2001) 039901]
  [arXiv:hep-ph/9701219];
  M.~Drees, Y.~G.~Kim, M.~M.~Nojiri, D.~Toya, K.~Hasuko and T.~Kobayashi,
  Phys.\ Rev.\  D {\bf 63} (2001) 035008
  [arXiv:hep-ph/0007202];
  P.~Nath and R.~Arnowitt,
  Phys.\ Rev.\  D {\bf 56} (1997) 2820
  [arXiv:hep-ph/9701301];
  J.~R.~Ellis, T.~Falk, G.~Ganis, K.~A.~Olive and M.~Schmitt,
  Phys.\ Rev.\  D {\bf 58} (1998) 095002
  [arXiv:hep-ph/9801445];
  J.~R.~Ellis, T.~Falk, G.~Ganis and K.~A.~Olive,
  Phys.\ Rev.\  D {\bf 62} (2000) 075010
  [arXiv:hep-ph/0004169];
  A.~Bottino, F.~Donato, N.~Fornengo and S.~Scopel,
  Phys.\ Rev.\  D {\bf 63} (2001) 12500
  [arXiv:hep-ph/0010203];
  S.~Profumo,
  Phys.\ Rev.\  D {\bf 68} (2003) 015006
  [arXiv:hep-ph/0304071];
  D.~G.~Cerdeno and C.~Munoz,
  JHEP {\bf 0410} (2004) 015
  [arXiv:hep-ph/0405057];
  H.~Baer, A.~Mustafayev, S.~Profumo, A.~Belyaev and X.~Tata,
  JHEP {\bf 0507} (2005) 065
  [arXiv:hep-ph/0504001];
  J.~R.~Ellis, S.~Heinemeyer, K.~A.~Olive and G.~Weiglein,
  arXiv:0706.0977 [hep-ph].


\bibitem{Hisano:1995cp}
  J.~Hisano, T.~Moroi, K.~Tobe and M.~Yamaguchi,
  Phys.\ Rev.\  D {\bf 53} (1996) 2442
  [arXiv:hep-ph/9510309].

\bibitem{Kitano:2003wn}
  R.~Kitano, M.~Koike, S.~Komine and Y.~Okada,
  Phys.\ Lett.\  B {\bf 575} (2003) 300
  [arXiv:hep-ph/0308021].

\bibitem{neutrinodata_fits}
  M.~C.~Gonz\'alez-Garcia and C.~Pe\~na-Garay,
  Phys.\ Rev.\ D {\bf 68} (2003) 093003
  [arXiv:hep-ph/0306001];
  M.~Maltoni, T.~Schwetz, M.~A.~Tortola and J.~W.~F.~Valle,
  New J.\ Phys.\  {\bf 6} (2004) 122
  [arXiv:hep-ph/0405172];
  G.~L.~Fogli, E.~Lisi, A.~Marrone and A.~Palazzo,
  Prog.\ Part.\ Nucl.\ Phys.\  {\bf 57} (2006) 742
  [arXiv:hep-ph/0506083].

\bibitem{Yaguna:2005qn}
  C.~E.~Yaguna,
  Int.\ J.\ Mod.\ Phys.\  A {\bf 21} (2006) 1283
  [arXiv:hep-ph/0502014].

\bibitem{stringent_mue}
  W.~J.~Marciano and A.~I.~Sanda,
  Phys.\ Rev.\ Lett.\  {\bf 38} (1977) 1512;
  G.~Altarelli, L.~Baulieu, N.~Cabibbo, L.~Maiani and R.~Petronzio,
  Nucl.\ Phys.\  B {\bf 125} (1977) 285
  [Erratum-ibid.\  B {\bf 130} (1977) 516];
  M.~Raidal and A.~Santamaria,
  Phys.\ Lett.\  B {\bf 421} (1998) 250
  [arXiv:hep-ph/9710389].

\bibitem{Maki:1962mu}
Z.~Maki, M.~Nakagawa and S.~Sakata,
Prog.\ Theor.\ Phys.\  {\bf 28} (1962) 870.

\bibitem{Pontecorvo:1957cp}
B.~Pontecorvo,
Sov.\ Phys.\ JETP {\bf 6} (1957) 429
[Zh.\ Eksp.\ Teor.\ Fiz.\  {\bf 33} (1957) 549];
Sov.\ Phys.\ JETP {\bf 7} (1958) 172
[Zh.\ Eksp.\ Teor.\ Fiz.\  {\bf 34} (1957) 247].

\bibitem{Casas:2001sr}
J.~A.~Casas and A.~Ibarra,
Nucl.\ Phys.\ B {\bf 618} (2001) 171
[arXiv:hep-ph/0103065].

\bibitem{nucleon_level}
  J.~D.~Vergados,
  Phys.\ Rept.\  {\bf 133} (1986) 1;
  J.~Bernabeu, E.~Nardi and D.~Tommasini,
  Nucl.\ Phys.\  B {\bf 409} (1993) 69
  [arXiv:hep-ph/9306251];
  A.~Faessler, T.~S.~Kosmas, S.~Kovalenko and J.~D.~Vergados,
  arXiv:hep-ph/9904335.

\bibitem{Kosmas:2001mv}
  T.~S.~Kosmas, S.~Kovalenko and I.~Schmidt,
  Phys.\ Lett.\  B {\bf 511} (2001) 203
  [arXiv:hep-ph/0102101].

\bibitem{Chiang:1993xz}
  H.~C.~Chiang, E.~Oset, T.~S.~Kosmas, A.~Faessler and J.~D.~Vergados,
  Nucl.\ Phys.\  A {\bf 559} (1993) 526.

\bibitem{Arganda:2004bz}
  E.~Arganda, A.~M.~Curiel, M.~J.~Herrero and D.~Temes,
  Phys.\ Rev.\  D {\bf 71} (2005) 035011
  [arXiv:hep-ph/0407302].

\bibitem{Brignole_Rossi}
  A.~Brignole and A.~Rossi,
  Phys.\ Lett.\  B {\bf 566} (2003) 217
  [arXiv:hep-ph/0304081];
  Nucl.\ Phys.\  B {\bf 701} (2004) 3
  [arXiv:hep-ph/0404211].

\bibitem{Paradisi:2005tk}
P.~Paradisi,
JHEP {\bf 0602} (2006) 050
[arXiv:hep-ph/0508054].

\bibitem{Babu:2002et}
K.~S.~Babu and C.~Kolda,
Phys.\ Rev.\ Lett.\  {\bf 89} (2002) 241802
[arXiv:hep-ph/0206310].

\bibitem{Allanach:2002nj}
  B.~C.~Allanach {\it et al.},
in {\it Proc. of the APS/DPF/DPB Summer Study on the Future of 
Particle Physics (Snowmass 2001) } ed. N.~Graf,
  Eur.\ Phys.\ J.\ C {\bf 25} (2002) 113
  [eConf {\bf C010630} (2001) P125]
  [arXiv:hep-ph/0202233].

\bibitem{Porod:2003um}
  W.~Porod,
  Comput.\ Phys.\ Commun.\  {\bf 153} (2003) 275
  [arXiv:hep-ph/0301101].

\bibitem{Arganda:2005ji}
  E.~Arganda and M.~J.~Herrero,
  Phys.\ Rev.\  D {\bf 73} (2006) 055003
  [arXiv:hep-ph/0510405].

\bibitem{Antusch:2006vw}
  S.~Antusch, E.~Arganda, M.~J.~Herrero and A.~M.~Teixeira,
  JHEP {\bf 0611} (2006) 090
  [arXiv:hep-ph/0607263].

\bibitem{Kitano:2002mt}
  R.~Kitano, M.~Koike and Y.~Okada,
  Phys.\ Rev.\  D {\bf 66} (2002) 096002
  [arXiv:hep-ph/0203110].

\bibitem{Passera:2007fk}
  M.~Passera,
  Nucl.\ Phys.\ Proc.\ Suppl.\  {\bf 169} (2007) 213
  [arXiv:hep-ph/0702027].

\bibitem{Calibbi:2006nq}
  L.~Calibbi, A.~Faccia, A.~Masiero and S.~K.~Vempati,
  arXiv:hep-ph/0605139.

\bibitem{Deppisch:2005zm}
  F.~Deppisch, T.~S.~Kosmas and J.~W.~F.~Valle,
  Nucl.\ Phys.\  B {\bf 752} (2006) 80
  [arXiv:hep-ph/0512360].

\bibitem{Blanke:2007db}
  M.~Blanke, A.~J.~Buras, B.~Duling, A.~Poschenrieder and C.~Tarantino,
  JHEP {\bf 0705} (2007) 013
  [arXiv:hep-ph/0702136].

\bibitem{theta13_future}
E. Ables et al. [MINOS Collaboration], 
Fermilab-proposal-0875; G.~S.~ Tzanakos [MINOS Collaboration], AIP
Conf. Proc. 721 (2004) 179;
  M.~Komatsu, P.~Migliozzi and F.~Terranova,
  J.\ Phys.\ G {\bf 29} (2003) 443
  [arXiv:hep-ph/0210043];
  P.~Migliozzi and F.~Terranova,
  Phys.\ Lett.\ B {\bf 563} (2003) 73
  [arXiv:hep-ph/0302274];
  P.~Huber, J.~Kopp, M.~Lindner, M.~Rolinec and W.~Winter,
  JHEP {\bf 0605} (2006) 072
  [arXiv:hep-ph/0601266];
  Y.~Itow {\it et al.},
  arXiv:hep-ex/0106019;
  A.~Blondel, A.~Cervera-Villanueva, A.~Donini, P.~Huber, M.~Mezzetto and P.~Strolin,
  arXiv:hep-ph/0606111.
  P.~Huber, M.~Lindner, M.~Rolinec and W.~Winter,
  arXiv:hep-ph/0606119;
  J.~Burguet-Castell, D.~Casper, E.~Couce, J.~J.~Gomez-Cadenas and P.~Hernandez,
  Nucl.\ Phys.\ B {\bf 725} (2005) 306
  [arXiv:hep-ph/0503021];
  J.~E.~Campagne, M.~Maltoni, M.~Mezzetto and T.~Schwetz,
  [arXiv:hep-ph/0603172].


\bibitem{Hollik}
W.~Hollik, in {\it Precision Tests of the Standard Electroweak
Model}, edited by P.~Langacker (World Scientific, Singapore,
1995), pp.~37--116;

\bibitem{Haber:1984rc}
  H.~E.~Haber and G.~L.~Kane,
  Phys.\ Rept.\  {\bf 117} (1985) 75.

\bibitem{Gunion:1984yn}
  J.~F.~Gunion and H.~E.~Haber,
  Nucl.\ Phys.\  B {\bf 272} (1986) 1
  [Erratum-ibid.\  B {\bf 402} (1993) 567].

\end{thebibliography}
\end{document}